\documentclass[%
 reprint,
superscriptaddress,
 amsmath,amssymb,
 aps,
 nofootinbib
]{revtex4-2}

\usepackage{graphicx}
\usepackage{dcolumn}
\usepackage{bm}

\usepackage{newtxtext,newtxmath}
\AtBeginDocument{\mathcode`v=\varv} 

\usepackage[T1]{fontenc}
\usepackage{ae,aecompl}

\usepackage{xcolor}
\usepackage{xspace}	    
\usepackage{caption}    
\usepackage{subcaption}
\usepackage{multirow}
\usepackage{hyperref}
\hypersetup{
    colorlinks=true,
    citecolor=blue,
    linkcolor=blue,
    filecolor=magenta,      
    urlcolor=cyan,
    }

\newcommand{\simgt}{\,\rlap{\lower 3.5 pt \hbox{$\mathchar \sim$}} \raise 1pt \hbox {$>$}\,}
\newcommand{\simlt}{\,\rlap{\lower 3.5 pt \hbox{$\mathchar \sim$}} \raise 1pt \hbox {$<$}\,}

\newcommand{\BE}{\begin{equation}}
\newcommand{\EE}{\end{equation}}
\newcommand{\BEA}{\begin{eqnarray}}
\newcommand{\EEA}{\end{eqnarray}}

\newcommand{\Ob}{\Omega_\textrm{b}}
\newcommand{\Om}{\Omega_\textrm{m}}

\newcommand{\DV}{\ifmmode{\Delta v}\else $\Delta v$\xspace\fi}

\newcommand{\HI}{\ifmmode{\textsc{hi}}\else H\textsc{i}\fi\xspace}
\newcommand{\HII}{\ifmmode{\textsc{hii}}\else H\textsc{ii}\fi\xspace}

\newcommand{\MUV}{\ifmmode{M_\textsc{uv}}\else $M_\textsc{uv}$\xspace\fi}
\newcommand{\fesc}{\ifmmode{f_\textrm{esc}}\else $f_\textrm{esc}$\xspace\fi}
\newcommand{\lya}{\ifmmode{\mathrm{Ly}\alpha}\else Ly$\alpha$\xspace\fi}

\newcommand{\Msun}{\ifmmode{M_\odot}\else $M_\odot$\xspace\fi}

\newcommand{\nh}[1][]{\ifmmode{\overline{n}_\textsc{h}^{#1}}\else $\overline{n}_\textsc{h}$\xspace\fi}

\newcommand{\xHI}{\ifmmode{x_\HI}\else $x_\HI$\xspace\fi}
\newcommand{\xHImean}{\ifmmode{\overline{x}_\HI}\else $\overline{x}_\HI$\xspace\fi}
\newcommand{\xHIImean}{\ifmmode{\overline{x}_\HII}\else $\overline{x}_\HII$\xspace\fi}
\newcommand{\trec}{\ifmmode{t_\textrm{rec}}\else $t_\textrm{rec}$\xspace\fi}
\newcommand{\clump}[1][]{\ifmmode{C_\HII^{#1}}\else $C_\HII$\xspace\fi}

\newcommand{\Nion}{\ifmmode{\dot{N}_{\mathrm{ion}}}\else $\dot{N}_\mathrm{ion}$\xspace\fi}
\newcommand{\Rion}[1][]{\ifmmode{R_\mathrm{ion}^{#1}} \else $R_\mathrm{ion}$\xspace\fi}

\newcommand{\mh}{\ifmmode{M_\mathrm{h}}\else $M_\mathrm{h}$\xspace\fi}
\newcommand{\mpiv}{\ifmmode{M_\textrm{pivot}}\else $M_\textrm{pivot}$\xspace\fi}
\newcommand{\mdotstar}{\ifmmode{\Dot{M}_\star}\else $\Dot{M}_\star$\xspace\fi}
\newcommand{\mdoth}{\ifmmode{\Dot{M}_\mathrm{h}}\else $\Dot{M}_\mathrm{h}$\xspace\fi}
\newcommand{\fstar}{\ifmmode{f_{\star}}\else $f_{\star}$\xspace\fi}
\newcommand{\fb}{\ifmmode{f_\mathrm{b}}\else $f_\mathrm{b}$\xspace\fi}
\newcommand{\astar}{\ifmmode{\alpha_\star}\else $\alpha_\star$\xspace\fi}
\newcommand{\bstar}{\ifmmode{\beta_\star}\else $\beta_\star$\xspace\fi}
\newcommand{\epsstar}{\ifmmode{\epsilon_\star}\else 
$\epsilon_\star$\xspace\fi}

\newcommand{\PS}{\ifmmode{\Delta^2_\mathrm{21}}\else $\Delta^2_\mathrm{21}$\xspace\fi}
\newcommand{\PSerrsq}{\ifmmode{\sigma_{{\Delta^2_{21}},i}^2}\else $\sigma_{{\Delta^2_{21}},i}^2$\xspace\fi}
\newcommand{\PSerr}{\ifmmode{\sigma_{{\Delta^2_{21}},i}}\else $\sigma_{{\Delta^2_{21}},i}$\xspace\fi}

\newcommand{\Tb}{\ifmmode{T_{21}}\else $T_{21}$\xspace\fi}
\newcommand{\aesc}{\ifmmode{\alpha_\mathrm{esc}}\else $\alpha_\mathrm{esc}$\xspace\fi}
\newcommand{\fescII}{\ifmmode{f_\mathrm{esc,10}^\textsc{ii}}\else $f_\mathrm{esc,10}^\textsc{ii}$\xspace\fi}
\newcommand{\fescIII}{\ifmmode{f_\mathrm{esc,7}^\textsc{ii}}\else $f_\mathrm{esc,7}^\textsc{iii}$\xspace\fi}
\newcommand{\astarII}{\ifmmode{\alpha_\star^\textsc{ii}}\else 
$\alpha_\starII$\xspace\fi}

\newcommand{\astarIII}{\ifmmode{\alpha_\star^\textsc{iii}}\else $\alpha_\star^\textsc{iii}$\xspace\fi}
\newcommand{\fstarII}{\ifmmode{f_{\star,10}^\textsc{ii}}\else $f_{\star,10}^\textsc{ii}$\xspace\fi}
\newcommand{\fstarIII}{\ifmmode{f_{\star,7}^\textsc{iii}}\else $f_{\star,7}^\textsc{iii}$\xspace\fi}
\newcommand{\tstar}{\ifmmode{t_\star}\else $t_\star$\xspace\fi}
\newcommand{\Mturn}{\ifmmode{M_\mathrm{turn}}\else $M_\mathrm{turn}$\xspace\fi}
\newcommand{\LX}{\ifmmode{L_X/{\dot{M}_\star}}\else $L_X/{\dot{M}_\star}$\xspace\fi}
\newcommand{\nuX}{\ifmmode{\nu_0}\else $\nu_0$\xspace\fi}
\newcommand{\AVCB}{\ifmmode{A_\mathrm{VCB}}\else $A_\mathrm{VCB}$\xspace\fi}
\newcommand{\ALW}{\ifmmode{A_\mathrm{LW}}\else $A_\mathrm{LW}$\xspace\fi}
\newcommand{\Mpcinv}{\ifmmode{\,\mathrm{Mpc}^{-1}}\else \,Mpc$^{-1}$\xspace\fi} 

\newcommand{\kp}{\ifmmode{k_\textrm{peak}}\else $k_\textrm{peak}$\xspace\fi}
\newcommand{\hp}{\ifmmode{h_\textrm{peak}}\else $h_\textrm{peak}$\xspace\fi}
\newcommand{\hMpc}{\ifmmode{\,h^{-1}\textrm{Mpc}}\else \,$h^{-1}$Mpc\xspace\fi}
\newcommand{\hMpcinv}{\ifmmode{\,h\,\textrm{Mpc}^{-1}}\else \,$h$\,Mpc$^{-1}$\xspace\fi}

\newcommand{\kms}{\,\ifmmode{\mathrm{km}\,\mathrm{s}^{-1}}\else km\,s${}^{-1}$\fi\xspace}
\newcommand{\cm}{\,\ifmmode{\mathrm{cm}}\else cm\fi\xspace}

\newcommand\aj{{AJ}}
\newcommand\araa{{ARA\&A}}
\newcommand\mnras{{MNRAS}}
\newcommand\pasp{{PASP}}
\newcommand\physrep{{Phys.~Rep.}}

\newcommand\jcap{{J. Cosmology Astropart. Phys.}}

\graphicspath{{./}{figures/}}
\defcitealias{PlanckCollaboration2018}{P18} 	

\begin{document}

\preprint{APS/PRD}

\title{Separating Dark Acoustic Oscillations from Astrophysics at Cosmic Dawn}

\author{Jo Verwohlt}
    \email{jo.verwohlt@nbi.ku.dk}
    \affiliation{DARK, Niels Bohr Institute, University of Copenhagen, Jagtvej 128, 2200 København N, Denmark}
\author{Charlotte A. Mason}
    \affiliation{Niels Bohr Institute, University of Copenhagen, Jagtvej 128, 2200 København N, Denmark}
    \affiliation{Cosmic Dawn Center (DAWN)}
\author{Julian B. Mu\~noz}
    \affiliation{The University of Texas at Austin, Department of Astronomy, Austin, TX 78712, USA}
\author{Francis-Yan Cyr-Racine}%
    \affiliation{Department of Physics and Astronomy, University of New Mexico, Albuquerque, New Mexico 87106, USA}
\author{Mark Vogelsberger}%
    \affiliation{Department of Physics, Kavli Institute for Astrophysics and Space Research, Massachusetts Institute of Technology, Cambridge, MA 02139, USA}
\author{Jes\'us Zavala}%
    \affiliation{Center for Astrophysics and Cosmology, Science Institute, University of Iceland, Dunhagi 5, 107 Reykjavik, Iceland}

\date{\today}

\begin{abstract}
The formation redshift and abundance of the first stars and galaxies is highly sensitive to the build up of low mass dark matter halos as well as astrophysical feedback effects which modulate star formation in these low mass halos. The 21-cm signal at cosmic dawn will depend strongly on the formation of these first luminous sources and thus can be used to constrain unknown astrophysical and dark matter properties in the early universe. In this paper, we explore how well we could measure properties of dark matter using the 21-cm power spectrum at $z>10$, given unconstrained astrophysical parameters. We create a generalizable form of the dark matter halo mass function for models with damped and/or oscillatory linear power spectra, finding a single "smooth-k" window function which describes a broad range of models including CDM. We use this to make forecasts for structure formation using the Effective Theory of Structure Formation (ETHOS) framework to explore a broad parameter space of dark matter models. We make predictions for the 21-cm power spectrum observed by HERA varying both cosmological ETHOS parameters as well as astrophysical parameters. Using a Markov Chain Monte Carlo forecast we find that the ETHOS dark matter parameters are degenerate with astrophysical parameters linked to star formation in low mass dark matter halos but not with X-ray heating produced by the first generation of stars. After marginalizing over uncertainties in astrophysical parameters we demonstrate that with just 540 days of HERA observations it should be possible to distinguish between CDM and a broad range of dark matter models with suppression at wavenumbers $k\lesssim 200\,h$Mpc$^{-1}$ assuming a moderate noise level. These results demonstrate the potential of 21-cm observations to constrain the matter power spectrum on scales smaller than current probes.

\end{abstract}

\maketitle





\section{Introduction}
\label{sec:intro}

Over the last century we have discovered that the majority of matter in our universe is dark \citep{2018RvMP...90d5002B}. Despite numerous observational constraints on the behavior of dark matter (DM) at a wide range of physical scales \citep[e.g.,][]{Viel2005,Hinshaw2012,PlanckCollaboration2018} its nature is still unknown.

Current observations find DM to be cold and collisionless at large scales.
However, an exciting possibility is that dark matter deviates from this cold dark matter (CDM) paradigm at small scales and leaves unique signatures in observations.
These have to be distinguished from baryonic effects, which are also likely to affect structure formation on small scales \citep[e.g., for a review see][]{Bullock2017, 2022NatAs...6..897S}. The existence or absence of such unique signatures could help constrain the nature of dark matter. In particular, DM which produce some kind of damping: either collisionless (free-streaming) as in Warm Dark Matter \citep[WDM, for a review see][]{Adhikari2017}; or collisional due to interactions between DM and relativistic particles (e.g.~dark radiation) in the early universe, will induce a suppression in the primordial linear matter power spectrum \citep{2002PhRvD..66h3505B, 2014PhRvD..90d3524B, 2015MNRAS.449.3587S}. 
The latter case can additionally produce 
dark acoustic oscillations (DAOs) due to the interactions between DM and dark radiation, providing a unique signature of dark-sector physics. Similar to the standard baryon acoustic oscillations, the dark matter is unable to form gravitationally bound structures until it kinematically decouples from the relativistic particles, and thus an imprint of the sound horizon of dark matter at the time of decoupling will be left on the matter density field at small scales \citep{Cyr-Racine2013}. However simulations of nonlinear structure evolution including this effects show that DAOs are gradually reduced at late times as power is ``regenerated'' at small scales \citep{Bohr2020}. Thus, high redshift observations can open a new window to search for the effect of DAOs on structure formation \citep{Schaeffer2021}.

The 21-cm cosmic dawn signal can probe non-linear structure evolution in the early universe \citep[e.g.,][]{Furlanetoo2006-review}. This signal is a net absorption (or emission) of 21-cm photons from the cosmic microwave background (CMB) by neutral hydrogen in the intergalactic medium (IGM) due to the hyperfine ``spin-flip" transition. At $z\sim30$, we expect that the neutral hydrogen is in equilibrium with the CMB, and as such its spin temperature $T_S=T_{\rm CMB}$ (a measure of the relative populations in each spin state), so there is no absorption or emission. After the first stars are formed, they emit Lyman-$\alpha$ photons that couple the spin temperature to that of the cooler gas \citep[the Wouthuysen-Field effect,][]{Wouthuysen1952, 1959ApJ...129..536, 1959ApJ...129..551F}, resulting in 21-cm absorption. The formation redshift and abundance of the first stars strongly depend on the build up of low mass dark matter halos ($\sim 10^{6-8}\Msun$). Thus the evolution of the 21-cm signal (both global and fluctuations) can be used to infer the presence of DM damping at small scales \citep[e.g.,][]{Munoz2020,Sitwell2014,Jones2021,Giri2022}.

The space of possible DM models that could produce a suppression and/or oscillations of DM power on small scales is large. To span this space, an ``effective theory of structure formation'' (ETHOS) has been proposed to map between different physical DM models and structure formation \citep{Cyr-Racine2016,Vogelsberger2016}. In particular, \citet{Bohr2020} demonstrated that the linear matter power spectrum of ETHOS models can be effectively described by two empirical parameters, which in turn can be related to physical properties of the dark matter model. \citet{Munoz2021} demonstrated that this parameterization can be used to model the effect of DM damping on the 21-cm global signal, and showed that DAOs leave a distinct signature, different from a pure suppression of WDM, even when accounting for star formation feedback which can also suppress star formation in low mass halos.

However, the impact of other astrophysical parameters on the ability to measure this signal has not yet been assessed. On top of the formation of the underlying dark matter halos, unknown astrophysics in the early universe determines the formation of the first stars and galaxies and the heating and ionization of the IGM which determines the strength of the 21-cm signal. For example: the strength of star formation feedback in low mass halos, Lyman-Werner feedback, the impact of streaming velocities between baryons and dark matter, the typical X-ray emission of early galaxies, and the escape fraction of hydrogen ionizing photons from galaxies all contribute to the strength of the 21-cm signal at $z\sim6-30$. The degeneracies between these effects have been explored in CDM cosmologies \citep[e.g.,][]{Park2019,Qin2020,Qin2021a, 2019PhRvD.100f3538M,  2013MNRAS.432.2909F} but due to computational inefficiencies there has not been a thorough investigation of the degeneracies between effects that govern star formation and dark matter models (see, however,~\citealt{Munoz2020} for a study in the context of measuring the high-$k$ power spectrum). It is thus unclear how much the astrophysical effects will hamper our efforts to understand the underlying DM physics. For example, \citet{Sitwell2014,Jones2021,Giri2022} explored the impact of WDM, fuzzy dark matter and a mixture of cold and non-cold dark matter respectively on the 21-cm signal, but did not explore the degeneracies between astrophysical parameters and the dark matter models.

In this paper, we explore how well we can measure and distinguish DM-induced damping of small-scale structure using 21-cm signals, marginalizing degeneracies with astrophysical parameters for the first time. We create a generalized form of the dark matter halo mass function (HMF) for DM models with damped and/or oscillatory linear power spectra (building upon the work of~\cite{Bohr2020}, which also works for CDM), and use this to make forecasts for structure formation in a broad DM parameter space. 
We use the effective model for the cosmic dawn 21-cm signal implemented in the public code \texttt{Zeus} \citet{Munoz2023}\footnote{\url{https://github.com/JulianBMunoz/Zeus21}} to explore the impact of DM models from the ETHOS framework on the 21-cm signal at $z\simgt10$. We find that the 21 cm signal is shifted towards later times making the redshift of absorption in the 21-cm global signal lower and the power spectrum both delayed and amplified with respect to CDM. We perform a forecast for HERA observations of the 21-cm power spectrum using an Markov Chain Monte Carlo (MCMC) approach allowing us to see the degeneracies between astrophysical and non-CDM dark matter properties for the first time. After marginalizing over the astrophysical parameters, we find that, with just 540 days of HERA observations assuming a moderate noise level, we should be able to constrain the scale at which the power spectrum is suppressed for a range of DAO models and distinguish between CDM and non-cold dark matter.

The paper is structured as follows. Section~\ref{sec:method} describes the ETHOS framework and our method for modeling halo mass functions for a broad range of DM models. Section~\ref{sec:method_zeus}
 describes our setup for computing the 21-cm signal and performing observational forecasts. We present the impact of ETHOS DM models on the 21-cm signal and forecast their detectability in Section~\ref{sec:results}. We discuss our results in Section~\ref{sec:disc} and present our conclusions in Section~\ref{sec:conc}.
Throughout we use the Planck 2018 cosmology~\cite{PlanckCollaboration2018}.

\section{Methods} 
\label{sec:method}

We provide a summary of the ETHOS framework and its generalized linear matter power spectrum in Section~\ref{sec:method_ethos}. Section~\ref{sec:method_hmf} provides an overview of how we calculate the dark matter halo mass function in the ETHOS framework, including the smooth-$k$ window function we use to obtain the matter variance, $\sigma(M)$, in ETHOS cosmologies.

\subsection{ETHOS}
\label{sec:method_ethos}

The ETHOS framework provides a mapping between DM microphysics and structure formation. It encapsulates the effects of a variety of DM models (including WDM, DAOs, and CDM) through two parameters: the amplitude \hp and the scale \kp  of the first DAO peak \citep[][see their Equation~3]{Bohr2020}. These parameters, showcased in Figure~\ref{fig:transfer}, are related to the physical size of the sound horizon and the Silk damping scale in the dark sector. ETHOS converges to CDM in the limit $\kp \rightarrow \infty$, and to WDM in the limit ($\hp \rightarrow 0$). For WDM the scale \kp results in a free-streaming cut off and \kp can be related to the WDM particle mass. 
ETHOS models with low \kp differ from CDM by suppressing lower-mass halos ($10^6-10^8$ \Msun) which is exactly where the first galaxies formed, allowing us to test them through the 21-cm signal \citep{Munoz2020}. 
Let us see how in detail.

\subsubsection{ETHOS power spectrum}
\label{sec:method_ETHOS_Pk}

\begin{figure}
\includegraphics[width=\columnwidth]{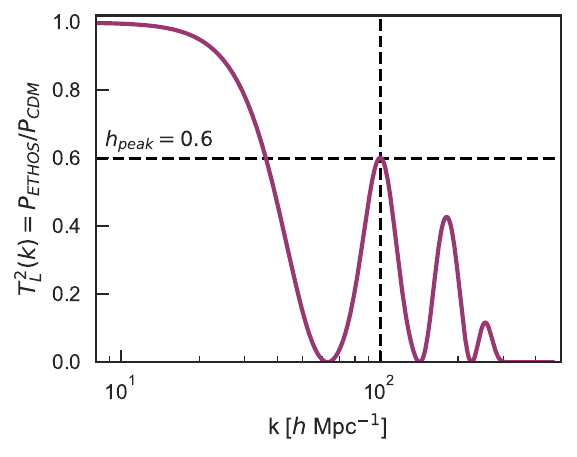}
\caption{Transfer function, defined as the ratio of the power spectrum to that of CDM, for an ETHOS scenario with parameters $\hp = 0.6$, which determines the height of the first peak at a location $\kp = 100$\,\hMpcinv.
\label{fig:transfer}}
\end{figure}

\citet{Bohr2020} demonstrated that the linear power spectrum of ETHOS models (given a cutoff with or without DAOs) can be approximated by the following transfer function (the ratio of the power spectrum to that of CDM, shown in Figure~\ref{fig:transfer}):

\BEA \label{eqn:Tk_ETHOS}
T_L(k) &=& \left[ 1+(\alpha k)^\beta \right]^\gamma - \sqrt{\hp}e^{-\frac{1}{2}\left(\frac{x-1}{\Sigma}\right)^2} \\
&& + \; \frac{\sqrt{h_2}}{4} \mathrm{erfc}{\left( \frac{x-x_0}{\tau} - 2 \right)} \nonumber \\ 
&& \times \; \mathrm{erfc}{\left(- \frac{x-x_0}{\Sigma} - 2 \right)} \cos{\left(1.1083 \pi x\right)} \nonumber
\EEA

with $x\equiv k/\kp$ and $x_0=1.805$, where $\mathrm{erfc}(y) = 1-\mathrm{erf}(y)$ is the complementary error function. 

Thus the ETHOS power spectrum is defined as:

\begin{equation} \label{eqn:ps}
P_{ETHOS}(k) = T^2_L(k)P_{CDM}(k)
\end{equation}

As described by \citet{Bohr2020} the parameters in this function describe the behavior of the power spectrum. $\alpha$ is a measure of the first cut-off scale length and is related to \kp; $\beta$ and $\gamma$ determine the shape of the cut-off; $\Sigma$ is the width of the first peak; $\tau$ determines the damping of the DAOs and
\BE \label{eqn:Tk_alpha}
\alpha = \frac{d}{\kp} \left( 2^{-1/2\gamma} - 1 \right)^{1/\beta}
\EE
\citet{Bohr2020} fit the parameters $h_2, \tau, \Sigma, \beta, d, \gamma$ for range of ETHOS model transfer functions computed using the cosmological code \texttt{CLASS} \citep{CLASS} for $\hp = 0 - 1$ in steps of 0.2, and $\kp=35-300$\,\hMpcinv with equidistant steps in $\log(\kp)$ on the intervals $[35,100]$\,\hMpcinv and $[100,300]$\,\hMpcinv.

To enable interpolation to a wider range of \hp,\kp we fit smooth functions of \hp and \kp to the parameters determined by \citet{Bohr2020}. We use the following parameterizations:
\BEA \label{eqn:Tk_params}
\eta, d, \tau &=& A e^{B\hp} + C \\
\Sigma &=& 0.2 \nonumber \\
h_2 &=& A_h(\hp) e^{B_h(\hp)\kp} + C_h(\hp) \nonumber
\EEA
where $A_h, B_h$ and $C_h$ are functions of \hp with the following parameterizations, defined such that $h_2(\hp = 0) = 0$:
\BEA \label{eqn:Tk_params_h2}
A_h &=& \frac{a e^{-\frac{(\hp - c)^2}{2b^2} }}{\sqrt{2\pi}b}  \left[1 + \mathrm{erf}\left(\frac{f (\hp - c)}{\sqrt{2}b}\right)\right]\\ 
 &-& \frac{a e^{-\frac{c^2}{2b^2} }}{\sqrt{2\pi}b}  \left[1 - \mathrm{erf}\left(\frac{fc}{\sqrt{2}b}\right)\right]\nonumber \\
B_h &=& a\left(\tanh{[b(\hp - c)]}+f\right) \nonumber \\
C_h &=& a\left(\tanh{[b(\hp - c)]}+f\right) \nonumber \\
&-& a\left(f -\tanh{bc}\right)\nonumber
\EEA
Our best-fit values of $A, B, C$ for the parameters $\eta, d, \tau$ are given in Table~\ref{tab:Tk_params1} and the values of $a, b, c, f$ which describe the functions in Equation~\eqref{eqn:Tk_params_h2} for $h_2$ are given in Table~\ref{tab:Tk_params2}.


\begin{table}
\caption{Parameters to describe the linear power spectrum with DAOs (Equations~\ref{eqn:Tk_ETHOS} and \ref{eqn:Tk_params}).}
\label{tab:Tk_params1}      
\centering       
\begin{tabular}{ cccc }
\hline\hline
$T_L(k)$ parameter & $A$ & $B$ & $C$\\
\hline\hline
$\eta$        &   $-2.1$ &   $-3.7$    & 4.1  \\
$d$&   1.8    &   $-6.7$    & 2.5  \\
$\tau$&   0.03  &   2.6     & 0.27  \\
\hline
\end{tabular}
\end{table}


\begin{table}
\caption{Parameters to describe the function $h_2$ (Equations~\ref{eqn:Tk_params} and \ref{eqn:Tk_params_h2}).}
\label{tab:Tk_params2}      
\centering       
\begin{tabular}{ ccccc }
\hline\hline
$h_2$ parameter & $a$ & $b$ & $c$ & $f$\\
\hline\hline
$A_h$  &   0.17 &   0.7  & 0.2   & $-3.0$  \\
$B_h$  &   1.0  &   3.0  & $-0.32$ & $-1.0$  \\
$C_h$  &   0.54 &   8.0   & 0.7   & 1.0  \\
\hline
\end{tabular}
\end{table}

\subsection{Halo Mass Function}
\label{sec:method_hmf}

\begin{figure*}
\includegraphics[width=\textwidth]{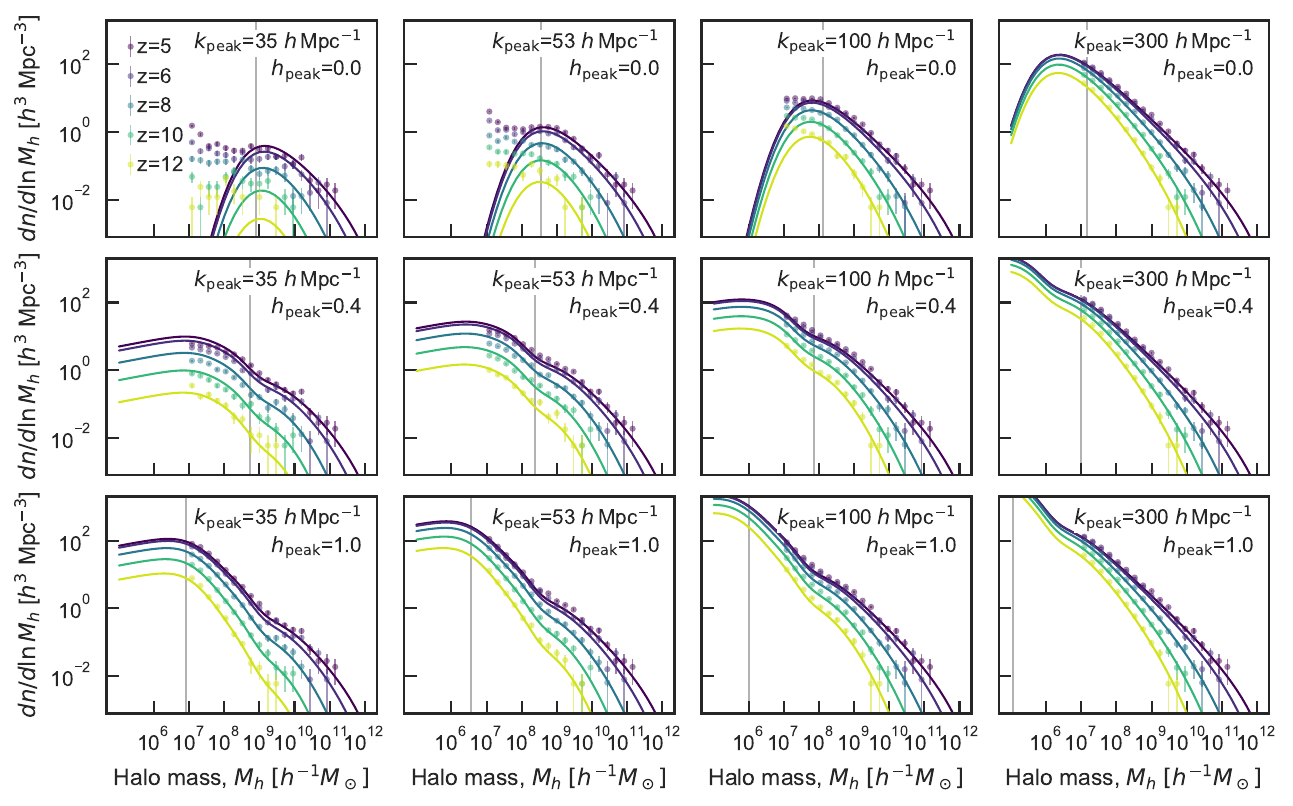}
\caption{Analytic HMFs at $z=5-12$ as a function of \hp and \kp created using the smooth-$k$ window function, with $\beta=50, c=2.8$ for WDM ($\hp=0$) and $\beta=3.6, c=3.6$ for all other DM models, as described in Sections~\ref{sec:method_hmf} and ~\ref{sec:method_W}. Points with error bars show the HMFs and their Poisson uncertainties from the ETHOS simulations of Ref.~\citep{Bohr2020}. Vertical grey lines show the mass limit due to Poisson noise in N-body simulations with suppressed power at small scales \citep{Wang2007}, for each of the plotted ETHOS models.
\label{fig:hmf}}
\end{figure*}

\begin{figure}
\includegraphics[width=\columnwidth]{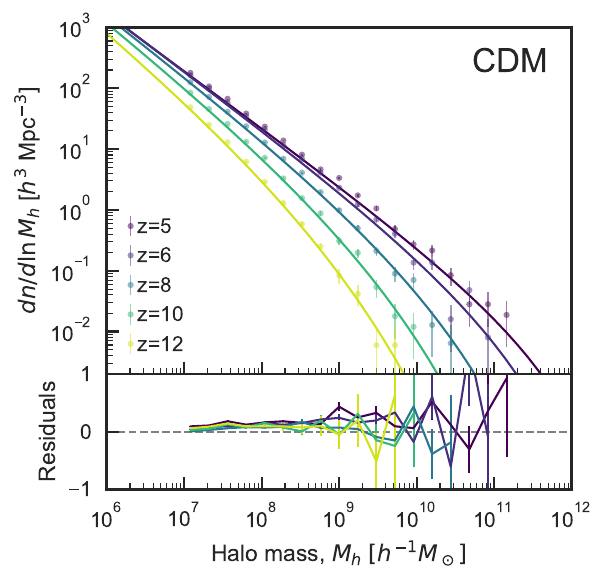}
\caption{Analytic HMFs at $z=5-12$ for CDM created using the smooth-$k$ window function with $\beta=3.6, c=3.6$ as described in Sections~\ref{sec:method_hmf} and ~\ref{sec:method_W}. Points with error bars show the HMFs and their Poisson uncertainties from the ETHOS simulations of Ref.~\citep{Bohr2020}. In the lower panels we show the relative residual between the simulated HMFs and the analytic EPS HMFs (analytic model - simulated HMF, normalised by the simulated HMF), demonstrating excellent agreement over a broad mass range.
\label{fig:hm_CDM}}
\end{figure}

Before we delve into an analysis of the 21-cm signal, we have to define the halo mass function (HMF).
Here we describe how to generate an analytic HMF using the ETHOS power spectrum, using extended Press-Schechter (EPS) formalism \citep[e.g.,][]{Sheth2001} with the smooth-$k$ window function \citep[Section~\ref{sec:method_W}][]{Leo2018}, which we find provides an excellent fit to both CDM and non-CDM halo mass functions.

Using EPS the halo mass function for halos of mass $M$ can be written as
\begin{equation} \label{eqn:EPS}
 \frac{d n }{d M} = \frac{\overline{\rho}}{M\sigma}f(\upsilon)\frac{d\sigma}{d M}
\end{equation}
where $\upsilon = [\delta_c/\sigma D(z)]^2$, with $\delta_c=1.686$ as the critical density, $\overline{\rho} = \Omega_{m,0} \rho_\mathrm{crit}$ is the mean matter density at $z=0$. $\sigma^2(M)$ is the variance of density perturbations and $f(\sigma)$ is the distribution of first-crossings: i.e. the probability of a density perturbation crossing a threshold for collapse.

The variance of density perturbations for halos of mass $M$, $\sigma^2(M)$, is given by:
\begin{equation} \label{eqn:sigma2}
    \sigma^2(M) = \int \frac{d^3\vec{k}}{(2\pi)^3} P(\vec{k}) |\widetilde{W}(kR)|^2 
\end{equation}
where $P(k)$ is the linear matter power spectrum as described above in Section~\ref{sec:method_ETHOS_Pk}, and $\widetilde{W}(\vec{k}R)$ is the Fourier transform of the window function. In the case of CDM, $\sigma^2(M)$ increases with decreasing $M$ and becomes infinite as $M\rightarrow0$, however, if free-streaming takes place and the power spectrum is suppressed at high $k$, $\sigma^2(M)$ plateaus for low $M$.

The distribution of first-crossings based on an ellipsoidal collapse barrier \citep{Sheth2001} is:
\begin{equation} \label{eqn:fsigma}
   f(\upsilon) = A \sqrt{\frac{2q\upsilon}{\pi}} \left[1 + (q\upsilon)^{-p} \right] e^{-q\upsilon/2}
\end{equation}

where $A$ is defined so that $\int d\upsilon f(\upsilon) = 1$ \citep{Sheth1999}. $A,  q$ and $p$ are free parameters usually fit to N-body simulations. We  describe our choice of these parameters in the next section.


\subsubsection{Window Function}
\label{sec:method_W}

The window function, $\widetilde{W}(kR)$, is used to define a spatial scale over which density fluctuations are smoothed (Equation~\ref{eqn:sigma2}). By comparing HMFs produced by N-body simulations and estimated from analytic theory it is possible to test how well a window function describes halo formation.

The commonly used top hat and sharp-$k$ window functions have been shown to overestimate and underestimate, respectively, the number of low mass halos in DM models with small-scale suppression of the power spectrum \citep{Leo2018}. Thus, in this work, following \citet{Leo2018,Bohr2021,Schaeffer2021}, we use the smooth-$k$ window function, which we generalize to describe a broad range of DM models. 
It is defined as:
\begin{equation} \label{eqn:W_smoothk}
    \widetilde{W}(k R) = \frac{1}{1 + \left(\frac{kR}{c}\right)^\beta}
\end{equation}
here $\beta$ controls the sharpness of the window function, with $\beta \rightarrow \infty$ converging to the sharp-$k$ window function, and $c$ rescales the size of the collapsing region (replacing the $q$ parameter in the usual Sheth-Tormen with a top hat window function).

As discussed by \citet{Schaeffer2021}, there is a lack of unique $\beta$ and $c$ parameters because $\beta$ controls simultaneously the smoothness of oscillations in the interacting dark matter models \textit{and} the low mass slope of the halo mass function for models with a cut-off in the power spectrum (i.e. WDM). For very low masses, using the smooth-$k$ window function, Equation~\eqref{eqn:EPS} can be approximated as:
\begin{equation} \label{eqn:EPS_WDM}
    \frac{d n_\textsc{wdm}}{d\ln{M}} \sim M^{(\beta-3)/3}
\end{equation}
Thus $\beta \gg 3$ is required to produce a HMF with a suppression of low mass halos as expected for e.g. WDM. Conversely, for CDM, where $P(k) \propto k^n$, with $n \approx -3$ for high $k$, the low mass-end of the halo mass function can be approximated as:
\begin{equation} \label{eqn:EPS_CDM}
    \frac{d n_\textsc{cdm}}{d\ln{M}} \sim M^{-1}
\end{equation}
Thus, as CDM has no turnover in the power spectrum at high $k$ the smooth-$k$ window function recovers the appropriate CDM low mass slope of the HMF, independently of $\beta$.

Here we aim to find a set of smooth-$k$ window function parameters, $\beta$ and $c$ which accurately fit HMFs spanning a broad range of DM models, aiming to produce a generalizable form of the halo mass function.
We fit for the window function parameters $\beta, c$ as a function of \hp and \kp, using the ETHOS HMFs produced using N-body simulations by \citet{Bohr2021}. Following \citet{Schneider2018a,Leo2018} we fix the $f(\sigma)$ parameters $q=0.707, A=0.3222, p=0.3$.

We obtain likelihoods for the simulated HMFs, $\frac{d\sigma}{d M}$, given HMFs calculated using Extended Press-Schechter theory with different smooth-$k$ window function parameters, $\phi_\mathrm{EPS}(M, \beta, c)$, on a grid of \hp and \kp. We assume $\beta$ and $c$ are redshift independent and include the Poisson uncertainty on the simulated number density in each redshift and mass bin.

As described above, a range of $\beta$ and $c$ parameters could describe HMFs, but here we aim to find a unique pair which describes HMFs spanning as broad a range of DM models as possible. There are two known edge cases which we use to guide our fitting: (1) as $\hp \rightarrow 0$, the models become WDM-like, so should be well described by a sharp-$k$ window function ($\beta \rightarrow \infty$), and (2) as $\kp \rightarrow \infty$ the models become CDM-like. 

To achieve a sharp turnover in the WDM HMFs, as seen using a sharp-$k$ window function we set $\beta=50$ for $\hp=0$, and fit for $c$ to maximize the likelihood. We find $c=2.8$. 

As described above the low-mass slope of the HMF in CDM is independent of $\beta$, therefore, for all other non-WDM cosmologies we fit for one unique set of $\beta$ and $c$ using the following procedure.
As the most distinctive feature of DAO cosmologies on the HMF are oscillations at mass scales corresponding to where the power spectra start to decline (\kp) we optimize our fits for $\beta$ and $c$ to most accurately describe these oscillations. We thus fit $\beta$ (which defines the smoothness of these oscillations) by combining likelihoods for $\hp \leq 0.6$, $\kp < 81.1\hMpcinv$, for $\hp < 0.6$, $\kp \leq 100\hMpcinv$ which is where the oscillations in the simulated ETHOS HMFs are most apparent \citep{Bohr2020}. We obtain the maximum likelihood value of $\beta=3.6$. We then fit for $c$ by fixing $\beta=3.6$ and combining the likelihoods for $c$ from \textit{all} the non-WDM HMFs, including CDM. We obtain $c=3.6$. The parameters are summarized in Table~\ref{tab:window_function}.

The analytic halo mass functions obtained using our maximum likelihood $\beta$ and $c$ are shown in Figure~\ref{fig:hmf} for ETHOS models and Figure~\ref{fig:hm_CDM} for CDM, along with HMFs from N-body simulations by \citet{Bohr2020}. We see that our one set of $\beta=3.6, c=3.6$ parameters accurately describes a very broad range of models.

\begin{table}
\caption{Smooth-$k$ window function parameters (Equation~\ref{eqn:W_smoothk})}
\label{tab:window_function}      
\centering       
\begin{tabular}{lccc }
\hline\hline
& $\hp$ & $\beta$ & $c$ \\
\hline\hline
WDM & 0 &   50      &   2.8 \\
All other DM models & $>0$ &   3.6     &   3.6 \\
\hline
\end{tabular}
\end{table}

\section{Simulating the 21-cm signal}
\label{sec:method_zeus}

Here we describe our method for generating the 21-cm signal using the \texttt{Zeus21} code. Section~\ref{sec:method_bayes} describes our Bayesian inference method for carrying out parameter forecasts and Section~\ref{sec:method_21-cmsense} describes our setup for observational forecasts with \texttt{21cmsense}.

The 21-cm signal at cosmic dawn is characterized by the spin temperature, i.e., by the relative populations of neutral hydrogen atoms in each spin states. 
The timing of the signal is determined by when the spin temperature decouples from the CMB and recouples to the temperature of the gas via the Wouthuysen effect of the Lyman-$\alpha$ flux produced by the first stars \citep{Wouthuysen1952}, which drives the signal into absorption. X-rays heat up the gas driving the signal into emission until the gas is fully ionized at the time of reionization, $z\simlt 6$. Therefore the timing of the onset of the absorption signal depends on how quickly halos big enough to host galaxies form, how efficient star formation is in the first galaxies as well as the strength of the Lyman-$\alpha$ background.

The 21-cm signal is therefore intimately linked to the star formation rate density (SFRD), which itself depends on the halo mass function (HMF) ${dn}/{d\mh}$ and the mean star formation rate (SFR) $\mdotstar(\mh)$ \citep{2007MNRAS.376.1680P},

\begin{equation} \label{eqn:SFRD}
    \overline{\mathrm{SFRD}}(z)= \int d\mh\frac{dn}{d\mh}\mdotstar(\mh),
\end{equation}
The HMF is defined in Equation \eqref{eqn:EPS}, and depends on the DM model, which modulates the abundance of halos.

We implement the ETHOS dark matter models described in Section~\ref{sec:method} into the Python package \texttt{Zeus21} to calculate the 21-cm signal for WDM and dark matter with DAOs. \texttt{Zeus21} is an analytical model for the 21-cm signal that predicts the global 21-cm signal and power spectra for CDM in a matter of seconds from the cosmic dawn ($z\sim30$) until the start of reionization ($z\sim10$, where we will stop). Here we briefly describe the astrophysical model focusing on the changes to include alternative dark matter models in \texttt{Zeus21}, and refer the reader to \citet{Munoz2023} for further details.

We implement ETHOS such that \kp and \hp are free parameters in \texttt{Zeus21}. By default, \texttt{Zeus21} uses the cosmological code \texttt{CLASS} \citep{CLASS} to calculate $\sigma(M,z)$ and $d\sigma/dM$ which is used to generate the halo mass function for CDM. Here, instead, we replace the \texttt{CLASS}-calculated $\sigma(M,z)$ and d$\sigma$/dM for non-CDM models with our calculations (see Equation \eqref{eqn:sigma2}) including the ETHOS transfer function (Equation \eqref{eqn:Tk_ETHOS}) and the smooth-$k$ window function (Equation \eqref{eqn:W_smoothk}, otherwise assumed top-hat in \texttt{CLASS} by default).
A further change is required for robustness in the case of strong suppression in the power spectrum (\hp $\approx 0$ and \kp $\lesssim 50\,\Mpcinv$), as this can produce significantly smaller values for $\sigma_M$, where the EPS approach that {\tt Zeus21} employs is expected to break down (as $\sigma_R \approx \sigma_M$). 
We note, however, that this strongly suppressed region of parameter space is likely already disfavored~\cite{Bohr2020}.
Nevertheless, we set a minimum $R_{\rm min}$ for non-linear calculations in {\tt Zeus21}, such that $\sigma_R<\sigma_{\rm max}(M,z)/3.8$. This make it possible to model all the ETHOS parameter range that we are interested in, while keeping the non-linear calculation in {\tt Zeus21} for observable scales $k\approx 0.1-1\,\Mpcinv$.
We verify that our implementation in \texttt{Zeus21} reproduces the HMFs shown in Sec.~\ref{sec:method_hmf}.

As the star-formation rate at $z>10$ is not well constrained, \texttt{Zeus21} uses a parametric model to describe the fraction of gas that is accreted by a galaxy and converted into stars. 
In particular, we assume that the SFR is

\begin{equation} \label{eqn:SFR}
    \mdotstar = \fstar \fb \mdoth,
\end{equation}
where $\mdoth$ is the mass accretion rate and $\fb$ the baryon fraction of the baryon and matter densities: $\fb=\Ob/\Om$.
Here \fstar is the star-formation efficiency, assumed to be a double-power law in halo mass \citep[see e.g.,][]{Moster2010,Behroozi2013a,Tacchella2018a,Park2019,Munoz2023}:

\begin{equation} \label{eqn:SF efficiency}
    \fstar(\mh) = \frac{2\epsstar}{(\mh/\mpiv)^{-\astar}+(\mh/\mpiv)^{-\bstar}}f_\mathrm{duty},
\end{equation}
where $\astar$, $\bstar$, $\epsstar$, and $\mpiv$ are free parameters, and $f_\mathrm{duty}=\exp\left(-M_\mathrm{turn}/\mh\right)$ parameterizes a duty fraction assumed to be unity above some turn-over mass $M_\mathrm{turn}$.
This captures the effects of stellar feedback that result in a less efficient star formation. We note that since $\mpiv \simgt 10^{11}\,\Msun$ \citep[determined by fitting to UV luminosity functions, e.g.,][]{Park2019,Sabti2021} the low mass end slope parameter, $\astar$, is most important, as halos with mass $\mh > \mpiv$ are extremely rare at $z>10$.

This would be enough to find how the ETHOS parameters affect the SFRD. The 21-cm signal, however, depends on the X-ray and UV emission from the first galaxies, which are related to the SFRD via an SED.
Thus, an essential component of the 21-cm signal is the X-ray flux produced by the first galaxies. The X-ray heating rate depends on the X-ray SED:

\begin{equation} \label{egn:Xray SED}
{\epsilon_X}(\nu)=L_{40}\times\frac{10^{40}\,\mathrm{erg}/\mathrm{s}}{M_\odot/\mathrm{yr}}\frac{I_\mathrm{X}(\nu)}{\nu}    
\end{equation}
where $I_\mathrm{X}(\nu)$ is the X-ray spectrum integrated over a band from $\nuX$ to $\nu_\mathrm{max}=2$ keV with $\nuX$ as a free parameter. $L_{40}$ is the X-ray luminosity in $10^{40}$ erg/s/$M_\odot$/yr and a free parameter in $\texttt{Zeus21}$. We assume the standard $\texttt{Zeus21}$ SED for UV light, from Ref.~\cite{Barkana2005}.

\subsection{Bayesian Inference}
\label{sec:method_bayes}

\begin{table*}
\caption{Priors for Bayesian Inference}
\label{tab:fiducial_params}      
\centering      
\begin{tabular}{llrcrcr}
\hline
\multicolumn{1}{c}{}& & \multicolumn{2}{l}{ } & \multicolumn{2}{l}{} &   \\ 
\hline
Parameter& Description & Minimum value      &    & Maximum value      &  & Fiducial \\  
\hline
\hp     & Amplitude of DAO first peak & $0.0$&  & $1.0$& & \\ 
$\log_{10}\kp $    & Position of DAO first peak [\hMpcinv] & 1.6 &  & $2.5$& &  \\ 
\hline
\astar   & SFR power law slope for $\mh<\mpiv$ & $0.0$  &   & $3.0$ & & 0.5 \\
$\log_{10}\epsstar$ & SFR normalization & $-5.0$ &   & $0.0$ & & $-1.0$ \\
\bstar  & SFR power law slope for $>\mpiv$ & $$  &   & $$ & & $-0.5$ \\
\mpiv & Halo mass at the pivot point [\Msun] & $$ &   & $$ &  & $3\times 10^{11}$  \\
\hline
$\log_{10}L_{40}/{\dot{M}_\star}$ & X-ray luminosity per SFR [$10^{40}$ erg s$^{-1}$ M$_\odot^{-1}$ yr] & $-3.0$ &  & $3.0$ & & 0.477 \\
$\log_{10}\nuX$ & Minimum X-ray energy which escapes galaxies [eV] & $2.0$ &  & $3.0$& & 2.70 \\ 
\hline

\end{tabular}
\end{table*}

To forecast the expected uncertainties that would be obtained from 21-cm observations, we use Bayesian inference to obtain a posterior distribution for our parameters of interest ($\vec{\theta}$) from our data (21-cm observables):

\begin{equation} \label{eqn: prior}
p(\vec{\theta} | \mathrm{data}) \propto \mathcal{L}(\mathrm{data} | \vec{\theta}) \times \Pi(\vec{\theta})
\end{equation}
where $\mathcal{L}(\mathrm{data} | \vec{\theta})$ is the likelihood and $\Pi(\vec{\theta})$ is the prior. 

Here we will focus on the 21-cm power spectrum, $\PS(k,z)$. We assume the likelihood in each $k$ and $z$ bin can be described as a Gaussian function of parameters $\theta$ and it is thus defined as: 
\begin{equation} \label{eqn: likelihood}
    \mathcal{L}_{i}(\PS_{,i} | \vec{\theta},\PSerrsq) = \frac{1}{\sqrt{2\pi \PSerrsq}} \exp\left[-\frac{(\PS_{,i} - \PS(\theta))^2}{2 \PSerrsq}\right]
\end{equation}
where \PS is the 21-cm power spectrum with $k$ bin $i_k$ and $z$ bin $i_z$ and \PSerr is the measurement error in each power spectrum bin. We assume each $k$ and $z$ bin is independent and that the errors between $k$ and $z$ bins are uncorrelated, following previous work, \citep[e.g.,][]{Park2019,Mason2022}. Thus the total likelihood is the product of these individual likelihoods: $\mathcal{L} = \prod_{{i_k},{i_z}} \mathcal{L}_{i}$.

We estimate \PSerr for HERA observations using {\tt 21-cmSense}\footnote{\url{https://github.com/jpober/21-cmSense}} \citep[][]{Pober2014} which we describe in Section~\ref{sec:method_21-cmsense}.

For our forecasts we vary the two cosmological ETHOS parameters \hp and \kp as well as four astrophysical parameters: \astar, \epsstar, $L_{40}$, and $\nuX$ that determine the SFR and X-ray SED. We choose a fiducial model (a set of parameters) to create mock observations, and then fit the mock observations as described above to see how well we can recover the input model. The fiducial values are listed in  Table \ref{tab:fiducial_params}. We fix the following parameters in our inference: the slope $\beta_\star$ of the bright end of the star-formation efficiency ($f_\star(M_h)$, Equation~\eqref{eqn:SF efficiency} at $-0.5$ and the pivot mass of the star formation efficiency $M_\mathrm{pivot}=3\times10^{11}\,\Msun$, since halos $M_h > M_\mathrm{pivot}$ are rare at $z>10$ these parameters do not affect the SFRD and thus our 21-cm signal forecasts significantly; and the UV SED amplitude $N_\alpha$ at $9690$ Lyman-$\alpha$ photons per baryon, as it is exactly degenerate with $f_\star$.

For our Bayesian inference forecast we use uniform priors on the varied parameters $\vec{\theta}$: 
$\hp \in (0,1)$,
$\log_{10} \kp/\hMpcinv \in (1.6,2.5)$,
$\alpha_\star \in (0,3)$, 
$\log_{10} \epsilon_\star \in (-5,0)$, 
$\log_{10} L_{40}/10^{40} $erg s$^{-1}$ M$_\odot^{-1}$ yr$\,\in (-3,3)$, 
$\log_{10} \nuX \in (2,3)$.

We use the public package \texttt{emcee}\citep{Foreman-Mackey2013} to run an Markov Chain Monte Carlo (MCMC) estimate of the posterior for the six parameters in 9000 steps of 80 walkers. In some cases some of the walkers get stuck in lower posteriors peaks. We examined the posterior chains and selected the chains with the highest posteriors based on visual inspection of the walkers. For future analysis, a nested sampler would be suggested to deal with the multimodality of the likelihood.  

We perform forecasts for four ETHOS models: Weak DAO, moderate DAO, strong DAO and WDM, to test how well these models can be constrained and distinguished each other and from CDM, after marginalizing over astrophysics. The \hp and \kp parameters for these models are given in Table~\ref{tab:fiducial_ETHOS}.
For our WDM run we set $\hp = 0.05$, to avoid numerical issues at the edge of the prior. 
The 21-cm power spectrum at $\hp = 0.05$ is very close to WDM, thus we do not expect our conclusions to change. 

We carry out forecasts using two noise models for HERA observations from {\tt 21cm-Sense}, described in Section~\ref{sec:method_21-cmsense} below.

\begin{table}
\caption{Fiducial sets of ETHOS parameters $\hp$ and $\kp$ for forecasts in Section~\ref{sec:res_forecast}}
\label{tab:fiducial_ETHOS}      
\centering       
\setlength{\tabcolsep}{8pt} 
\begin{tabular}{lccc}
\hline\
& $\hp$ & $\log_{10}$\kp $[\hMpcinv]$& \\
\hline
Weak DAO & 0.3 & 2.2 &\\
Moderate DAO & 0.5  &2.0& \\
Strong DAO & 0.8 & 1.7 &\\
WDM & 0.05 &  2.0 &\\
\hline
\end{tabular}
\end{table}

\subsection{Observational Forecasts}
\label{sec:method_21-cmsense}

To make forecasts for upcoming observations with the Hydrogen Epoch of Reionization Array (HERA) \citep{DeBoer2017, 2022ApJ...925..221A, 2024arXiv240104304B} we use the python package {\tt 21-cmSense} \citep[][]{Pober2014}. We assume two different observing times; one case of 180 days of observations and one case of 540 days of observations, each in equal bandwidths of 8\,MHz across the redshift range $z\sim5-28$ ($50-250$\,MHz), using 331 antennae.

While \texttt{zeus21} produces continuous power spectra as a function of wavenumber and redshift (see Figure \ref{fig:ps}), HERA will observe the 21-cm background in specific wave number and redshift bins. To reflect in the mock data how HERA will observe the 21-cm background, we bin the power spectra by taking the central value in the wave number and redshift bins described above. The wavenumber bins are determined by the bandwidth and the separation between the antennas \citep{DeBoer2017}. {\tt 21-cmsense} is used to calculate the wavenumber bins for the MCMC analysis using this information. 

We assume a `moderate' noise level for these observations, following \citet{Mason2022}. For this we use the `moderate' foregrounds model in {\tt 21-cmsense}, with a wedge super-horizon buffer $a=0.1\,h \Mpcinv$~\citep{2009MNRAS.398..401L, 2010ApJ...724..526D} and a HERA system temperature $T_{\rm sys} (\nu) = 100\,\mathrm K + 120 \,\mathrm K \left(\nu/150\rm MHz \right)^{-2.55}$ \citep{DeBoer2017}.

Additionally, we follow \citet{Park2019,Mason2022} and add Poisson noise (from the finite-size simulations) and a 20\% modeling error on the power spectrum. These terms are added in quadrature to the {\tt 21-cmSense} noise. We calculate the likelihood in the window $k = 0.1-1$ $h \Mpcinv$, corresponding roughly to limits on the foreground noise and the Poisson noise, respectively.

Thus we produce two final observing times -- one case assuming 540 days of observations, and one caseassuming 180 days of observations, both assuming moderate foreground noise.

\section{Results} 
\label{sec:results}

Here we describe the impact of ETHOS dark matter models on the 21-cm global signal (Section~\ref{sec:res_gs}) and power spectrum (Section~\ref{sec:res_ps}). We then describe our forecasts for the prospects of distinguishing ETHOS dark matter models from astrophysics in Section~\ref{sec:res_forecast}.

\subsection{21-cm Global Signal}
\label{sec:res_gs}

While our forecasts will not include the redshift evolution of the sky-averaged 21-cm temperature (i.e., the global signal), its behavior is enlightening so we begin by describing  it. 
Different ETHOS models result in different structure formation histories, affecting the timing of the global 21-cm signal as a function of \kp and \hp. 

\begin{figure}
\includegraphics[width=\columnwidth]{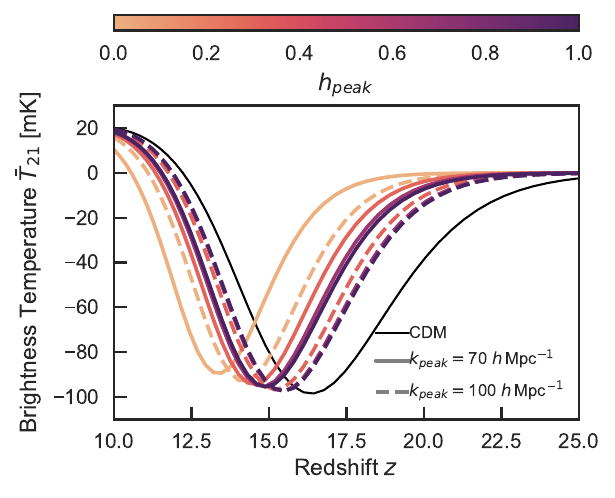}
\caption{21-cm global signal for CDM (black thin line) and ETHOS models, with $\kp=70$ and 200 $\hMpcinv$ (solid and dashed lines respectively) for a range of $\hp \in \{0-1\}$ (different colors). Models with most suppression in the power spectrum at small scales (low \kp and low \hp) have significantly delayed global signals relative to CDM.
\label{fig:gs}}
\end{figure}

\begin{figure}
\includegraphics[width=\columnwidth]{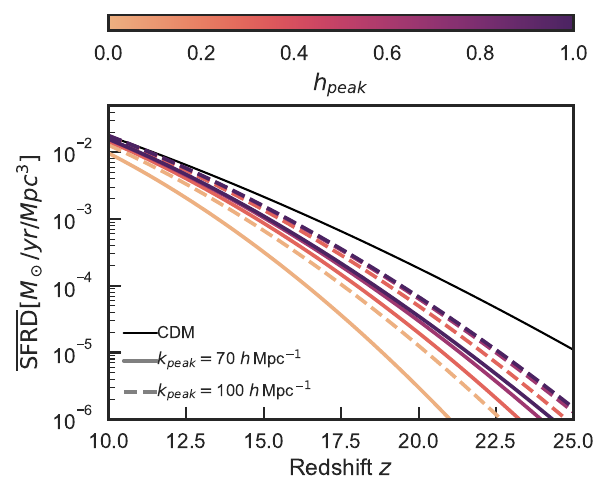}
\caption{Average star formation rate density (SFRD) for CDM (black thin line) and ETHOS models  with \kp=70 and 200 $h$Mpc$^{-1}$ (solid and dashed lines respectively) for a range of \hp (colored lines). ETHOS models result in a delayed onset of the SFRD relative to CDM.
\label{fig:sfrd}}
\end{figure}

Figure~\ref{fig:gs} shows the redshift evolution of the global signal for a range of ETHOS models. We compare models at two different $\kp=[70, 100]$ \hMpcinv for a range of \hp.
We find that models with lower \kp have a delayed global signal compared to CDM, due to the suppression in the abundance of low mass halos, thus reducing the Lyman-$\alpha$ and X-ray flux at early times. In Figure~\ref{fig:sfrd} we plot the star formation rate density (SFRD) for these models where the impact of this suppression is clear in the later build up of the SFRD.

Models with lower \hp at fixed \kp have more delayed global signals due to the enhanced suppression in those models. WDM ($\hp=0$) at fixed \kp has the most delayed global signal caused by the strong suppression of low mass halos. 

The signal also happens more rapidly as a function of redshift compared to CDM as low mass halos are suppressed meaning there isless of an early `tail' to the Lyman-$\alpha$ background. As seen in Figure~\ref{fig:sfrd} once high mass halos are formed the SFRD builds up quickly, resulting in a more rapid period of Lyman-$\alpha$ and X-ray heating compared to CDM.

\subsection{21-cm Power Spectrum}
\label{sec:res_ps}

\begin{figure}
\includegraphics[width=\columnwidth]{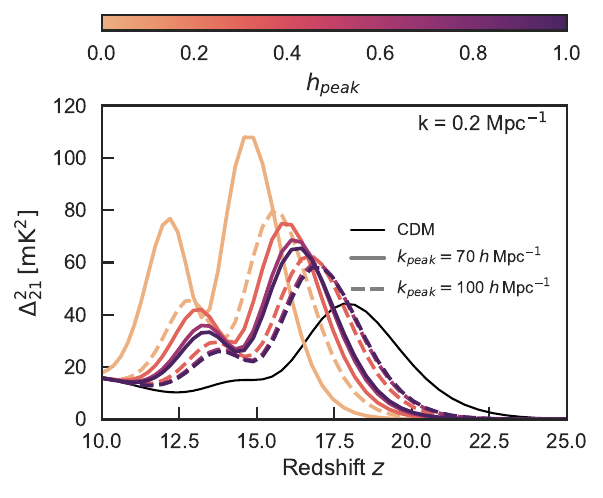}
\caption{21-cm power spectrum for CDM (black line) and ETHOS models at $k=0.2$ Mpc$^{-1}$, with \kp= 70 and 200 $h$Mpc$^{-1}$ (solid and dashed lines respectively) for a range of \hp (colored lines). ETHOS models result in a delayed power spectrum with stronger amplitude and stronger separation between the two peaks (corresponding to periods of Lyman-$\alpha$ and X-ray heating). This is due to the delayed onset, but faster build-up, of the SFRD, relative to CDM.
\label{fig:ps}}
\end{figure}

Our main analysis will be of the 21-cm fluctuations, parameterized through their Fourier-space two-point function: the 21-cm power spectrum.
Figure~\ref{fig:ps} shows the 21-cm power spectrum $\PS$ (technically ``reduced'', with units of mK$^2$) for a range of ETHOS models as a function of redshift at fixed wavenumber $k=0.2$\,\Mpcinv. We see a similar trend to the global signal above where models with low \kp and \hp have a more delayed 21-cm signal. The amplitude of the fluctuations is also stronger for the ETHOS models compared to CDM. At fixed \kp the amplitude of the fluctuations becomes stronger as \hp decreases with WDM ($\hp=0$) resulting in the highest amplitude. The increased amplitude is because the suppression of small-scale structure means that low density regions of the matter field have a weaker 21-cm signal compared to CDM. This thus increases the contrast in the 21-cm signal between low and high density regions and therefore boosts the amplitude of the power spectrum.
Such a boost, at lower $z$ makes WDM models potentially easier to detect than their CDM counterparts (akin to the fuzzy DM case in~\cite{Sarkar:2022dvl}).

Models with strongest power spectrum suppression ($\hp < 0.5, \kp < 100\,\hMpcinv$) also have a more defined separation in the two peaks in the power spectrum compared to CDM. This is a result of a later, but more rapid build-up of the SFRD (see Figure~\ref{fig:sfrd}). The two peaks come from when the signal is dominated by Lyman-$\alpha$ and X-ray fluxes respectively, i.e. as the global signal goes into absorption and again when the global signal starts to increase (note that we assume the gas is neutral all the way to $z\sim10$, so there is no impact of reionization). The delayed structure formation in ETHOS results in a delayed onset but more rapid build-up of Lyman-$\alpha$ and X-ray backgrounds, which increases the contrast between the two peaks. This in contrast to CDM where halos are formed earlier, such that X-ray heating happens at early times and resulting in the Lyman-$\alpha$ and X-ray heating epochs overlapping.

\subsection{HERA forecasts}
\label{sec:res_forecast}

To forecast how well upcoming 21-cm experiments will be able to constrain astrophysics and dark matter properties we perform an MCMC forecast using the 21-cm power spectrum, as described in Section~\ref{sec:method_bayes}. We make forecasts with HERA \citep{DeBoer2017} assuming moderate foreground contamination (see Section~\ref{sec:method_21-cmsense}). 

\begin{figure*}[t!]
\includegraphics[width=\textwidth]{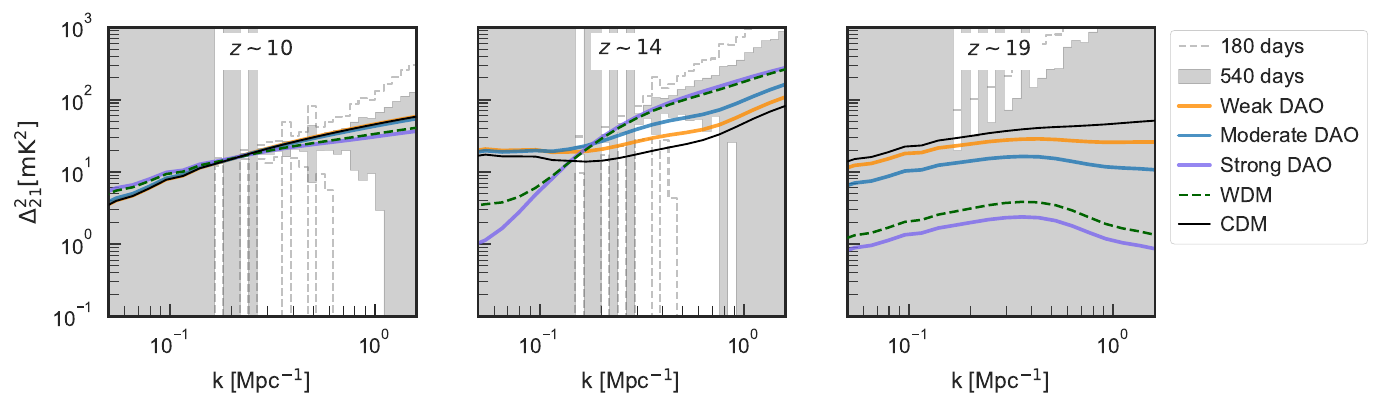}
\caption{21-cm power spectrum for CDM (black line) and the fiducial ETHOS model with $\kp=70$ and $200 \,h$Mpc$^{-1}$ at different redshifts $z=10, 14, 19$. The plot shows both the signal for 180 days of observations (dashed line) and for 540 days of observations (shaded region).
\label{fig:PS_bins}}
\end{figure*}

We examine the four cases of ETHOS dark matter as listed in Table~\ref{tab:fiducial_ETHOS}, corresponding to WDM, weak, moderate and strong DAOs. 
We show the 21-cm power spectra as a function of wavenumber for these models in three redshift bins in Figure~\ref{fig:PS_bins}. The figure shows that amongst the DAO models (weak, moderate, and strong, with $\hp \gtrsim 0.3$), the value of \kp is most important. The DAO models with the lowest \kp result in the most suppression in the power spectrum at early times: the strong DAO case is most suppressed with $\log_{10}\kp = 1.7 \hMpcinv$. The WDM case with $\hp \approx 0$ shows much lower power at early times.

\begin{figure*}[t!]
\includegraphics[width=\textwidth]{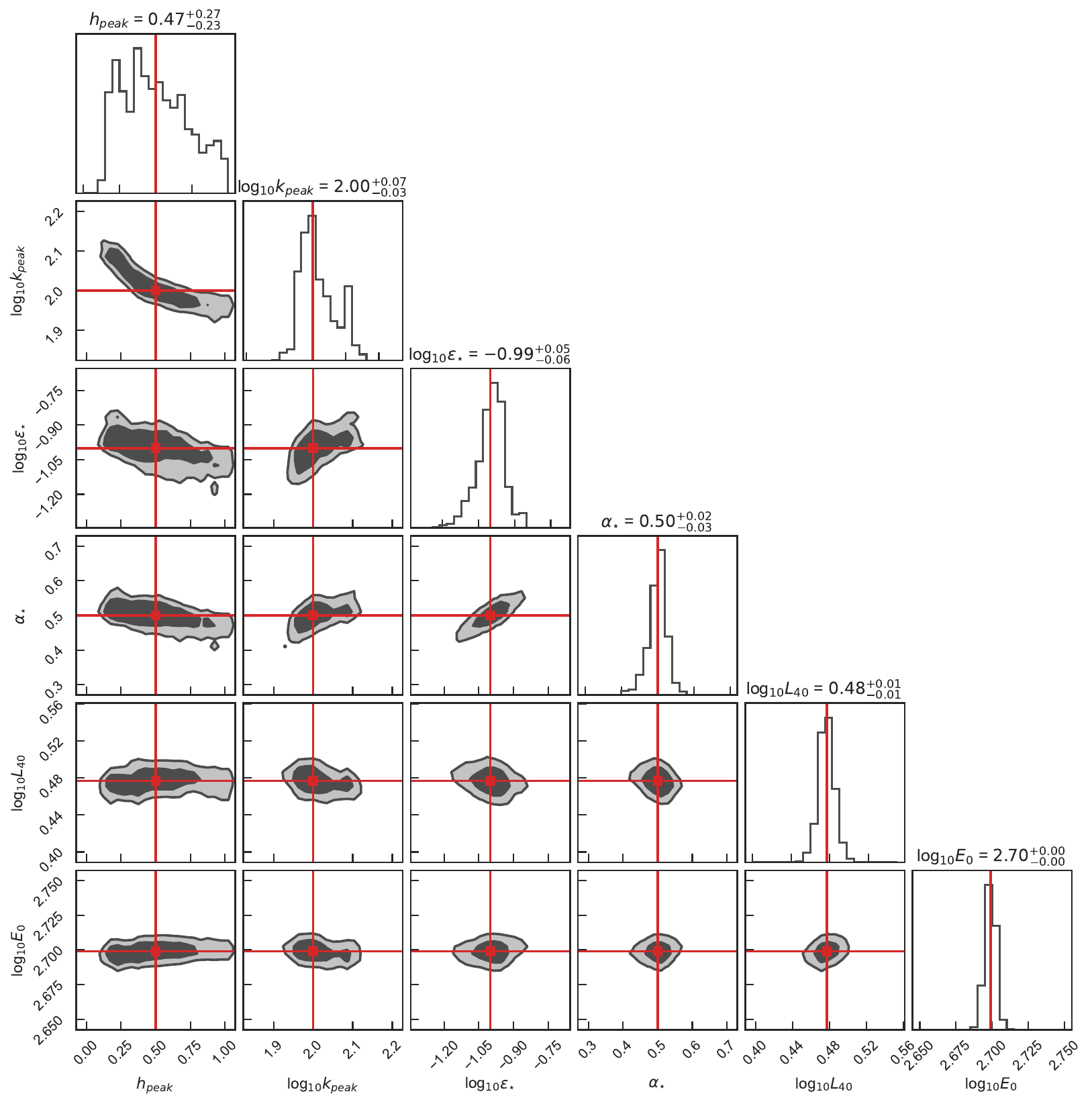}
\caption{Corner plot assuming a fiducial ETHOS model with \hp = 0.5, \kp = 100 \hMpcinv with 5$\%$ noise. Contours mark 1$\sigma$ and 2$\sigma$ confidence intervals, and the red lines show the fiducial input values, in all case well recovered. 
\label{fig:corner}}
\end{figure*}

Figure~\ref{fig:corner} shows the resulting parameter constraint forecasts for the moderate DAO case assuming 5$\%$ noise. This figure shows, for the first time, the degeneracies between astrophysical parameters and dark matter properties.
Excitingly, the dark matter parameters can be recovered well, especially \kp. All astrophysical parameters can be recovered well.

We see a degeneracy between \hp and \kp, as the power spectrum could be explained by high \hp and low \kp (i.e. strong DAO but suppression at large scales) or low \hp and high \kp (i.e. weak DAO but suppression at smaller scales). There is little degeneracy between the X-ray parameters $E_0$ and $L_{40}$ and the star formation parameters \epsstar and \astar. There is no degeneracy between \kp and \hp and the X-ray parameters. The star formation efficiency normalization, \epsstar, and low-mass slope, \astar, are strongly degenerate with each other as they both describe how efficiently the first low-mass galaxies formed stars. As \epsstar is defined as the normalization of the star formation efficiency at the pivot mass $\mpiv = 3 \times 10^{11}\,\Msun$, producing a given star formation efficiency at $\sim10^{6-8}\,\Msun$ requires either high values of both \epsstar and \astar or low values of both parameters.

The cosmological ETHOS parameters \hp and \kp affect the formation time of the first galaxies, so they are also degenerate with the star-formation parameters \epsstar and \astar. Delayed galaxy formation (from a lower halo abundance) will result in a lower Lyman-$\alpha$ flux at a fixed redshift, degenerate with a lower star-formation efficiency. The strongest degeneracy between the ETHOS and astrophysical parameters is thus between \kp and \epsstar.

\begin{figure*}[t!]
\includegraphics[width=0.43\textwidth]{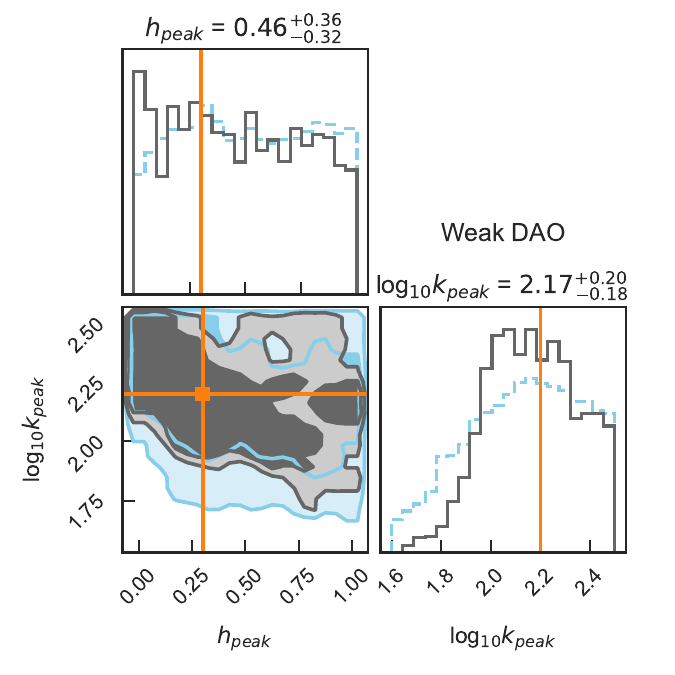} 
\includegraphics[width=0.43\textwidth]{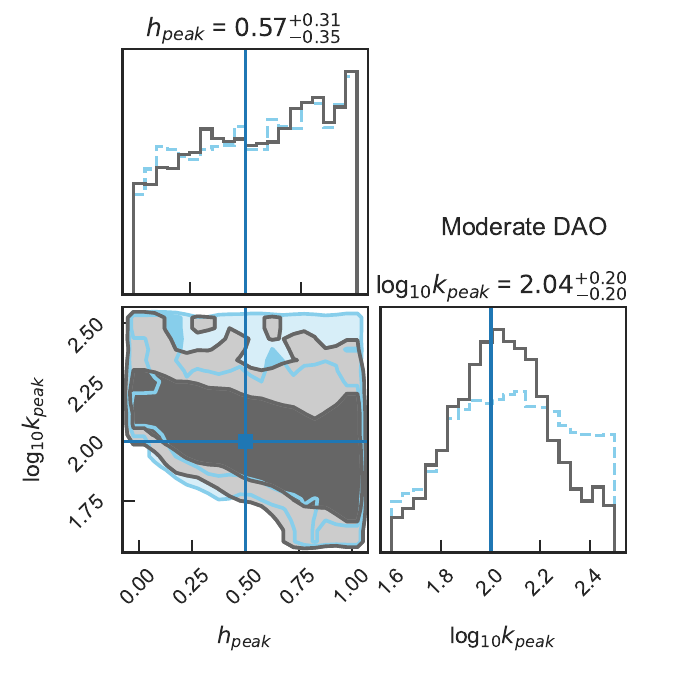} \\
\includegraphics[width=0.43\textwidth]{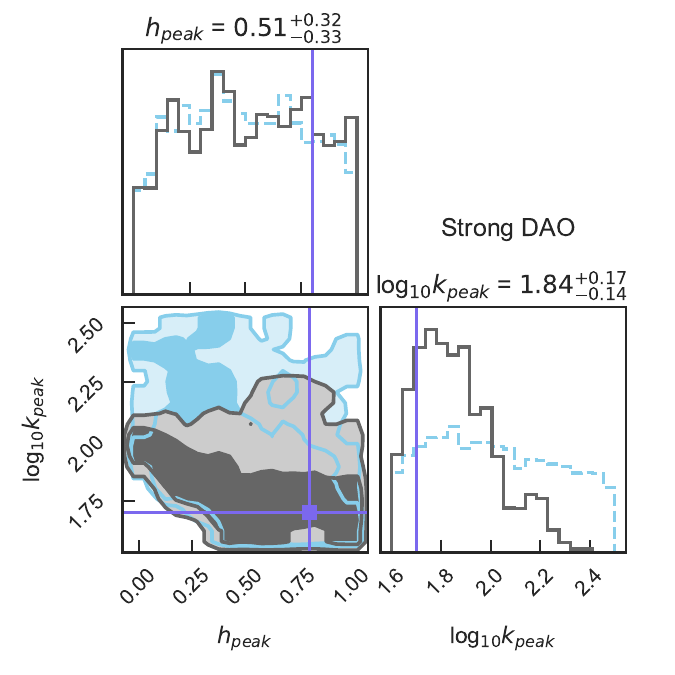}
\includegraphics[width=0.43\textwidth]{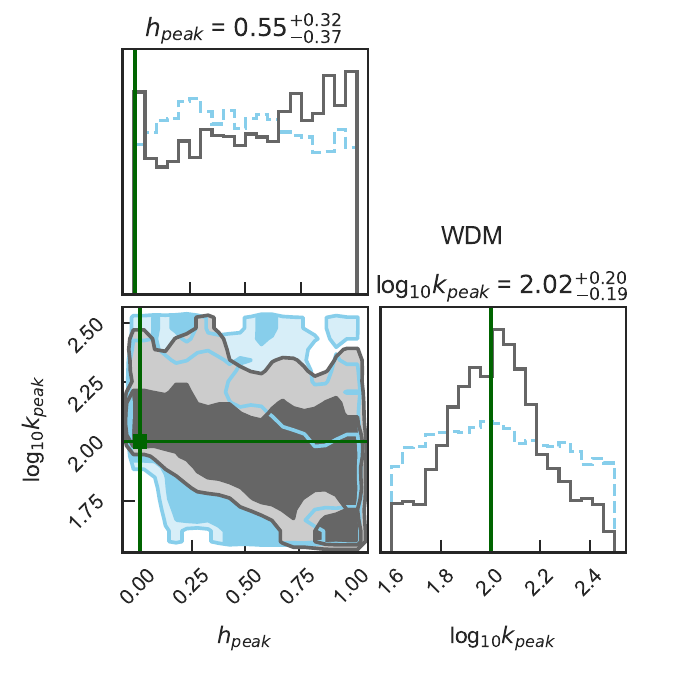}
\caption{Marginalized 2D posteriors for \hp and \kp for the four DM models listed in Table~\ref{tab:fiducial_ETHOS}; Weak DAO, Moderate DAO, Strong DAO and WDM. We show the posteriors for 540 days(solid gray) and 180 days (dashed blue) of observation with HERA respectively. Contours mark 1$\sigma$ and 2$\sigma$ confidence interval of the posterior. The colored lines show the input (fiducial) values. 
\label{fig:corner_models}}
\end{figure*}

Figure~\ref{fig:corner_models} shows the resulting marginalized parameter constraint forecasts on \kp and \hp for the four different ETHOS dark matter models listed in Table~\ref{tab:fiducial_ETHOS}. Excitingly, \kp can be well-constrained (within 10\% in $\log_{10}\kp$) in all cases for 540 days of observations. 
Thus, HERA will be able to place strong constraints on a cut-off in the DM power spectrum if there is a significant deviation from CDM (i.e. $\kp \gg 100$\,\hMpcinv).
In the moderate noise case, the peak of the posterior is still at the input \kp value, but the posteriors are broader, making it unlikely to rule out CDM at $>1\sigma$ significance. 

As mentioned in Section~\ref{sec:method_bayes}, we wanted to see how well we can distinguish between dark matter with DAOs and WDM. Both our WDM model and moderate DAO model have $\log_{10}\kp = 2.0$ but different \hp. While we can constrain \kp in both cases, unfortunately, even in the case of 540 days of observations, we cannot constrain \hp. 
This can be understood by examining the power spectra (Figures~\ref{fig:ps} and ~\ref{fig:PS_bins}): \hp produces a smaller difference in the power spectra compared to \kp making it more difficult to constrain.

\section{Discussion} 
\label{sec:disc}

Here we have presented the first MCMC-based forecast on the ETHOS parameter space using the 21-cm signal at cosmic dawn.
We have explored how different values of \kp and \hp, which parameterize the DM power spectrum at small scales, can be distinguished from astrophysical effects, and from each other. 
We find only small degeneracies between the X-ray parameters $\nu_0$ and $L_{40}$ and the Lyman-$\alpha$ parameters \epsstar and \astar, recovering the results of \citet{Mason23} who used a Fisher-matrix approach, and \citet{Park2019} who used a MCMC to explore degeneracies of astrophysical parameters with HERA (assuming CDM). 

The key advances of our approach have been, first, to use  {\tt Zeus21} to efficiently model the 21-cm signal relative to e.g. {\tt 21cmFAST} \citep{Mesinger2007,2011MNRAS.411..955M,Park2019,Murray2021}. By introducing the ETHOS framework into {\tt Zeus21}, changing the various cosmological parameters does not incur a large computational cost. 
Our MCMC ran $10^3-10^4$ times faster than if using \texttt{21cmfast}, allowing us to expand the parameter space to include non-CDM cosmologies. 
Secondly, by using MCMCs instead of  Fisher matrices we are able to unveil more complex degeneracies, such as the banana shaped degeneracy between \kp and \hp (see Figure \ref{fig:corner}) which would not have been caught with a Fisher matrix approach. 

We investigated two scenarios with HERA of 180 and 540 days of observations respectively. 
Excitingly, we find that in the case of 540 days of observations \kp can be well-measured and CDM can be ruled out with high significance cases where $\kp \simlt 100$\,\hMpcinv. 
We found \hp is difficult to constrain even in the case of 540 days of observations, as the strength of the DAOs is less important than their location (\kp) in changing the 21-cm signal.
Nevertheless, our forecasts indicate that \kp can be constrained even with moderate noise, showing that non-cold dark matter models are within the reach of current HERA observations. 

Our results are qualitatively similar to those of \citet{Giri2022} who performed MCMC parameter space forecasts for mixed (cold + non-cold) dark matter scenarios with SKA, finding that 21-cm signal observations should place strong constraints on the nature of dark matter.

We have not included molecular cooling mini-halos $M_h \simlt 10^8\,\Msun$, which would increase early star formation and push the onset of the 21-cm signal to higher redshifts and reduce its amplitude (at least for CDM, \citep{Qin2021a,Munoz2022}). The inclusion of molecular cooling halos would thus make the difference between CDM and non-CDM structure formation even stronger as the suppression of low mass halos will be more apparent. 
In this sense, our forecasts are conservative, as more subtle departures from CDM can be found when including molecular-cooling galaxies.

We have also only considered a one-to-one mapping between halo mass and star formation rate (Equation~\ref{eqn:SF efficiency}). Recent JWST observations hint that star formation may be highly stochastic in high redshift, low mass dark matter halos \citep[e.g.,][]{Endsley2023}. As the 21-cm signal depends on the total integrated SFRD in the universe we expect the effects of stochasticity should be averaged out, but this may add some additional uncertainty to the inferred star formation parameters which could be explored in future work.
Furthermore, we have focused on $z>10$ and not modeled hydrogen reionization, as it is currently not implemented in {\tt Zeus21} beyond a global neutral fraction. However, the sensitivity of 21-cm experiments is best at high frequencies ($\simgt 150$\,MHz, $z\simlt 8$) where the radio background is lowest, thus the improved signal-to-noise at lower redshifts would likely improve our parameter constraints.

\section{Conclusions} 
\label{sec:conc}

In this work, we have explored the impact of non-CDM dark matter on the 21-cm signal at cosmic dawn. We have implemented the ETHOS dark matter framework in the \texttt{Zeus21} 21-cm signal code and performed MCMC parameter space forecasts to, for the first time, test the prospects for constraining ETHOS dark matter with the 21-cm power spectrum, marginalizing over unknown astrophysics. Our main results are as follows:
\begin{enumerate}
    \item We find a generalized form of the smooth-$k$ window function in Extended Press-Schechter theory which provides a good description of halo mass functions spanning a broad range of dark matter models, including CDM.
    \item We create a generalizable form of the ETHOS power spectrum as a function of the parameters \hp and \kp which set the amplitude and wavenumber scale of dark acoustic oscillations. This enables broad parameter space exploration interpolating between previous results.
    \item We have added ETHOS to the 21-cm signal code \texttt{Zeus21}, making it possible to investigate how different dark matter models affect the global 21-cm signal and the power spectrum. As \texttt{Zeus21} currently only includes atomic cooling halos, the effect we find of ETHOS on the global signal and the power spectrum is conservative. Star formation in molecular-cooling halos can make it easier to detect deviations from CDM at small scales.
    \item Non-CDM dark matter models, like  ETHOS, produce a delayed, but more rapid redshift evolution of the 21-cm global signal. This is because the suppression of small-scale power results in a delayed formation, but a more rapid build-up of halos and thus galaxies, compared to CDM. The slower build up of the SFRD delays the onset of Lyman-$\alpha$ heating, shifting the absorption in the 21-cm global signal to lower redshifts than in CDM.
    \item ETHOS models produce 21-cm power spectra which are also delayed relative to CDM (due to the later onset of Lyman-$\alpha$ heating), and show a higher amplitude of fluctuations and two strong peaks in redshift corresponding to the epoch of Lyman-$\alpha$ and X-ray heating respectively.
    The higher amplitude of the power spectrum is due to the increased contrast in the 21-cm signal between high and low density regions relative to CDM 
    \item We present the first MCMC forecast of the 21-cm power spectrum varying ETHOS cosmology and astrophysical parameters. We find the ETHOS parameters \kp and \hp to be degenerate with astrophysical parameters that determine the star formation efficiency (the normalization, $\epsstar$, and low-mass end slope, $\astar$, of the star formation efficiency) as they both affect when stars are formed. Low \kp and \hp describe a HMF with fewer low-mass halos, and as a consequence the formation of the first stars will be delayed. 
    \item We forecast it will be possible measure the ETHOS parameter \kp, the scale at which power is suppressed, for a broad range of DAO models with HERA with just 540 days of observations, assuming a moderate-foreground noise model.
    Our results imply we can rule out CDM with high significance in the case of models with $\hp \simgt 0.5$ (i.e. moderate and strong DAOs), by constraining suppression at wavenumbers several times larger than probed by other observations (e.g., the Lyman-alpha forest).
    \item The amplitude of DAOs, \hp, is more difficult to constrain in all models as it has a smaller impact on the 21-cm power spectrum compared to \kp. We find it is not possible to distinguish between a moderate DAO and a WDM case with the same $\kp = 100$\,\hMpcinv, but different \hp, in the case of 180 or 540 days of observations.
    It is possible that future work including reionization ($z\lesssim 10$), molecular-cooling galaxies ($\mh\lesssim 10^8\,\Msun$) or combining with information from the high redshift UV luminosity functions \citep[][]{Park2019}, will demonstrate it is possible to measure \hp.
\end{enumerate}

In summary, upcoming HERA 21-cm observations will be able to measure the ETHOS parameter \kp, though likely not \hp. 
This will open the window to detecting non-CDM in the 21-cm signal.
Our results demonstrate that 21-cm observations have the potential to constrain the matter power spectrum on scales smaller than current probes ($k \simlt 200$\,\hMpcinv).
We expect that future observations with lower noise may be able to detect and distinguish both ETHOS parameters. Our results suggest that this is possible regardless of the degeneracies between the ETHOS parameters and astrophysics, showing that we can separate dark acoustic oscillations from astrophysics at cosmic dawn.

\begin{acknowledgments}
We thank Sebastian Bohr for providing the HMFs from the N-body simulations presented in \citet{Bohr2021}
JV was supported by a VILLUM FONDEN Investigator grant (project number 16599) and by a research grant (VIL54489) from VILLUM FONDEN.
CAM acknowledges support by the VILLUM FONDEN under grant 37459 and the Carlsberg Foundation under grant CF22-1322. The Cosmic Dawn Center (DAWN) is funded by the Danish National Research Foundation under grant DNRF140.
JBM acknowledges support from the National Science Foundation (NSF) under Grant No.~2307354. FYCR acknowledges the support of program HST-AR-17061.001-A whose support was provided by the National Aeronautical and Space Administration (NASA) through a grant from the Space Telescope Science Institute, which is operated by the Association of Universities for Research in Astronomy, Incorporated, under NASA contract NAS5-26555. FYCR also acknowledges support from the NSF under Grant No.~2327192.
\end{acknowledgments}

\section*{Software}
{\tt Zeus21} \citep{Munoz2023}, {\tt 21-cmSense} \citep{Pober2014}, {\tt CLASS} \citep{CLASS}, {\tt emcee} \citep{Foreman-Mackey2013}, {\tt corner} \citep{corner}, {\tt IPython} \citep{Perez2007a}, {\tt matplotlib} \citep{Hunter2007a}, {\tt NumPy} \citep{VanderWalt2011a}, {\tt SciPy} \citep{Oliphant2007a}, {\tt Astropy} \citep{Robitaille2013}.




\bibliography{library}

\begin{thebibliography}{64}%
\makeatletter
\providecommand \@ifxundefined [1]{%
 \@ifx{#1\undefined}
}%
\providecommand \@ifnum [1]{%
 \ifnum #1\expandafter \@firstoftwo
 \else \expandafter \@secondoftwo
 \fi
}%
\providecommand \@ifx [1]{%
 \ifx #1\expandafter \@firstoftwo
 \else \expandafter \@secondoftwo
 \fi
}%
\providecommand \natexlab [1]{#1}%
\providecommand \enquote  [1]{``#1''}%
\providecommand \bibnamefont  [1]{#1}%
\providecommand \bibfnamefont [1]{#1}%
\providecommand \citenamefont [1]{#1}%
\providecommand \href@noop [0]{\@secondoftwo}%
\providecommand \href [0]{\begingroup \@sanitize@url \@href}%
\providecommand \@href[1]{\@@startlink{#1}\@@href}%
\providecommand \@@href[1]{\endgroup#1\@@endlink}%
\providecommand \@sanitize@url [0]{\catcode `\\12\catcode `\$12\catcode `\&12\catcode `\#12\catcode `\^12\catcode `\_12\catcode `\%12\relax}%
\providecommand \@@startlink[1]{}%
\providecommand \@@endlink[0]{}%
\providecommand \url  [0]{\begingroup\@sanitize@url \@url }%
\providecommand \@url [1]{\endgroup\@href {#1}{\urlprefix }}%
\providecommand \urlprefix  [0]{URL }%
\providecommand \Eprint [0]{\href }%
\providecommand \doibase [0]{https://doi.org/}%
\providecommand \selectlanguage [0]{\@gobble}%
\providecommand \bibinfo  [0]{\@secondoftwo}%
\providecommand \bibfield  [0]{\@secondoftwo}%
\providecommand \translation [1]{[#1]}%
\providecommand \BibitemOpen [0]{}%
\providecommand \bibitemStop [0]{}%
\providecommand \bibitemNoStop [0]{.\EOS\space}%
\providecommand \EOS [0]{\spacefactor3000\relax}%
\providecommand \BibitemShut  [1]{\csname bibitem#1\endcsname}%
\let\auto@bib@innerbib\@empty
\bibitem [{\citenamefont {{Bertone}}\ and\ \citenamefont {{Hooper}}(2018)}]{2018RvMP...90d5002B}%
  \BibitemOpen
  \bibfield  {author} {\bibinfo {author} {\bibfnamefont {G.}~\bibnamefont {{Bertone}}}\ and\ \bibinfo {author} {\bibfnamefont {D.}~\bibnamefont {{Hooper}}},\ }\bibfield  {title} {\bibinfo {title} {{History of dark matter}},\ }\href {https://doi.org/10.1103/RevModPhys.90.045002} {\bibfield  {journal} {\bibinfo  {journal} {Reviews of Modern Physics}\ }\textbf {\bibinfo {volume} {90}},\ \bibinfo {eid} {045002} (\bibinfo {year} {2018})},\ \Eprint {https://arxiv.org/abs/1605.04909} {arXiv:1605.04909 [astro-ph.CO]} \BibitemShut {NoStop}%
\bibitem [{\citenamefont {{Viel}}\ \emph {et~al.}(2005)\citenamefont {{Viel}}, \citenamefont {{Lesgourgues}}, \citenamefont {{Haehnelt}}, \citenamefont {{Matarrese}},\ and\ \citenamefont {{Riotto}}}]{Viel2005}%
  \BibitemOpen
  \bibfield  {author} {\bibinfo {author} {\bibfnamefont {M.}~\bibnamefont {{Viel}}}, \bibinfo {author} {\bibfnamefont {J.}~\bibnamefont {{Lesgourgues}}}, \bibinfo {author} {\bibfnamefont {M.~G.}\ \bibnamefont {{Haehnelt}}}, \bibinfo {author} {\bibfnamefont {S.}~\bibnamefont {{Matarrese}}},\ and\ \bibinfo {author} {\bibfnamefont {A.}~\bibnamefont {{Riotto}}},\ }\bibfield  {title} {\bibinfo {title} {{Constraining warm dark matter candidates including sterile neutrinos and light gravitinos with WMAP and the Lyman-{\ensuremath{\alpha}} forest}},\ }\href {https://doi.org/10.1103/PhysRevD.71.063534} {\bibfield  {journal} {\bibinfo  {journal} {\prd}\ }\textbf {\bibinfo {volume} {71}},\ \bibinfo {eid} {063534} (\bibinfo {year} {2005})},\ \Eprint {https://arxiv.org/abs/astro-ph/0501562} {arXiv:astro-ph/0501562 [astro-ph]} \BibitemShut {NoStop}%
\bibitem [{\citenamefont {Hinshaw}\ \emph {et~al.}(2012)\citenamefont {Hinshaw}, \citenamefont {Larson}, \citenamefont {Komatsu}, \citenamefont {Spergel}, \citenamefont {Bennett}, \citenamefont {Dunkley}, \citenamefont {Nolta}, \citenamefont {Halpern}, \citenamefont {Hill}, \citenamefont {Odegard}, \citenamefont {Page}, \citenamefont {Smith}, \citenamefont {Weiland}, \citenamefont {Gold}, \citenamefont {Jarosik}, \citenamefont {Kogut}, \citenamefont {Limon}, \citenamefont {Meyer}, \citenamefont {Tucker}, \citenamefont {Wollack},\ and\ \citenamefont {Wright}}]{Hinshaw2012}%
  \BibitemOpen
  \bibfield  {author} {\bibinfo {author} {\bibfnamefont {G.}~\bibnamefont {Hinshaw}}, \bibinfo {author} {\bibfnamefont {D.}~\bibnamefont {Larson}}, \bibinfo {author} {\bibfnamefont {E.}~\bibnamefont {Komatsu}}, \bibinfo {author} {\bibfnamefont {D.~N.}\ \bibnamefont {Spergel}}, \bibinfo {author} {\bibfnamefont {C.~L.}\ \bibnamefont {Bennett}}, \bibinfo {author} {\bibfnamefont {J.}~\bibnamefont {Dunkley}}, \bibinfo {author} {\bibfnamefont {M.~R.}\ \bibnamefont {Nolta}}, \bibinfo {author} {\bibfnamefont {M.}~\bibnamefont {Halpern}}, \bibinfo {author} {\bibfnamefont {R.~S.}\ \bibnamefont {Hill}}, \bibinfo {author} {\bibfnamefont {N.}~\bibnamefont {Odegard}}, \bibinfo {author} {\bibfnamefont {L.}~\bibnamefont {Page}}, \bibinfo {author} {\bibfnamefont {K.~M.}\ \bibnamefont {Smith}}, \bibinfo {author} {\bibfnamefont {J.~L.}\ \bibnamefont {Weiland}}, \bibinfo {author} {\bibfnamefont {B.}~\bibnamefont {Gold}}, \bibinfo {author} {\bibfnamefont {N.}~\bibnamefont {Jarosik}}, \bibinfo {author} {\bibfnamefont
  {A.}~\bibnamefont {Kogut}}, \bibinfo {author} {\bibfnamefont {M.}~\bibnamefont {Limon}}, \bibinfo {author} {\bibfnamefont {S.~S.}\ \bibnamefont {Meyer}}, \bibinfo {author} {\bibfnamefont {G.~S.}\ \bibnamefont {Tucker}}, \bibinfo {author} {\bibfnamefont {E.}~\bibnamefont {Wollack}},\ and\ \bibinfo {author} {\bibfnamefont {E.~L.}\ \bibnamefont {Wright}},\ }\bibfield  {title} {\bibinfo {title} {Nine-{{Year Wilkinson Microwave Anisotropy Probe}} ({{WMAP}}) {{Observations}}: {{Cosmological Parameter Results}}},\ }\href@noop {} {\bibfield  {journal} {\bibinfo  {journal} {Astrophys. J. Suppl. Ser.}\ }\textbf {\bibinfo {volume} {192}},\ \bibinfo {pages} {18} (\bibinfo {year} {2012})}\BibitemShut {NoStop}%
\bibitem [{\citenamefont {{Planck Collaboration}}\ \emph {et~al.}(2018)\citenamefont {{Planck Collaboration}}, \citenamefont {Aghanim}, \citenamefont {Akrami}, \citenamefont {Ashdown}, \citenamefont {Aumont}, \citenamefont {Baccigalupi}, \citenamefont {Ballardini}, \citenamefont {Banday}, \citenamefont {Barreiro}, \citenamefont {Bartolo}, \citenamefont {Basak}, \citenamefont {Battye}, \citenamefont {Benabed}, \citenamefont {Bernard}, \citenamefont {Bersanelli}, \citenamefont {Bielewicz}, \citenamefont {Bock}, \citenamefont {Bond}, \citenamefont {Borrill}, \citenamefont {Bouchet}, \citenamefont {Boulanger}, \citenamefont {Bucher}, \citenamefont {Burigana}, \citenamefont {Butler}, \citenamefont {Calabrese}, \citenamefont {Cardoso}, \citenamefont {Carron}, \citenamefont {Challinor}, \citenamefont {Chiang}, \citenamefont {Chluba}, \citenamefont {Colombo}, \citenamefont {Combet}, \citenamefont {Contreras}, \citenamefont {Crill}, \citenamefont {Cuttaia}, \citenamefont {{de Bernardis}}, \citenamefont {{de Zotti}},
  \citenamefont {Delabrouille}, \citenamefont {Delouis}, \citenamefont {Di~Valentino}, \citenamefont {Diego}, \citenamefont {Dor{\'e}}, \citenamefont {Douspis}, \citenamefont {Ducout}, \citenamefont {Dupac}, \citenamefont {Dusini}, \citenamefont {Efstathiou}, \citenamefont {Elsner}, \citenamefont {En{\ss}lin}, \citenamefont {Eriksen}, \citenamefont {Fantaye}, \citenamefont {Farhang}, \citenamefont {Fergusson}, \citenamefont {{Fernandez-Cobos}}, \citenamefont {Finelli}, \citenamefont {Forastieri}, \citenamefont {Frailis}, \citenamefont {Franceschi}, \citenamefont {Frolov}, \citenamefont {Galeotta}, \citenamefont {Galli}, \citenamefont {Ganga}, \citenamefont {{G{\'e}nova-Santos}}, \citenamefont {Gerbino}, \citenamefont {Ghosh}, \citenamefont {{Gonz{\'a}lez-Nuevo}}, \citenamefont {G{\'o}rski}, \citenamefont {Gratton}, \citenamefont {Gruppuso}, \citenamefont {Gudmundsson}, \citenamefont {Hamann}, \citenamefont {Handley}, \citenamefont {Herranz}, \citenamefont {Hivon}, \citenamefont {Huang}, \citenamefont {Jaffe},
  \citenamefont {Jones}, \citenamefont {Karakci}, \citenamefont {Keih{\"a}nen}, \citenamefont {Keskitalo}, \citenamefont {Kiiveri}, \citenamefont {Kim}, \citenamefont {Kisner}, \citenamefont {Knox}, \citenamefont {Krachmalnicoff}, \citenamefont {Kunz}, \citenamefont {{Kurki-Suonio}}, \citenamefont {Lagache}, \citenamefont {Lamarre}, \citenamefont {Lasenby}, \citenamefont {Lattanzi}, \citenamefont {Lawrence}, \citenamefont {Jeune}, \citenamefont {Lemos}, \citenamefont {Lesgourgues}, \citenamefont {Levrier}, \citenamefont {Lewis}, \citenamefont {Liguori}, \citenamefont {Lilje}, \citenamefont {Lilley}, \citenamefont {Lindholm}, \citenamefont {{L{\'o}pez-Caniego}}, \citenamefont {Lubin}, \citenamefont {Ma}, \citenamefont {{Mac{\'i}as-P{\'e}rez}}, \citenamefont {Maggio}, \citenamefont {Maino}, \citenamefont {Mandolesi}, \citenamefont {Mangilli}, \citenamefont {{Marcos-Caballero}}, \citenamefont {Maris}, \citenamefont {Martin}, \citenamefont {Martinelli}, \citenamefont {{Mart{\'i}nez-Gonz{\'a}lez}}, \citenamefont
  {Matarrese}, \citenamefont {Mauri}, \citenamefont {McEwen}, \citenamefont {Meinhold}, \citenamefont {Melchiorri}, \citenamefont {Mennella}, \citenamefont {Migliaccio}, \citenamefont {Millea}, \citenamefont {Mitra}, \citenamefont {{Miville-Desch{\^e}nes}}, \citenamefont {Molinari}, \citenamefont {Montier}, \citenamefont {Morgante}, \citenamefont {Moss}, \citenamefont {Natoli}, \citenamefont {{N{\o}rgaard-Nielsen}}, \citenamefont {Pagano}, \citenamefont {Paoletti}, \citenamefont {Partridge}, \citenamefont {Patanchon}, \citenamefont {Peiris}, \citenamefont {Perrotta}, \citenamefont {Pettorino}, \citenamefont {Piacentini}, \citenamefont {Polastri}, \citenamefont {Polenta}, \citenamefont {Puget}, \citenamefont {Rachen}, \citenamefont {Reinecke}, \citenamefont {Remazeilles}, \citenamefont {Renzi}, \citenamefont {Rocha}, \citenamefont {Rosset}, \citenamefont {Roudier}, \citenamefont {{Rubi{\~n}o-Mart{\'i}n}}, \citenamefont {{Ruiz-Granados}}, \citenamefont {Salvati}, \citenamefont {Sandri}, \citenamefont
  {Savelainen}, \citenamefont {Scott}, \citenamefont {Shellard}, \citenamefont {Sirignano}, \citenamefont {Sirri}, \citenamefont {Spencer}, \citenamefont {Sunyaev}, \citenamefont {{Suur-Uski}}, \citenamefont {Tauber}, \citenamefont {Tavagnacco}, \citenamefont {Tenti}, \citenamefont {Toffolatti}, \citenamefont {Tomasi}, \citenamefont {Trombetti}, \citenamefont {Valenziano}, \citenamefont {Valiviita}, \citenamefont {Van~Tent}, \citenamefont {Vibert}, \citenamefont {Vielva}, \citenamefont {Villa}, \citenamefont {Vittorio}, \citenamefont {Wandelt}, \citenamefont {Wehus}, \citenamefont {White}, \citenamefont {White}, \citenamefont {Zacchei},\ and\ \citenamefont {Zonca}}]{PlanckCollaboration2018}%
  \BibitemOpen
  \bibfield  {author} {\bibinfo {author} {\bibnamefont {{Planck Collaboration}}}, \bibinfo {author} {\bibfnamefont {N.}~\bibnamefont {Aghanim}}, \bibinfo {author} {\bibfnamefont {Y.}~\bibnamefont {Akrami}}, \bibinfo {author} {\bibfnamefont {M.}~\bibnamefont {Ashdown}}, \bibinfo {author} {\bibfnamefont {J.}~\bibnamefont {Aumont}}, \bibinfo {author} {\bibfnamefont {C.}~\bibnamefont {Baccigalupi}}, \bibinfo {author} {\bibfnamefont {M.}~\bibnamefont {Ballardini}}, \bibinfo {author} {\bibfnamefont {A.~J.}\ \bibnamefont {Banday}}, \bibinfo {author} {\bibfnamefont {R.~B.}\ \bibnamefont {Barreiro}}, \bibinfo {author} {\bibfnamefont {N.}~\bibnamefont {Bartolo}}, \bibinfo {author} {\bibfnamefont {S.}~\bibnamefont {Basak}}, \bibinfo {author} {\bibfnamefont {R.}~\bibnamefont {Battye}}, \bibinfo {author} {\bibfnamefont {K.}~\bibnamefont {Benabed}}, \bibinfo {author} {\bibfnamefont {J.-P.}\ \bibnamefont {Bernard}}, \bibinfo {author} {\bibfnamefont {M.}~\bibnamefont {Bersanelli}}, \bibinfo {author} {\bibfnamefont
  {P.}~\bibnamefont {Bielewicz}}, \bibinfo {author} {\bibfnamefont {J.~J.}\ \bibnamefont {Bock}}, \bibinfo {author} {\bibfnamefont {J.~R.}\ \bibnamefont {Bond}}, \bibinfo {author} {\bibfnamefont {J.}~\bibnamefont {Borrill}}, \bibinfo {author} {\bibfnamefont {F.~R.}\ \bibnamefont {Bouchet}}, \bibinfo {author} {\bibfnamefont {F.}~\bibnamefont {Boulanger}}, \bibinfo {author} {\bibfnamefont {M.}~\bibnamefont {Bucher}}, \bibinfo {author} {\bibfnamefont {C.}~\bibnamefont {Burigana}}, \bibinfo {author} {\bibfnamefont {R.~C.}\ \bibnamefont {Butler}}, \bibinfo {author} {\bibfnamefont {E.}~\bibnamefont {Calabrese}}, \bibinfo {author} {\bibfnamefont {J.-F.}\ \bibnamefont {Cardoso}}, \bibinfo {author} {\bibfnamefont {J.}~\bibnamefont {Carron}}, \bibinfo {author} {\bibfnamefont {A.}~\bibnamefont {Challinor}}, \bibinfo {author} {\bibfnamefont {H.~C.}\ \bibnamefont {Chiang}}, \bibinfo {author} {\bibfnamefont {J.}~\bibnamefont {Chluba}}, \bibinfo {author} {\bibfnamefont {L.~P.~L.}\ \bibnamefont {Colombo}}, \bibinfo {author}
  {\bibfnamefont {C.}~\bibnamefont {Combet}}, \bibinfo {author} {\bibfnamefont {D.}~\bibnamefont {Contreras}}, \bibinfo {author} {\bibfnamefont {B.~P.}\ \bibnamefont {Crill}}, \bibinfo {author} {\bibfnamefont {F.}~\bibnamefont {Cuttaia}}, \bibinfo {author} {\bibfnamefont {P.}~\bibnamefont {{de Bernardis}}}, \bibinfo {author} {\bibfnamefont {G.}~\bibnamefont {{de Zotti}}}, \bibinfo {author} {\bibfnamefont {J.}~\bibnamefont {Delabrouille}}, \bibinfo {author} {\bibfnamefont {J.-M.}\ \bibnamefont {Delouis}}, \bibinfo {author} {\bibfnamefont {E.}~\bibnamefont {Di~Valentino}}, \bibinfo {author} {\bibfnamefont {J.~M.}\ \bibnamefont {Diego}}, \bibinfo {author} {\bibfnamefont {O.}~\bibnamefont {Dor{\'e}}}, \bibinfo {author} {\bibfnamefont {M.}~\bibnamefont {Douspis}}, \bibinfo {author} {\bibfnamefont {A.}~\bibnamefont {Ducout}}, \bibinfo {author} {\bibfnamefont {X.}~\bibnamefont {Dupac}}, \bibinfo {author} {\bibfnamefont {S.}~\bibnamefont {Dusini}}, \bibinfo {author} {\bibfnamefont {G.}~\bibnamefont {Efstathiou}},
  \bibinfo {author} {\bibfnamefont {F.}~\bibnamefont {Elsner}}, \bibinfo {author} {\bibfnamefont {T.~A.}\ \bibnamefont {En{\ss}lin}}, \bibinfo {author} {\bibfnamefont {H.~K.}\ \bibnamefont {Eriksen}}, \bibinfo {author} {\bibfnamefont {Y.}~\bibnamefont {Fantaye}}, \bibinfo {author} {\bibfnamefont {M.}~\bibnamefont {Farhang}}, \bibinfo {author} {\bibfnamefont {J.}~\bibnamefont {Fergusson}}, \bibinfo {author} {\bibfnamefont {R.}~\bibnamefont {{Fernandez-Cobos}}}, \bibinfo {author} {\bibfnamefont {F.}~\bibnamefont {Finelli}}, \bibinfo {author} {\bibfnamefont {F.}~\bibnamefont {Forastieri}}, \bibinfo {author} {\bibfnamefont {M.}~\bibnamefont {Frailis}}, \bibinfo {author} {\bibfnamefont {E.}~\bibnamefont {Franceschi}}, \bibinfo {author} {\bibfnamefont {A.}~\bibnamefont {Frolov}}, \bibinfo {author} {\bibfnamefont {S.}~\bibnamefont {Galeotta}}, \bibinfo {author} {\bibfnamefont {S.}~\bibnamefont {Galli}}, \bibinfo {author} {\bibfnamefont {K.}~\bibnamefont {Ganga}}, \bibinfo {author} {\bibfnamefont {R.~T.}\
  \bibnamefont {{G{\'e}nova-Santos}}}, \bibinfo {author} {\bibfnamefont {M.}~\bibnamefont {Gerbino}}, \bibinfo {author} {\bibfnamefont {T.}~\bibnamefont {Ghosh}}, \bibinfo {author} {\bibfnamefont {J.}~\bibnamefont {{Gonz{\'a}lez-Nuevo}}}, \bibinfo {author} {\bibfnamefont {K.~M.}\ \bibnamefont {G{\'o}rski}}, \bibinfo {author} {\bibfnamefont {S.}~\bibnamefont {Gratton}}, \bibinfo {author} {\bibfnamefont {A.}~\bibnamefont {Gruppuso}}, \bibinfo {author} {\bibfnamefont {J.~E.}\ \bibnamefont {Gudmundsson}}, \bibinfo {author} {\bibfnamefont {J.}~\bibnamefont {Hamann}}, \bibinfo {author} {\bibfnamefont {W.}~\bibnamefont {Handley}}, \bibinfo {author} {\bibfnamefont {D.}~\bibnamefont {Herranz}}, \bibinfo {author} {\bibfnamefont {E.}~\bibnamefont {Hivon}}, \bibinfo {author} {\bibfnamefont {Z.}~\bibnamefont {Huang}}, \bibinfo {author} {\bibfnamefont {A.~H.}\ \bibnamefont {Jaffe}}, \bibinfo {author} {\bibfnamefont {W.~C.}\ \bibnamefont {Jones}}, \bibinfo {author} {\bibfnamefont {A.}~\bibnamefont {Karakci}}, \bibinfo
  {author} {\bibfnamefont {E.}~\bibnamefont {Keih{\"a}nen}}, \bibinfo {author} {\bibfnamefont {R.}~\bibnamefont {Keskitalo}}, \bibinfo {author} {\bibfnamefont {K.}~\bibnamefont {Kiiveri}}, \bibinfo {author} {\bibfnamefont {J.}~\bibnamefont {Kim}}, \bibinfo {author} {\bibfnamefont {T.~S.}\ \bibnamefont {Kisner}}, \bibinfo {author} {\bibfnamefont {L.}~\bibnamefont {Knox}}, \bibinfo {author} {\bibfnamefont {N.}~\bibnamefont {Krachmalnicoff}}, \bibinfo {author} {\bibfnamefont {M.}~\bibnamefont {Kunz}}, \bibinfo {author} {\bibfnamefont {H.}~\bibnamefont {{Kurki-Suonio}}}, \bibinfo {author} {\bibfnamefont {G.}~\bibnamefont {Lagache}}, \bibinfo {author} {\bibfnamefont {J.-M.}\ \bibnamefont {Lamarre}}, \bibinfo {author} {\bibfnamefont {A.}~\bibnamefont {Lasenby}}, \bibinfo {author} {\bibfnamefont {M.}~\bibnamefont {Lattanzi}}, \bibinfo {author} {\bibfnamefont {C.~R.}\ \bibnamefont {Lawrence}}, \bibinfo {author} {\bibfnamefont {M.~L.}\ \bibnamefont {Jeune}}, \bibinfo {author} {\bibfnamefont {P.}~\bibnamefont {Lemos}},
  \bibinfo {author} {\bibfnamefont {J.}~\bibnamefont {Lesgourgues}}, \bibinfo {author} {\bibfnamefont {F.}~\bibnamefont {Levrier}}, \bibinfo {author} {\bibfnamefont {A.}~\bibnamefont {Lewis}}, \bibinfo {author} {\bibfnamefont {M.}~\bibnamefont {Liguori}}, \bibinfo {author} {\bibfnamefont {P.~B.}\ \bibnamefont {Lilje}}, \bibinfo {author} {\bibfnamefont {M.}~\bibnamefont {Lilley}}, \bibinfo {author} {\bibfnamefont {V.}~\bibnamefont {Lindholm}}, \bibinfo {author} {\bibfnamefont {M.}~\bibnamefont {{L{\'o}pez-Caniego}}}, \bibinfo {author} {\bibfnamefont {P.~M.}\ \bibnamefont {Lubin}}, \bibinfo {author} {\bibfnamefont {Y.-Z.}\ \bibnamefont {Ma}}, \bibinfo {author} {\bibfnamefont {J.~F.}\ \bibnamefont {{Mac{\'i}as-P{\'e}rez}}}, \bibinfo {author} {\bibfnamefont {G.}~\bibnamefont {Maggio}}, \bibinfo {author} {\bibfnamefont {D.}~\bibnamefont {Maino}}, \bibinfo {author} {\bibfnamefont {N.}~\bibnamefont {Mandolesi}}, \bibinfo {author} {\bibfnamefont {A.}~\bibnamefont {Mangilli}}, \bibinfo {author} {\bibfnamefont
  {A.}~\bibnamefont {{Marcos-Caballero}}}, \bibinfo {author} {\bibfnamefont {M.}~\bibnamefont {Maris}}, \bibinfo {author} {\bibfnamefont {P.~G.}\ \bibnamefont {Martin}}, \bibinfo {author} {\bibfnamefont {M.}~\bibnamefont {Martinelli}}, \bibinfo {author} {\bibfnamefont {E.}~\bibnamefont {{Mart{\'i}nez-Gonz{\'a}lez}}}, \bibinfo {author} {\bibfnamefont {S.}~\bibnamefont {Matarrese}}, \bibinfo {author} {\bibfnamefont {N.}~\bibnamefont {Mauri}}, \bibinfo {author} {\bibfnamefont {J.~D.}\ \bibnamefont {McEwen}}, \bibinfo {author} {\bibfnamefont {P.~R.}\ \bibnamefont {Meinhold}}, \bibinfo {author} {\bibfnamefont {A.}~\bibnamefont {Melchiorri}}, \bibinfo {author} {\bibfnamefont {A.}~\bibnamefont {Mennella}}, \bibinfo {author} {\bibfnamefont {M.}~\bibnamefont {Migliaccio}}, \bibinfo {author} {\bibfnamefont {M.}~\bibnamefont {Millea}}, \bibinfo {author} {\bibfnamefont {S.}~\bibnamefont {Mitra}}, \bibinfo {author} {\bibfnamefont {M.-A.}\ \bibnamefont {{Miville-Desch{\^e}nes}}}, \bibinfo {author} {\bibfnamefont
  {D.}~\bibnamefont {Molinari}}, \bibinfo {author} {\bibfnamefont {L.}~\bibnamefont {Montier}}, \bibinfo {author} {\bibfnamefont {G.}~\bibnamefont {Morgante}}, \bibinfo {author} {\bibfnamefont {A.}~\bibnamefont {Moss}}, \bibinfo {author} {\bibfnamefont {P.}~\bibnamefont {Natoli}}, \bibinfo {author} {\bibfnamefont {H.~U.}\ \bibnamefont {{N{\o}rgaard-Nielsen}}}, \bibinfo {author} {\bibfnamefont {L.}~\bibnamefont {Pagano}}, \bibinfo {author} {\bibfnamefont {D.}~\bibnamefont {Paoletti}}, \bibinfo {author} {\bibfnamefont {B.}~\bibnamefont {Partridge}}, \bibinfo {author} {\bibfnamefont {G.}~\bibnamefont {Patanchon}}, \bibinfo {author} {\bibfnamefont {H.~V.}\ \bibnamefont {Peiris}}, \bibinfo {author} {\bibfnamefont {F.}~\bibnamefont {Perrotta}}, \bibinfo {author} {\bibfnamefont {V.}~\bibnamefont {Pettorino}}, \bibinfo {author} {\bibfnamefont {F.}~\bibnamefont {Piacentini}}, \bibinfo {author} {\bibfnamefont {L.}~\bibnamefont {Polastri}}, \bibinfo {author} {\bibfnamefont {G.}~\bibnamefont {Polenta}}, \bibinfo {author}
  {\bibfnamefont {J.-L.}\ \bibnamefont {Puget}}, \bibinfo {author} {\bibfnamefont {J.~P.}\ \bibnamefont {Rachen}}, \bibinfo {author} {\bibfnamefont {M.}~\bibnamefont {Reinecke}}, \bibinfo {author} {\bibfnamefont {M.}~\bibnamefont {Remazeilles}}, \bibinfo {author} {\bibfnamefont {A.}~\bibnamefont {Renzi}}, \bibinfo {author} {\bibfnamefont {G.}~\bibnamefont {Rocha}}, \bibinfo {author} {\bibfnamefont {C.}~\bibnamefont {Rosset}}, \bibinfo {author} {\bibfnamefont {G.}~\bibnamefont {Roudier}}, \bibinfo {author} {\bibfnamefont {J.~A.}\ \bibnamefont {{Rubi{\~n}o-Mart{\'i}n}}}, \bibinfo {author} {\bibfnamefont {B.}~\bibnamefont {{Ruiz-Granados}}}, \bibinfo {author} {\bibfnamefont {L.}~\bibnamefont {Salvati}}, \bibinfo {author} {\bibfnamefont {M.}~\bibnamefont {Sandri}}, \bibinfo {author} {\bibfnamefont {M.}~\bibnamefont {Savelainen}}, \bibinfo {author} {\bibfnamefont {D.}~\bibnamefont {Scott}}, \bibinfo {author} {\bibfnamefont {E.~P.~S.}\ \bibnamefont {Shellard}}, \bibinfo {author} {\bibfnamefont {C.}~\bibnamefont
  {Sirignano}}, \bibinfo {author} {\bibfnamefont {G.}~\bibnamefont {Sirri}}, \bibinfo {author} {\bibfnamefont {L.~D.}\ \bibnamefont {Spencer}}, \bibinfo {author} {\bibfnamefont {R.}~\bibnamefont {Sunyaev}}, \bibinfo {author} {\bibfnamefont {A.-S.}\ \bibnamefont {{Suur-Uski}}}, \bibinfo {author} {\bibfnamefont {J.~A.}\ \bibnamefont {Tauber}}, \bibinfo {author} {\bibfnamefont {D.}~\bibnamefont {Tavagnacco}}, \bibinfo {author} {\bibfnamefont {M.}~\bibnamefont {Tenti}}, \bibinfo {author} {\bibfnamefont {L.}~\bibnamefont {Toffolatti}}, \bibinfo {author} {\bibfnamefont {M.}~\bibnamefont {Tomasi}}, \bibinfo {author} {\bibfnamefont {T.}~\bibnamefont {Trombetti}}, \bibinfo {author} {\bibfnamefont {L.}~\bibnamefont {Valenziano}}, \bibinfo {author} {\bibfnamefont {J.}~\bibnamefont {Valiviita}}, \bibinfo {author} {\bibfnamefont {B.}~\bibnamefont {Van~Tent}}, \bibinfo {author} {\bibfnamefont {L.}~\bibnamefont {Vibert}}, \bibinfo {author} {\bibfnamefont {P.}~\bibnamefont {Vielva}}, \bibinfo {author} {\bibfnamefont
  {F.}~\bibnamefont {Villa}}, \bibinfo {author} {\bibfnamefont {N.}~\bibnamefont {Vittorio}}, \bibinfo {author} {\bibfnamefont {B.~D.}\ \bibnamefont {Wandelt}}, \bibinfo {author} {\bibfnamefont {I.~K.}\ \bibnamefont {Wehus}}, \bibinfo {author} {\bibfnamefont {M.}~\bibnamefont {White}}, \bibinfo {author} {\bibfnamefont {S.~D.~M.}\ \bibnamefont {White}}, \bibinfo {author} {\bibfnamefont {A.}~\bibnamefont {Zacchei}},\ and\ \bibinfo {author} {\bibfnamefont {A.}~\bibnamefont {Zonca}},\ }\bibfield  {title} {\bibinfo {title} {Planck 2018 results. {{VI}}. {{Cosmological}} parameters},\ }\href@noop {} {\bibfield  {journal} {\bibinfo  {journal} {arXiv:1807.06209 [astro-ph]}\ } (\bibinfo {year} {2018})},\ \Eprint {https://arxiv.org/abs/1807.06209} {arXiv:1807.06209 [astro-ph]} \BibitemShut {NoStop}%
\bibitem [{\citenamefont {{Bullock}}\ and\ \citenamefont {{Boylan-Kolchin}}(2017)}]{Bullock2017}%
  \BibitemOpen
  \bibfield  {author} {\bibinfo {author} {\bibfnamefont {J.~S.}\ \bibnamefont {{Bullock}}}\ and\ \bibinfo {author} {\bibfnamefont {M.}~\bibnamefont {{Boylan-Kolchin}}},\ }\bibfield  {title} {\bibinfo {title} {{Small-Scale Challenges to the {\ensuremath{\Lambda}}CDM Paradigm}},\ }\href {https://doi.org/10.1146/annurev-astro-091916-055313} {\bibfield  {journal} {\bibinfo  {journal} {\araa}\ }\textbf {\bibinfo {volume} {55}},\ \bibinfo {pages} {343} (\bibinfo {year} {2017})},\ \Eprint {https://arxiv.org/abs/1707.04256} {arXiv:1707.04256 [astro-ph.CO]} \BibitemShut {NoStop}%
\bibitem [{\citenamefont {{Sales}}\ \emph {et~al.}(2022)\citenamefont {{Sales}}, \citenamefont {{Wetzel}},\ and\ \citenamefont {{Fattahi}}}]{2022NatAs...6..897S}%
  \BibitemOpen
  \bibfield  {author} {\bibinfo {author} {\bibfnamefont {L.~V.}\ \bibnamefont {{Sales}}}, \bibinfo {author} {\bibfnamefont {A.}~\bibnamefont {{Wetzel}}},\ and\ \bibinfo {author} {\bibfnamefont {A.}~\bibnamefont {{Fattahi}}},\ }\bibfield  {title} {\bibinfo {title} {{Baryonic solutions and challenges for cosmological models of dwarf galaxies}},\ }\href {https://doi.org/10.1038/s41550-022-01689-w} {\bibfield  {journal} {\bibinfo  {journal} {Nature Astronomy}\ }\textbf {\bibinfo {volume} {6}},\ \bibinfo {pages} {897} (\bibinfo {year} {2022})},\ \Eprint {https://arxiv.org/abs/2206.05295} {arXiv:2206.05295 [astro-ph.GA]} \BibitemShut {NoStop}%
\bibitem [{\citenamefont {{Adhikari}}\ \emph {et~al.}(2017)\citenamefont {{Adhikari}}, \citenamefont {{Agostini}}, \citenamefont {{Ky}}, \citenamefont {{Araki}}, \citenamefont {{Archidiacono}}, \citenamefont {{Bahr}}, \citenamefont {{Baur}}, \citenamefont {{Behrens}}, \citenamefont {{Bezrukov}}, \citenamefont {{Bhupal Dev}}, \citenamefont {{Borah}}, \citenamefont {{Boyarsky}}, \citenamefont {{de Gouvea}}, \citenamefont {{Pires}}, \citenamefont {{de Vega}}, \citenamefont {{Dias}}, \citenamefont {{Di Bari}}, \citenamefont {{Djurcic}}, \citenamefont {{Dolde}}, \citenamefont {{Dorrer}}, \citenamefont {{Durero}}, \citenamefont {{Dragoun}}, \citenamefont {{Drewes}}, \citenamefont {{Drexlin}}, \citenamefont {{D{\"u}llmann}}, \citenamefont {{Eberhardt}}, \citenamefont {{Eliseev}}, \citenamefont {{Enss}}, \citenamefont {{Evans}}, \citenamefont {{Faessler}}, \citenamefont {{Filianin}}, \citenamefont {{Fischer}}, \citenamefont {{Fleischmann}}, \citenamefont {{Formaggio}}, \citenamefont {{Franse}}, \citenamefont
  {{Fraenkle}}, \citenamefont {{Frenk}}, \citenamefont {{Fuller}}, \citenamefont {{Gastaldo}}, \citenamefont {{Garzilli}}, \citenamefont {{Giunti}}, \citenamefont {{Gl{\"u}ck}}, \citenamefont {{Goodman}}, \citenamefont {{Gonzalez-Garcia}}, \citenamefont {{Gorbunov}}, \citenamefont {{Hamann}}, \citenamefont {{Hannen}}, \citenamefont {{Hannestad}}, \citenamefont {{Hansen}}, \citenamefont {{Hassel}}, \citenamefont {{Heeck}}, \citenamefont {{Hofmann}}, \citenamefont {{Houdy}}, \citenamefont {{Huber}}, \citenamefont {{Iakubovskyi}}, \citenamefont {{Ianni}}, \citenamefont {{Ibarra}}, \citenamefont {{Jacobsson}}, \citenamefont {{Jeltema}}, \citenamefont {{Jochum}}, \citenamefont {{Kempf}}, \citenamefont {{Kieck}}, \citenamefont {{Korzeczek}}, \citenamefont {{Kornoukhov}}, \citenamefont {{Lachenmaier}}, \citenamefont {{Laine}}, \citenamefont {{Langacker}}, \citenamefont {{Lasserre}}, \citenamefont {{Lesgourgues}}, \citenamefont {{Lhuillier}}, \citenamefont {{Li}}, \citenamefont {{Liao}}, \citenamefont {{Long}},
  \citenamefont {{Maltoni}}, \citenamefont {{Mangano}}, \citenamefont {{Mavromatos}}, \citenamefont {{Menci}}, \citenamefont {{Merle}}, \citenamefont {{Mertens}}, \citenamefont {{Mirizzi}}, \citenamefont {{Monreal}}, \citenamefont {{Nozik}}, \citenamefont {{Neronov}}, \citenamefont {{Niro}}, \citenamefont {{Novikov}}, \citenamefont {{Oberauer}}, \citenamefont {{Otten}}, \citenamefont {{Palanque-Delabrouille}}, \citenamefont {{Pallavicini}}, \citenamefont {{Pantuev}}, \citenamefont {{Papastergis}}, \citenamefont {{Parke}}, \citenamefont {{Pascoli}}, \citenamefont {{Pastor}}, \citenamefont {{Patwardhan}}, \citenamefont {{Pilaftsis}}, \citenamefont {{Radford}}, \citenamefont {{Ranitzsch}}, \citenamefont {{Rest}}, \citenamefont {{Robinson}}, \citenamefont {{Rodrigues da Silva}}, \citenamefont {{Ruchayskiy}}, \citenamefont {{Sanchez}}, \citenamefont {{Sasaki}}, \citenamefont {{Saviano}}, \citenamefont {{Schneider}}, \citenamefont {{Schneider}}, \citenamefont {{Schwetz}}, \citenamefont {{Sch{\"o}nert}},
  \citenamefont {{Scholl}}, \citenamefont {{Shankar}}, \citenamefont {{Shrock}}, \citenamefont {{Steinbrink}}, \citenamefont {{Strigari}}, \citenamefont {{Suekane}}, \citenamefont {{Suerfu}}, \citenamefont {{Takahashi}}, \citenamefont {{Van}}, \citenamefont {{Tkachev}}, \citenamefont {{Totzauer}}, \citenamefont {{Tsai}}, \citenamefont {{Tully}}, \citenamefont {{Valerius}}, \citenamefont {{Valle}}, \citenamefont {{Venos}}, \citenamefont {{Viel}}, \citenamefont {{Vivier}}, \citenamefont {{Wang}}, \citenamefont {{Weinheimer}}, \citenamefont {{Wendt}}, \citenamefont {{Winslow}}, \citenamefont {{Wolf}}, \citenamefont {{Wurm}}, \citenamefont {{Xing}}, \citenamefont {{Zhou}},\ and\ \citenamefont {{Zuber}}}]{Adhikari2017}%
  \BibitemOpen
  \bibfield  {author} {\bibinfo {author} {\bibfnamefont {R.}~\bibnamefont {{Adhikari}}}, \bibinfo {author} {\bibfnamefont {M.}~\bibnamefont {{Agostini}}}, \bibinfo {author} {\bibfnamefont {N.~A.}\ \bibnamefont {{Ky}}}, \bibinfo {author} {\bibfnamefont {T.}~\bibnamefont {{Araki}}}, \bibinfo {author} {\bibfnamefont {M.}~\bibnamefont {{Archidiacono}}}, \bibinfo {author} {\bibfnamefont {M.}~\bibnamefont {{Bahr}}}, \bibinfo {author} {\bibfnamefont {J.}~\bibnamefont {{Baur}}}, \bibinfo {author} {\bibfnamefont {J.}~\bibnamefont {{Behrens}}}, \bibinfo {author} {\bibfnamefont {F.}~\bibnamefont {{Bezrukov}}}, \bibinfo {author} {\bibfnamefont {P.~S.}\ \bibnamefont {{Bhupal Dev}}}, \bibinfo {author} {\bibfnamefont {D.}~\bibnamefont {{Borah}}}, \bibinfo {author} {\bibfnamefont {A.}~\bibnamefont {{Boyarsky}}}, \bibinfo {author} {\bibfnamefont {A.}~\bibnamefont {{de Gouvea}}}, \bibinfo {author} {\bibfnamefont {C.~A. d.~S.}\ \bibnamefont {{Pires}}}, \bibinfo {author} {\bibfnamefont {H.~J.}\ \bibnamefont {{de Vega}}},
  \bibinfo {author} {\bibfnamefont {A.~G.}\ \bibnamefont {{Dias}}}, \bibinfo {author} {\bibfnamefont {P.}~\bibnamefont {{Di Bari}}}, \bibinfo {author} {\bibfnamefont {Z.}~\bibnamefont {{Djurcic}}}, \bibinfo {author} {\bibfnamefont {K.}~\bibnamefont {{Dolde}}}, \bibinfo {author} {\bibfnamefont {H.}~\bibnamefont {{Dorrer}}}, \bibinfo {author} {\bibfnamefont {M.}~\bibnamefont {{Durero}}}, \bibinfo {author} {\bibfnamefont {O.}~\bibnamefont {{Dragoun}}}, \bibinfo {author} {\bibfnamefont {M.}~\bibnamefont {{Drewes}}}, \bibinfo {author} {\bibfnamefont {G.}~\bibnamefont {{Drexlin}}}, \bibinfo {author} {\bibfnamefont {C.~E.}\ \bibnamefont {{D{\"u}llmann}}}, \bibinfo {author} {\bibfnamefont {K.}~\bibnamefont {{Eberhardt}}}, \bibinfo {author} {\bibfnamefont {S.}~\bibnamefont {{Eliseev}}}, \bibinfo {author} {\bibfnamefont {C.}~\bibnamefont {{Enss}}}, \bibinfo {author} {\bibfnamefont {N.~W.}\ \bibnamefont {{Evans}}}, \bibinfo {author} {\bibfnamefont {A.}~\bibnamefont {{Faessler}}}, \bibinfo {author} {\bibfnamefont
  {P.}~\bibnamefont {{Filianin}}}, \bibinfo {author} {\bibfnamefont {V.}~\bibnamefont {{Fischer}}}, \bibinfo {author} {\bibfnamefont {A.}~\bibnamefont {{Fleischmann}}}, \bibinfo {author} {\bibfnamefont {J.~A.}\ \bibnamefont {{Formaggio}}}, \bibinfo {author} {\bibfnamefont {J.}~\bibnamefont {{Franse}}}, \bibinfo {author} {\bibfnamefont {F.~M.}\ \bibnamefont {{Fraenkle}}}, \bibinfo {author} {\bibfnamefont {C.~S.}\ \bibnamefont {{Frenk}}}, \bibinfo {author} {\bibfnamefont {G.}~\bibnamefont {{Fuller}}}, \bibinfo {author} {\bibfnamefont {L.}~\bibnamefont {{Gastaldo}}}, \bibinfo {author} {\bibfnamefont {A.}~\bibnamefont {{Garzilli}}}, \bibinfo {author} {\bibfnamefont {C.}~\bibnamefont {{Giunti}}}, \bibinfo {author} {\bibfnamefont {F.}~\bibnamefont {{Gl{\"u}ck}}}, \bibinfo {author} {\bibfnamefont {M.~C.}\ \bibnamefont {{Goodman}}}, \bibinfo {author} {\bibfnamefont {M.~C.}\ \bibnamefont {{Gonzalez-Garcia}}}, \bibinfo {author} {\bibfnamefont {D.}~\bibnamefont {{Gorbunov}}}, \bibinfo {author} {\bibfnamefont
  {J.}~\bibnamefont {{Hamann}}}, \bibinfo {author} {\bibfnamefont {V.}~\bibnamefont {{Hannen}}}, \bibinfo {author} {\bibfnamefont {S.}~\bibnamefont {{Hannestad}}}, \bibinfo {author} {\bibfnamefont {S.~H.}\ \bibnamefont {{Hansen}}}, \bibinfo {author} {\bibfnamefont {C.}~\bibnamefont {{Hassel}}}, \bibinfo {author} {\bibfnamefont {J.}~\bibnamefont {{Heeck}}}, \bibinfo {author} {\bibfnamefont {F.}~\bibnamefont {{Hofmann}}}, \bibinfo {author} {\bibfnamefont {T.}~\bibnamefont {{Houdy}}}, \bibinfo {author} {\bibfnamefont {A.}~\bibnamefont {{Huber}}}, \bibinfo {author} {\bibfnamefont {D.}~\bibnamefont {{Iakubovskyi}}}, \bibinfo {author} {\bibfnamefont {A.}~\bibnamefont {{Ianni}}}, \bibinfo {author} {\bibfnamefont {A.}~\bibnamefont {{Ibarra}}}, \bibinfo {author} {\bibfnamefont {R.}~\bibnamefont {{Jacobsson}}}, \bibinfo {author} {\bibfnamefont {T.}~\bibnamefont {{Jeltema}}}, \bibinfo {author} {\bibfnamefont {J.}~\bibnamefont {{Jochum}}}, \bibinfo {author} {\bibfnamefont {S.}~\bibnamefont {{Kempf}}}, \bibinfo {author}
  {\bibfnamefont {T.}~\bibnamefont {{Kieck}}}, \bibinfo {author} {\bibfnamefont {M.}~\bibnamefont {{Korzeczek}}}, \bibinfo {author} {\bibfnamefont {V.}~\bibnamefont {{Kornoukhov}}}, \bibinfo {author} {\bibfnamefont {T.}~\bibnamefont {{Lachenmaier}}}, \bibinfo {author} {\bibfnamefont {M.}~\bibnamefont {{Laine}}}, \bibinfo {author} {\bibfnamefont {P.}~\bibnamefont {{Langacker}}}, \bibinfo {author} {\bibfnamefont {T.}~\bibnamefont {{Lasserre}}}, \bibinfo {author} {\bibfnamefont {J.}~\bibnamefont {{Lesgourgues}}}, \bibinfo {author} {\bibfnamefont {D.}~\bibnamefont {{Lhuillier}}}, \bibinfo {author} {\bibfnamefont {Y.~F.}\ \bibnamefont {{Li}}}, \bibinfo {author} {\bibfnamefont {W.}~\bibnamefont {{Liao}}}, \bibinfo {author} {\bibfnamefont {A.~W.}\ \bibnamefont {{Long}}}, \bibinfo {author} {\bibfnamefont {M.}~\bibnamefont {{Maltoni}}}, \bibinfo {author} {\bibfnamefont {G.}~\bibnamefont {{Mangano}}}, \bibinfo {author} {\bibfnamefont {N.~E.}\ \bibnamefont {{Mavromatos}}}, \bibinfo {author} {\bibfnamefont
  {N.}~\bibnamefont {{Menci}}}, \bibinfo {author} {\bibfnamefont {A.}~\bibnamefont {{Merle}}}, \bibinfo {author} {\bibfnamefont {S.}~\bibnamefont {{Mertens}}}, \bibinfo {author} {\bibfnamefont {A.}~\bibnamefont {{Mirizzi}}}, \bibinfo {author} {\bibfnamefont {B.}~\bibnamefont {{Monreal}}}, \bibinfo {author} {\bibfnamefont {A.}~\bibnamefont {{Nozik}}}, \bibinfo {author} {\bibfnamefont {A.}~\bibnamefont {{Neronov}}}, \bibinfo {author} {\bibfnamefont {V.}~\bibnamefont {{Niro}}}, \bibinfo {author} {\bibfnamefont {Y.}~\bibnamefont {{Novikov}}}, \bibinfo {author} {\bibfnamefont {L.}~\bibnamefont {{Oberauer}}}, \bibinfo {author} {\bibfnamefont {E.}~\bibnamefont {{Otten}}}, \bibinfo {author} {\bibfnamefont {N.}~\bibnamefont {{Palanque-Delabrouille}}}, \bibinfo {author} {\bibfnamefont {M.}~\bibnamefont {{Pallavicini}}}, \bibinfo {author} {\bibfnamefont {V.~S.}\ \bibnamefont {{Pantuev}}}, \bibinfo {author} {\bibfnamefont {E.}~\bibnamefont {{Papastergis}}}, \bibinfo {author} {\bibfnamefont {S.}~\bibnamefont {{Parke}}},
  \bibinfo {author} {\bibfnamefont {S.}~\bibnamefont {{Pascoli}}}, \bibinfo {author} {\bibfnamefont {S.}~\bibnamefont {{Pastor}}}, \bibinfo {author} {\bibfnamefont {A.}~\bibnamefont {{Patwardhan}}}, \bibinfo {author} {\bibfnamefont {A.}~\bibnamefont {{Pilaftsis}}}, \bibinfo {author} {\bibfnamefont {D.~C.}\ \bibnamefont {{Radford}}}, \bibinfo {author} {\bibfnamefont {P.~C.~O.}\ \bibnamefont {{Ranitzsch}}}, \bibinfo {author} {\bibfnamefont {O.}~\bibnamefont {{Rest}}}, \bibinfo {author} {\bibfnamefont {D.~J.}\ \bibnamefont {{Robinson}}}, \bibinfo {author} {\bibfnamefont {P.~S.}\ \bibnamefont {{Rodrigues da Silva}}}, \bibinfo {author} {\bibfnamefont {O.}~\bibnamefont {{Ruchayskiy}}}, \bibinfo {author} {\bibfnamefont {N.~G.}\ \bibnamefont {{Sanchez}}}, \bibinfo {author} {\bibfnamefont {M.}~\bibnamefont {{Sasaki}}}, \bibinfo {author} {\bibfnamefont {N.}~\bibnamefont {{Saviano}}}, \bibinfo {author} {\bibfnamefont {A.}~\bibnamefont {{Schneider}}}, \bibinfo {author} {\bibfnamefont {F.}~\bibnamefont {{Schneider}}},
  \bibinfo {author} {\bibfnamefont {T.}~\bibnamefont {{Schwetz}}}, \bibinfo {author} {\bibfnamefont {S.}~\bibnamefont {{Sch{\"o}nert}}}, \bibinfo {author} {\bibfnamefont {S.}~\bibnamefont {{Scholl}}}, \bibinfo {author} {\bibfnamefont {F.}~\bibnamefont {{Shankar}}}, \bibinfo {author} {\bibfnamefont {R.}~\bibnamefont {{Shrock}}}, \bibinfo {author} {\bibfnamefont {N.}~\bibnamefont {{Steinbrink}}}, \bibinfo {author} {\bibfnamefont {L.}~\bibnamefont {{Strigari}}}, \bibinfo {author} {\bibfnamefont {F.}~\bibnamefont {{Suekane}}}, \bibinfo {author} {\bibfnamefont {B.}~\bibnamefont {{Suerfu}}}, \bibinfo {author} {\bibfnamefont {R.}~\bibnamefont {{Takahashi}}}, \bibinfo {author} {\bibfnamefont {N.~T.~H.}\ \bibnamefont {{Van}}}, \bibinfo {author} {\bibfnamefont {I.}~\bibnamefont {{Tkachev}}}, \bibinfo {author} {\bibfnamefont {M.}~\bibnamefont {{Totzauer}}}, \bibinfo {author} {\bibfnamefont {Y.}~\bibnamefont {{Tsai}}}, \bibinfo {author} {\bibfnamefont {C.~G.}\ \bibnamefont {{Tully}}}, \bibinfo {author} {\bibfnamefont
  {K.}~\bibnamefont {{Valerius}}}, \bibinfo {author} {\bibfnamefont {J.~W.~F.}\ \bibnamefont {{Valle}}}, \bibinfo {author} {\bibfnamefont {D.}~\bibnamefont {{Venos}}}, \bibinfo {author} {\bibfnamefont {M.}~\bibnamefont {{Viel}}}, \bibinfo {author} {\bibfnamefont {M.}~\bibnamefont {{Vivier}}}, \bibinfo {author} {\bibfnamefont {M.~Y.}\ \bibnamefont {{Wang}}}, \bibinfo {author} {\bibfnamefont {C.}~\bibnamefont {{Weinheimer}}}, \bibinfo {author} {\bibfnamefont {K.}~\bibnamefont {{Wendt}}}, \bibinfo {author} {\bibfnamefont {L.}~\bibnamefont {{Winslow}}}, \bibinfo {author} {\bibfnamefont {J.}~\bibnamefont {{Wolf}}}, \bibinfo {author} {\bibfnamefont {M.}~\bibnamefont {{Wurm}}}, \bibinfo {author} {\bibfnamefont {Z.}~\bibnamefont {{Xing}}}, \bibinfo {author} {\bibfnamefont {S.}~\bibnamefont {{Zhou}}},\ and\ \bibinfo {author} {\bibfnamefont {K.}~\bibnamefont {{Zuber}}},\ }\bibfield  {title} {\bibinfo {title} {{A White Paper on keV sterile neutrino Dark Matter}},\ }\href {https://doi.org/10.1088/1475-7516/2017/01/025}
  {\bibfield  {journal} {\bibinfo  {journal} {\jcap}\ }\textbf {\bibinfo {volume} {2017}},\ \bibinfo {eid} {025} (\bibinfo {year} {2017})},\ \Eprint {https://arxiv.org/abs/1602.04816} {arXiv:1602.04816 [hep-ph]} \BibitemShut {NoStop}%
\bibitem [{\citenamefont {{B{\oe}hm}}\ \emph {et~al.}(2002)\citenamefont {{B{\oe}hm}}, \citenamefont {{Riazuelo}}, \citenamefont {{Hansen}},\ and\ \citenamefont {{Schaeffer}}}]{2002PhRvD..66h3505B}%
  \BibitemOpen
  \bibfield  {author} {\bibinfo {author} {\bibfnamefont {C.}~\bibnamefont {{B{\oe}hm}}}, \bibinfo {author} {\bibfnamefont {A.}~\bibnamefont {{Riazuelo}}}, \bibinfo {author} {\bibfnamefont {S.~H.}\ \bibnamefont {{Hansen}}},\ and\ \bibinfo {author} {\bibfnamefont {R.}~\bibnamefont {{Schaeffer}}},\ }\bibfield  {title} {\bibinfo {title} {{Interacting dark matter disguised as warm dark matter}},\ }\href {https://doi.org/10.1103/PhysRevD.66.083505} {\bibfield  {journal} {\bibinfo  {journal} {\prd}\ }\textbf {\bibinfo {volume} {66}},\ \bibinfo {eid} {083505} (\bibinfo {year} {2002})},\ \Eprint {https://arxiv.org/abs/astro-ph/0112522} {arXiv:astro-ph/0112522 [astro-ph]} \BibitemShut {NoStop}%
\bibitem [{\citenamefont {{Buckley}}\ \emph {et~al.}(2014)\citenamefont {{Buckley}}, \citenamefont {{Zavala}}, \citenamefont {{Cyr-Racine}}, \citenamefont {{Sigurdson}},\ and\ \citenamefont {{Vogelsberger}}}]{2014PhRvD..90d3524B}%
  \BibitemOpen
  \bibfield  {author} {\bibinfo {author} {\bibfnamefont {M.~R.}\ \bibnamefont {{Buckley}}}, \bibinfo {author} {\bibfnamefont {J.}~\bibnamefont {{Zavala}}}, \bibinfo {author} {\bibfnamefont {F.-Y.}\ \bibnamefont {{Cyr-Racine}}}, \bibinfo {author} {\bibfnamefont {K.}~\bibnamefont {{Sigurdson}}},\ and\ \bibinfo {author} {\bibfnamefont {M.}~\bibnamefont {{Vogelsberger}}},\ }\bibfield  {title} {\bibinfo {title} {{Scattering, damping, and acoustic oscillations: Simulating the structure of dark matter halos with relativistic force carriers}},\ }\href {https://doi.org/10.1103/PhysRevD.90.043524} {\bibfield  {journal} {\bibinfo  {journal} {\prd}\ }\textbf {\bibinfo {volume} {90}},\ \bibinfo {eid} {043524} (\bibinfo {year} {2014})},\ \Eprint {https://arxiv.org/abs/1405.2075} {arXiv:1405.2075 [astro-ph.CO]} \BibitemShut {NoStop}%
\bibitem [{\citenamefont {{Schewtschenko}}\ \emph {et~al.}(2015)\citenamefont {{Schewtschenko}}, \citenamefont {{Wilkinson}}, \citenamefont {{Baugh}}, \citenamefont {{B{\oe}hm}},\ and\ \citenamefont {{Pascoli}}}]{2015MNRAS.449.3587S}%
  \BibitemOpen
  \bibfield  {author} {\bibinfo {author} {\bibfnamefont {J.~A.}\ \bibnamefont {{Schewtschenko}}}, \bibinfo {author} {\bibfnamefont {R.~J.}\ \bibnamefont {{Wilkinson}}}, \bibinfo {author} {\bibfnamefont {C.~M.}\ \bibnamefont {{Baugh}}}, \bibinfo {author} {\bibfnamefont {C.}~\bibnamefont {{B{\oe}hm}}},\ and\ \bibinfo {author} {\bibfnamefont {S.}~\bibnamefont {{Pascoli}}},\ }\bibfield  {title} {\bibinfo {title} {{Dark matter-radiation interactions: the impact on dark matter haloes}},\ }\href {https://doi.org/10.1093/mnras/stv431} {\bibfield  {journal} {\bibinfo  {journal} {\mnras}\ }\textbf {\bibinfo {volume} {449}},\ \bibinfo {pages} {3587} (\bibinfo {year} {2015})},\ \Eprint {https://arxiv.org/abs/1412.4905} {arXiv:1412.4905 [astro-ph.CO]} \BibitemShut {NoStop}%
\bibitem [{\citenamefont {{Cyr-Racine}}\ and\ \citenamefont {{Sigurdson}}(2013)}]{Cyr-Racine2013}%
  \BibitemOpen
  \bibfield  {author} {\bibinfo {author} {\bibfnamefont {F.-Y.}\ \bibnamefont {{Cyr-Racine}}}\ and\ \bibinfo {author} {\bibfnamefont {K.}~\bibnamefont {{Sigurdson}}},\ }\bibfield  {title} {\bibinfo {title} {{Cosmology of atomic dark matter}},\ }\href {https://doi.org/10.1103/PhysRevD.87.103515} {\bibfield  {journal} {\bibinfo  {journal} {\prd}\ }\textbf {\bibinfo {volume} {87}},\ \bibinfo {eid} {103515} (\bibinfo {year} {2013})},\ \Eprint {https://arxiv.org/abs/1209.5752} {arXiv:1209.5752 [astro-ph.CO]} \BibitemShut {NoStop}%
\bibitem [{\citenamefont {{Bohr}}\ \emph {et~al.}(2020)\citenamefont {{Bohr}}, \citenamefont {{Zavala}}, \citenamefont {{Cyr-Racine}}, \citenamefont {{Vogelsberger}}, \citenamefont {{Bringmann}},\ and\ \citenamefont {{Pfrommer}}}]{Bohr2020}%
  \BibitemOpen
  \bibfield  {author} {\bibinfo {author} {\bibfnamefont {S.}~\bibnamefont {{Bohr}}}, \bibinfo {author} {\bibfnamefont {J.}~\bibnamefont {{Zavala}}}, \bibinfo {author} {\bibfnamefont {F.-Y.}\ \bibnamefont {{Cyr-Racine}}}, \bibinfo {author} {\bibfnamefont {M.}~\bibnamefont {{Vogelsberger}}}, \bibinfo {author} {\bibfnamefont {T.}~\bibnamefont {{Bringmann}}},\ and\ \bibinfo {author} {\bibfnamefont {C.}~\bibnamefont {{Pfrommer}}},\ }\bibfield  {title} {\bibinfo {title} {{ETHOS - an effective parametrization and classification for structure formation: the non-linear regime at z {\ensuremath{\gtrsim}} 5}},\ }\href {https://doi.org/10.1093/mnras/staa2579} {\bibfield  {journal} {\bibinfo  {journal} {\mnras}\ }\textbf {\bibinfo {volume} {498}},\ \bibinfo {pages} {3403} (\bibinfo {year} {2020})},\ \Eprint {https://arxiv.org/abs/2006.01842} {arXiv:2006.01842 [astro-ph.CO]} \BibitemShut {NoStop}%
\bibitem [{\citenamefont {{Schaeffer}}\ and\ \citenamefont {{Schneider}}(2021)}]{Schaeffer2021}%
  \BibitemOpen
  \bibfield  {author} {\bibinfo {author} {\bibfnamefont {T.}~\bibnamefont {{Schaeffer}}}\ and\ \bibinfo {author} {\bibfnamefont {A.}~\bibnamefont {{Schneider}}},\ }\bibfield  {title} {\bibinfo {title} {{Dark acoustic oscillations: imprints on the matter power spectrum and the halo mass function}},\ }\href {https://doi.org/10.1093/mnras/stab1116} {\bibfield  {journal} {\bibinfo  {journal} {\mnras}\ }\textbf {\bibinfo {volume} {504}},\ \bibinfo {pages} {3773} (\bibinfo {year} {2021})},\ \Eprint {https://arxiv.org/abs/2101.12229} {arXiv:2101.12229 [astro-ph.CO]} \BibitemShut {NoStop}%
\bibitem [{\citenamefont {{Furlanetto}}\ \emph {et~al.}(2006)\citenamefont {{Furlanetto}}, \citenamefont {{Oh}},\ and\ \citenamefont {{Briggs}}}]{Furlanetoo2006-review}%
  \BibitemOpen
  \bibfield  {author} {\bibinfo {author} {\bibfnamefont {S.~R.}\ \bibnamefont {{Furlanetto}}}, \bibinfo {author} {\bibfnamefont {S.~P.}\ \bibnamefont {{Oh}}},\ and\ \bibinfo {author} {\bibfnamefont {F.~H.}\ \bibnamefont {{Briggs}}},\ }\bibfield  {title} {\bibinfo {title} {{Cosmology at low frequencies: The 21 cm transition and the high-redshift Universe}},\ }\href {https://doi.org/10.1016/j.physrep.2006.08.002} {\bibfield  {journal} {\bibinfo  {journal} {\physrep}\ }\textbf {\bibinfo {volume} {433}},\ \bibinfo {pages} {181} (\bibinfo {year} {2006})},\ \Eprint {https://arxiv.org/abs/astro-ph/0608032} {arXiv:astro-ph/0608032 [astro-ph]} \BibitemShut {NoStop}%
\bibitem [{\citenamefont {{Wouthuysen}}(1952)}]{Wouthuysen1952}%
  \BibitemOpen
  \bibfield  {author} {\bibinfo {author} {\bibfnamefont {S.~A.}\ \bibnamefont {{Wouthuysen}}},\ }\bibfield  {title} {\bibinfo {title} {{On the excitation mechanism of the 21-cm (radio-frequency) interstellar hydrogen emission line.}},\ }\href {https://doi.org/10.1086/106661} {\bibfield  {journal} {\bibinfo  {journal} {\aj}\ }\textbf {\bibinfo {volume} {57}},\ \bibinfo {pages} {31} (\bibinfo {year} {1952})}\BibitemShut {NoStop}%
\bibitem [{\citenamefont {{Field}}(1959{\natexlab{a}})}]{1959ApJ...129..536}%
  \BibitemOpen
  \bibfield  {author} {\bibinfo {author} {\bibfnamefont {G.~B.}\ \bibnamefont {{Field}}},\ }\bibfield  {title} {\bibinfo {title} {{The Spin Temperature of Intergalactic Neutral Hydrogen.}},\ }\href {https://doi.org/10.1086/146653} {\bibfield  {journal} {\bibinfo  {journal} {\apj}\ }\textbf {\bibinfo {volume} {129}},\ \bibinfo {pages} {536} (\bibinfo {year} {1959}{\natexlab{a}})}\BibitemShut {NoStop}%
\bibitem [{\citenamefont {{Field}}(1959{\natexlab{b}})}]{1959ApJ...129..551F}%
  \BibitemOpen
  \bibfield  {author} {\bibinfo {author} {\bibfnamefont {G.~B.}\ \bibnamefont {{Field}}},\ }\bibfield  {title} {\bibinfo {title} {{The Time Relaxation of a Resonance-Line Profile.}},\ }\href {https://doi.org/10.1086/146654} {\bibfield  {journal} {\bibinfo  {journal} {\apj}\ }\textbf {\bibinfo {volume} {129}},\ \bibinfo {pages} {551} (\bibinfo {year} {1959}{\natexlab{b}})}\BibitemShut {NoStop}%
\bibitem [{\citenamefont {{Mu{\~n}oz}}\ \emph {et~al.}(2020)\citenamefont {{Mu{\~n}oz}}, \citenamefont {{Dvorkin}},\ and\ \citenamefont {{Cyr-Racine}}}]{Munoz2020}%
  \BibitemOpen
  \bibfield  {author} {\bibinfo {author} {\bibfnamefont {J.~B.}\ \bibnamefont {{Mu{\~n}oz}}}, \bibinfo {author} {\bibfnamefont {C.}~\bibnamefont {{Dvorkin}}},\ and\ \bibinfo {author} {\bibfnamefont {F.-Y.}\ \bibnamefont {{Cyr-Racine}}},\ }\bibfield  {title} {\bibinfo {title} {{Probing the small-scale matter power spectrum with large-scale 21-cm data}},\ }\href {https://doi.org/10.1103/PhysRevD.101.063526} {\bibfield  {journal} {\bibinfo  {journal} {\prd}\ }\textbf {\bibinfo {volume} {101}},\ \bibinfo {eid} {063526} (\bibinfo {year} {2020})},\ \Eprint {https://arxiv.org/abs/1911.11144} {arXiv:1911.11144 [astro-ph.CO]} \BibitemShut {NoStop}%
\bibitem [{\citenamefont {{Sitwell}}\ \emph {et~al.}(2014)\citenamefont {{Sitwell}}, \citenamefont {{Mesinger}}, \citenamefont {{Ma}},\ and\ \citenamefont {{Sigurdson}}}]{Sitwell2014}%
  \BibitemOpen
  \bibfield  {author} {\bibinfo {author} {\bibfnamefont {M.}~\bibnamefont {{Sitwell}}}, \bibinfo {author} {\bibfnamefont {A.}~\bibnamefont {{Mesinger}}}, \bibinfo {author} {\bibfnamefont {Y.-Z.}\ \bibnamefont {{Ma}}},\ and\ \bibinfo {author} {\bibfnamefont {K.}~\bibnamefont {{Sigurdson}}},\ }\bibfield  {title} {\bibinfo {title} {{The imprint of warm dark matter on the cosmological 21-cm signal}},\ }\href {https://doi.org/10.1093/mnras/stt2392} {\bibfield  {journal} {\bibinfo  {journal} {\mnras}\ }\textbf {\bibinfo {volume} {438}},\ \bibinfo {pages} {2664} (\bibinfo {year} {2014})},\ \Eprint {https://arxiv.org/abs/1310.0029} {arXiv:1310.0029 [astro-ph.CO]} \BibitemShut {NoStop}%
\bibitem [{\citenamefont {{Jones}}\ \emph {et~al.}(2021)\citenamefont {{Jones}}, \citenamefont {{Palatnick}}, \citenamefont {{Chen}}, \citenamefont {{Beane}},\ and\ \citenamefont {{Lidz}}}]{Jones2021}%
  \BibitemOpen
  \bibfield  {author} {\bibinfo {author} {\bibfnamefont {D.}~\bibnamefont {{Jones}}}, \bibinfo {author} {\bibfnamefont {S.}~\bibnamefont {{Palatnick}}}, \bibinfo {author} {\bibfnamefont {R.}~\bibnamefont {{Chen}}}, \bibinfo {author} {\bibfnamefont {A.}~\bibnamefont {{Beane}}},\ and\ \bibinfo {author} {\bibfnamefont {A.}~\bibnamefont {{Lidz}}},\ }\bibfield  {title} {\bibinfo {title} {{Fuzzy Dark Matter and the 21 cm Power Spectrum}},\ }\href {https://doi.org/10.3847/1538-4357/abf0a9} {\bibfield  {journal} {\bibinfo  {journal} {\apj}\ }\textbf {\bibinfo {volume} {913}},\ \bibinfo {eid} {7} (\bibinfo {year} {2021})},\ \Eprint {https://arxiv.org/abs/2101.07177} {arXiv:2101.07177 [astro-ph.CO]} \BibitemShut {NoStop}%
\bibitem [{\citenamefont {{Giri}}\ and\ \citenamefont {{Schneider}}(2022)}]{Giri2022}%
  \BibitemOpen
  \bibfield  {author} {\bibinfo {author} {\bibfnamefont {S.~K.}\ \bibnamefont {{Giri}}}\ and\ \bibinfo {author} {\bibfnamefont {A.}~\bibnamefont {{Schneider}}},\ }\bibfield  {title} {\bibinfo {title} {{Imprints of fermionic and bosonic mixed dark matter on the 21-cm signal at cosmic dawn}},\ }\href {https://doi.org/10.1103/PhysRevD.105.083011} {\bibfield  {journal} {\bibinfo  {journal} {\prd}\ }\textbf {\bibinfo {volume} {105}},\ \bibinfo {eid} {083011} (\bibinfo {year} {2022})},\ \Eprint {https://arxiv.org/abs/2201.02210} {arXiv:2201.02210 [astro-ph.CO]} \BibitemShut {NoStop}%
\bibitem [{\citenamefont {{Cyr-Racine}}\ \emph {et~al.}(2016)\citenamefont {{Cyr-Racine}}, \citenamefont {{Sigurdson}}, \citenamefont {{Zavala}}, \citenamefont {{Bringmann}}, \citenamefont {{Vogelsberger}},\ and\ \citenamefont {{Pfrommer}}}]{Cyr-Racine2016}%
  \BibitemOpen
  \bibfield  {author} {\bibinfo {author} {\bibfnamefont {F.-Y.}\ \bibnamefont {{Cyr-Racine}}}, \bibinfo {author} {\bibfnamefont {K.}~\bibnamefont {{Sigurdson}}}, \bibinfo {author} {\bibfnamefont {J.}~\bibnamefont {{Zavala}}}, \bibinfo {author} {\bibfnamefont {T.}~\bibnamefont {{Bringmann}}}, \bibinfo {author} {\bibfnamefont {M.}~\bibnamefont {{Vogelsberger}}},\ and\ \bibinfo {author} {\bibfnamefont {C.}~\bibnamefont {{Pfrommer}}},\ }\bibfield  {title} {\bibinfo {title} {{ETHOS{\textemdash}an effective theory of structure formation: From dark particle physics to the matter distribution of the Universe}},\ }\href {https://doi.org/10.1103/PhysRevD.93.123527} {\bibfield  {journal} {\bibinfo  {journal} {\prd}\ }\textbf {\bibinfo {volume} {93}},\ \bibinfo {eid} {123527} (\bibinfo {year} {2016})},\ \Eprint {https://arxiv.org/abs/1512.05344} {arXiv:1512.05344 [astro-ph.CO]} \BibitemShut {NoStop}%
\bibitem [{\citenamefont {{Vogelsberger}}\ \emph {et~al.}(2016)\citenamefont {{Vogelsberger}}, \citenamefont {{Zavala}}, \citenamefont {{Cyr-Racine}}, \citenamefont {{Pfrommer}}, \citenamefont {{Bringmann}},\ and\ \citenamefont {{Sigurdson}}}]{Vogelsberger2016}%
  \BibitemOpen
  \bibfield  {author} {\bibinfo {author} {\bibfnamefont {M.}~\bibnamefont {{Vogelsberger}}}, \bibinfo {author} {\bibfnamefont {J.}~\bibnamefont {{Zavala}}}, \bibinfo {author} {\bibfnamefont {F.-Y.}\ \bibnamefont {{Cyr-Racine}}}, \bibinfo {author} {\bibfnamefont {C.}~\bibnamefont {{Pfrommer}}}, \bibinfo {author} {\bibfnamefont {T.}~\bibnamefont {{Bringmann}}},\ and\ \bibinfo {author} {\bibfnamefont {K.}~\bibnamefont {{Sigurdson}}},\ }\bibfield  {title} {\bibinfo {title} {{ETHOS - an effective theory of structure formation: dark matter physics as a possible explanation of the small-scale CDM problems}},\ }\href {https://doi.org/10.1093/mnras/stw1076} {\bibfield  {journal} {\bibinfo  {journal} {\mnras}\ }\textbf {\bibinfo {volume} {460}},\ \bibinfo {pages} {1399} (\bibinfo {year} {2016})},\ \Eprint {https://arxiv.org/abs/1512.05349} {arXiv:1512.05349 [astro-ph.CO]} \BibitemShut {NoStop}%
\bibitem [{\citenamefont {{Mu{\~n}oz}}\ \emph {et~al.}(2021)\citenamefont {{Mu{\~n}oz}}, \citenamefont {{Bohr}}, \citenamefont {{Cyr-Racine}}, \citenamefont {{Zavala}},\ and\ \citenamefont {{Vogelsberger}}}]{Munoz2021}%
  \BibitemOpen
  \bibfield  {author} {\bibinfo {author} {\bibfnamefont {J.~B.}\ \bibnamefont {{Mu{\~n}oz}}}, \bibinfo {author} {\bibfnamefont {S.}~\bibnamefont {{Bohr}}}, \bibinfo {author} {\bibfnamefont {F.-Y.}\ \bibnamefont {{Cyr-Racine}}}, \bibinfo {author} {\bibfnamefont {J.}~\bibnamefont {{Zavala}}},\ and\ \bibinfo {author} {\bibfnamefont {M.}~\bibnamefont {{Vogelsberger}}},\ }\bibfield  {title} {\bibinfo {title} {{ETHOS - an effective theory of structure formation: Impact of dark acoustic oscillations on cosmic dawn}},\ }\href {https://doi.org/10.1103/PhysRevD.103.043512} {\bibfield  {journal} {\bibinfo  {journal} {\prd}\ }\textbf {\bibinfo {volume} {103}},\ \bibinfo {eid} {043512} (\bibinfo {year} {2021})},\ \Eprint {https://arxiv.org/abs/2011.05333} {arXiv:2011.05333 [astro-ph.CO]} \BibitemShut {NoStop}%
\bibitem [{\citenamefont {Park}\ \emph {et~al.}(2019)\citenamefont {Park}, \citenamefont {Mesinger}, \citenamefont {Greig},\ and\ \citenamefont {Gillet}}]{Park2019}%
  \BibitemOpen
  \bibfield  {author} {\bibinfo {author} {\bibfnamefont {J.}~\bibnamefont {Park}}, \bibinfo {author} {\bibfnamefont {A.}~\bibnamefont {Mesinger}}, \bibinfo {author} {\bibfnamefont {B.}~\bibnamefont {Greig}},\ and\ \bibinfo {author} {\bibfnamefont {N.}~\bibnamefont {Gillet}},\ }\bibfield  {title} {\bibinfo {title} {Inferring the astrophysics of reionization and cosmic dawn from galaxy luminosity functions and the 21-cm signal},\ }\href {https://doi.org/10.1093/mnras/stz032} {\bibfield  {journal} {\bibinfo  {journal} {Monthly Notices of the Royal Astronomical Society}\ }\textbf {\bibinfo {volume} {484}},\ \bibinfo {pages} {933} (\bibinfo {year} {2019})},\ \Eprint {https://arxiv.org/abs/1809.08995} {arXiv:1809.08995} \BibitemShut {NoStop}%
\bibitem [{\citenamefont {{Qin}}\ \emph {et~al.}(2020)\citenamefont {{Qin}}, \citenamefont {{Mesinger}}, \citenamefont {{Park}}, \citenamefont {{Greig}},\ and\ \citenamefont {{Mu{\~n}oz}}}]{Qin2020}%
  \BibitemOpen
  \bibfield  {author} {\bibinfo {author} {\bibfnamefont {Y.}~\bibnamefont {{Qin}}}, \bibinfo {author} {\bibfnamefont {A.}~\bibnamefont {{Mesinger}}}, \bibinfo {author} {\bibfnamefont {J.}~\bibnamefont {{Park}}}, \bibinfo {author} {\bibfnamefont {B.}~\bibnamefont {{Greig}}},\ and\ \bibinfo {author} {\bibfnamefont {J.~B.}\ \bibnamefont {{Mu{\~n}oz}}},\ }\bibfield  {title} {\bibinfo {title} {{A tale of two sites - I. Inferring the properties of minihalo-hosted galaxies from current observations}},\ }\href {https://doi.org/10.1093/mnras/staa1131} {\bibfield  {journal} {\bibinfo  {journal} {\mnras}\ }\textbf {\bibinfo {volume} {495}},\ \bibinfo {pages} {123} (\bibinfo {year} {2020})},\ \Eprint {https://arxiv.org/abs/2003.04442} {arXiv:2003.04442 [astro-ph.CO]} \BibitemShut {NoStop}%
\bibitem [{\citenamefont {{Qin}}\ \emph {et~al.}(2021)\citenamefont {{Qin}}, \citenamefont {{Mesinger}}, \citenamefont {{Greig}},\ and\ \citenamefont {{Park}}}]{Qin2021a}%
  \BibitemOpen
  \bibfield  {author} {\bibinfo {author} {\bibfnamefont {Y.}~\bibnamefont {{Qin}}}, \bibinfo {author} {\bibfnamefont {A.}~\bibnamefont {{Mesinger}}}, \bibinfo {author} {\bibfnamefont {B.}~\bibnamefont {{Greig}}},\ and\ \bibinfo {author} {\bibfnamefont {J.}~\bibnamefont {{Park}}},\ }\bibfield  {title} {\bibinfo {title} {{A tale of two sites - II. Inferring the properties of minihalo-hosted galaxies with upcoming 21-cm interferometers}},\ }\href {https://doi.org/10.1093/mnras/staa3408} {\bibfield  {journal} {\bibinfo  {journal} {\mnras}\ }\textbf {\bibinfo {volume} {501}},\ \bibinfo {pages} {4748} (\bibinfo {year} {2021})},\ \Eprint {https://arxiv.org/abs/2009.11493} {arXiv:2009.11493 [astro-ph.CO]} \BibitemShut {NoStop}%
\bibitem [{\citenamefont {{Mu{\~n}oz}}(2019)}]{2019PhRvD.100f3538M}%
  \BibitemOpen
  \bibfield  {author} {\bibinfo {author} {\bibfnamefont {J.~B.}\ \bibnamefont {{Mu{\~n}oz}}},\ }\bibfield  {title} {\bibinfo {title} {{Robust velocity-induced acoustic oscillations at cosmic dawn}},\ }\href {https://doi.org/10.1103/PhysRevD.100.063538} {\bibfield  {journal} {\bibinfo  {journal} {\prd}\ }\textbf {\bibinfo {volume} {100}},\ \bibinfo {eid} {063538} (\bibinfo {year} {2019})},\ \Eprint {https://arxiv.org/abs/1904.07881} {arXiv:1904.07881 [astro-ph.CO]} \BibitemShut {NoStop}%
\bibitem [{\citenamefont {{Fialkov}}\ \emph {et~al.}(2013)\citenamefont {{Fialkov}}, \citenamefont {{Barkana}}, \citenamefont {{Visbal}}, \citenamefont {{Tseliakhovich}},\ and\ \citenamefont {{Hirata}}}]{2013MNRAS.432.2909F}%
  \BibitemOpen
  \bibfield  {author} {\bibinfo {author} {\bibfnamefont {A.}~\bibnamefont {{Fialkov}}}, \bibinfo {author} {\bibfnamefont {R.}~\bibnamefont {{Barkana}}}, \bibinfo {author} {\bibfnamefont {E.}~\bibnamefont {{Visbal}}}, \bibinfo {author} {\bibfnamefont {D.}~\bibnamefont {{Tseliakhovich}}},\ and\ \bibinfo {author} {\bibfnamefont {C.~M.}\ \bibnamefont {{Hirata}}},\ }\bibfield  {title} {\bibinfo {title} {{The 21-cm signature of the first stars during the Lyman-Werner feedback era}},\ }\href {https://doi.org/10.1093/mnras/stt650} {\bibfield  {journal} {\bibinfo  {journal} {\mnras}\ }\textbf {\bibinfo {volume} {432}},\ \bibinfo {pages} {2909} (\bibinfo {year} {2013})},\ \Eprint {https://arxiv.org/abs/1212.0513} {arXiv:1212.0513 [astro-ph.CO]} \BibitemShut {NoStop}%
\bibitem [{\citenamefont {{Mu{\~n}oz}}(2023)}]{Munoz2023}%
  \BibitemOpen
  \bibfield  {author} {\bibinfo {author} {\bibfnamefont {J.~B.}\ \bibnamefont {{Mu{\~n}oz}}},\ }\bibfield  {title} {\bibinfo {title} {{An effective model for the cosmic-dawn 21-cm signal}},\ }\href {https://doi.org/10.1093/mnras/stad1512} {\bibfield  {journal} {\bibinfo  {journal} {\mnras}\ }\textbf {\bibinfo {volume} {523}},\ \bibinfo {pages} {2587} (\bibinfo {year} {2023})},\ \Eprint {https://arxiv.org/abs/2302.08506} {arXiv:2302.08506 [astro-ph.CO]} \BibitemShut {NoStop}%
\bibitem [{\citenamefont {{Watts}}\ and\ \citenamefont {{CLASS Collaboration}}(2018)}]{CLASS}%
  \BibitemOpen
  \bibfield  {author} {\bibinfo {author} {\bibfnamefont {D.}~\bibnamefont {{Watts}}}\ and\ \bibinfo {author} {\bibnamefont {{CLASS Collaboration}}},\ }\bibfield  {title} {\bibinfo {title} {{Cosmology with CLASS}},\ }in\ \href@noop {} {\emph {\bibinfo {booktitle} {American Astronomical Society Meeting Abstracts \#231}}},\ \bibinfo {series} {American Astronomical Society Meeting Abstracts}, Vol.\ \bibinfo {volume} {231}\ (\bibinfo {year} {2018})\ p.\ \bibinfo {pages} {261.02}\BibitemShut {NoStop}%
\bibitem [{\citenamefont {{Wang}}\ and\ \citenamefont {{White}}(2007)}]{Wang2007}%
  \BibitemOpen
  \bibfield  {author} {\bibinfo {author} {\bibfnamefont {J.}~\bibnamefont {{Wang}}}\ and\ \bibinfo {author} {\bibfnamefont {S.~D.~M.}\ \bibnamefont {{White}}},\ }\bibfield  {title} {\bibinfo {title} {{Discreteness effects in simulations of hot/warm dark matter}},\ }\href {https://doi.org/10.1111/j.1365-2966.2007.12053.x} {\bibfield  {journal} {\bibinfo  {journal} {\mnras}\ }\textbf {\bibinfo {volume} {380}},\ \bibinfo {pages} {93} (\bibinfo {year} {2007})},\ \Eprint {https://arxiv.org/abs/astro-ph/0702575} {arXiv:astro-ph/0702575 [astro-ph]} \BibitemShut {NoStop}%
\bibitem [{\citenamefont {Sheth}\ \emph {et~al.}(2001)\citenamefont {Sheth}, \citenamefont {Mo},\ and\ \citenamefont {Tormen}}]{Sheth2001}%
  \BibitemOpen
  \bibfield  {author} {\bibinfo {author} {\bibfnamefont {R.~K.}\ \bibnamefont {Sheth}}, \bibinfo {author} {\bibfnamefont {H.~J.}\ \bibnamefont {Mo}},\ and\ \bibinfo {author} {\bibfnamefont {G.}~\bibnamefont {Tormen}},\ }\bibfield  {title} {\bibinfo {title} {Ellipsoidal collapse and an improved model for the number and spatial distribution of dark matter haloes},\ }\href {https://doi.org/10.1046/j.1365-8711.2001.04006.x} {\bibfield  {journal} {\bibinfo  {journal} {Mon. Not. R. Astron. Soc.}\ }\textbf {\bibinfo {volume} {323}},\ \bibinfo {pages} {1} (\bibinfo {year} {2001})}\BibitemShut {NoStop}%
\bibitem [{\citenamefont {{Leo}}\ \emph {et~al.}(2018)\citenamefont {{Leo}}, \citenamefont {{Baugh}}, \citenamefont {{Li}},\ and\ \citenamefont {{Pascoli}}}]{Leo2018}%
  \BibitemOpen
  \bibfield  {author} {\bibinfo {author} {\bibfnamefont {M.}~\bibnamefont {{Leo}}}, \bibinfo {author} {\bibfnamefont {C.~M.}\ \bibnamefont {{Baugh}}}, \bibinfo {author} {\bibfnamefont {B.}~\bibnamefont {{Li}}},\ and\ \bibinfo {author} {\bibfnamefont {S.}~\bibnamefont {{Pascoli}}},\ }\bibfield  {title} {\bibinfo {title} {{A new smooth-k space filter approach to calculate halo abundances}},\ }\href {https://doi.org/10.1088/1475-7516/2018/04/010} {\bibfield  {journal} {\bibinfo  {journal} {\jcap}\ }\textbf {\bibinfo {volume} {2018}},\ \bibinfo {eid} {010} (\bibinfo {year} {2018})},\ \Eprint {https://arxiv.org/abs/1801.02547} {arXiv:1801.02547 [astro-ph.CO]} \BibitemShut {NoStop}%
\bibitem [{\citenamefont {Sheth}\ and\ \citenamefont {Tormen}(1999)}]{Sheth1999}%
  \BibitemOpen
  \bibfield  {author} {\bibinfo {author} {\bibfnamefont {R.~K.}\ \bibnamefont {Sheth}}\ and\ \bibinfo {author} {\bibfnamefont {G.}~\bibnamefont {Tormen}},\ }\bibfield  {title} {\bibinfo {title} {Large-scale bias and the peak background split},\ }\href {https://doi.org/10.1046/j.1365-8711.1999.02692.x} {\bibfield  {journal} {\bibinfo  {journal} {Mon. Not. R. Astron. Soc.}\ }\textbf {\bibinfo {volume} {308}},\ \bibinfo {pages} {119} (\bibinfo {year} {1999})}\BibitemShut {NoStop}%
\bibitem [{\citenamefont {{Bohr}}\ \emph {et~al.}(2021)\citenamefont {{Bohr}}, \citenamefont {{Zavala}}, \citenamefont {{Cyr-Racine}},\ and\ \citenamefont {{Vogelsberger}}}]{Bohr2021}%
  \BibitemOpen
  \bibfield  {author} {\bibinfo {author} {\bibfnamefont {S.}~\bibnamefont {{Bohr}}}, \bibinfo {author} {\bibfnamefont {J.}~\bibnamefont {{Zavala}}}, \bibinfo {author} {\bibfnamefont {F.-Y.}\ \bibnamefont {{Cyr-Racine}}},\ and\ \bibinfo {author} {\bibfnamefont {M.}~\bibnamefont {{Vogelsberger}}},\ }\bibfield  {title} {\bibinfo {title} {{The halo mass function and inner structure of ETHOS haloes at high redshift}},\ }\href@noop {} {\bibfield  {journal} {\bibinfo  {journal} {arXiv e-prints}\ ,\ \bibinfo {eid} {arXiv:2101.08790}} (\bibinfo {year} {2021})},\ \Eprint {https://arxiv.org/abs/2101.08790} {arXiv:2101.08790 [astro-ph.CO]} \BibitemShut {NoStop}%
\bibitem [{\citenamefont {Schneider}(2018)}]{Schneider2018a}%
  \BibitemOpen
  \bibfield  {author} {\bibinfo {author} {\bibfnamefont {A.}~\bibnamefont {Schneider}},\ }\bibfield  {title} {\bibinfo {title} {Constraining {{Non}}-{{Cold Dark Matter Models}} with the {{Global}} 21-cm {{Signal}}},\ }\href {https://doi.org/10.1103/PhysRevD.98.063021} {\bibfield  {journal} {\bibinfo  {journal} {Physical Review D}\ }\textbf {\bibinfo {volume} {98}},\ \bibinfo {pages} {063021} (\bibinfo {year} {2018})},\ \Eprint {https://arxiv.org/abs/1805.00021} {arXiv:1805.00021} \BibitemShut {NoStop}%
\bibitem [{\citenamefont {{Pritchard}}\ and\ \citenamefont {{Furlanetto}}(2007)}]{2007MNRAS.376.1680P}%
  \BibitemOpen
  \bibfield  {author} {\bibinfo {author} {\bibfnamefont {J.~R.}\ \bibnamefont {{Pritchard}}}\ and\ \bibinfo {author} {\bibfnamefont {S.~R.}\ \bibnamefont {{Furlanetto}}},\ }\bibfield  {title} {\bibinfo {title} {{21-cm fluctuations from inhomogeneous X-ray heating before reionization}},\ }\href {https://doi.org/10.1111/j.1365-2966.2007.11519.x} {\bibfield  {journal} {\bibinfo  {journal} {\mnras}\ }\textbf {\bibinfo {volume} {376}},\ \bibinfo {pages} {1680} (\bibinfo {year} {2007})},\ \Eprint {https://arxiv.org/abs/astro-ph/0607234} {arXiv:astro-ph/0607234 [astro-ph]} \BibitemShut {NoStop}%
\bibitem [{\citenamefont {{Moster}}\ \emph {et~al.}(2010)\citenamefont {{Moster}}, \citenamefont {{Somerville}}, \citenamefont {{Maulbetsch}}, \citenamefont {{van den Bosch}}, \citenamefont {{Macci{\`o}}}, \citenamefont {{Naab}},\ and\ \citenamefont {{Oser}}}]{Moster2010}%
  \BibitemOpen
  \bibfield  {author} {\bibinfo {author} {\bibfnamefont {B.~P.}\ \bibnamefont {{Moster}}}, \bibinfo {author} {\bibfnamefont {R.~S.}\ \bibnamefont {{Somerville}}}, \bibinfo {author} {\bibfnamefont {C.}~\bibnamefont {{Maulbetsch}}}, \bibinfo {author} {\bibfnamefont {F.~C.}\ \bibnamefont {{van den Bosch}}}, \bibinfo {author} {\bibfnamefont {A.~V.}\ \bibnamefont {{Macci{\`o}}}}, \bibinfo {author} {\bibfnamefont {T.}~\bibnamefont {{Naab}}},\ and\ \bibinfo {author} {\bibfnamefont {L.}~\bibnamefont {{Oser}}},\ }\bibfield  {title} {\bibinfo {title} {{Constraints on the Relationship between Stellar Mass and Halo Mass at Low and High Redshift}},\ }\href {https://doi.org/10.1088/0004-637X/710/2/903} {\bibfield  {journal} {\bibinfo  {journal} {\apj}\ }\textbf {\bibinfo {volume} {710}},\ \bibinfo {pages} {903} (\bibinfo {year} {2010})},\ \Eprint {https://arxiv.org/abs/0903.4682} {arXiv:0903.4682 [astro-ph.CO]} \BibitemShut {NoStop}%
\bibitem [{\citenamefont {Behroozi}\ \emph {et~al.}(2013)\citenamefont {Behroozi}, \citenamefont {Marchesini}, \citenamefont {Wechsler}, \citenamefont {Muzzin}, \citenamefont {Papovich},\ and\ \citenamefont {Stefanon}}]{Behroozi2013a}%
  \BibitemOpen
  \bibfield  {author} {\bibinfo {author} {\bibfnamefont {P.~S.}\ \bibnamefont {Behroozi}}, \bibinfo {author} {\bibfnamefont {D.}~\bibnamefont {Marchesini}}, \bibinfo {author} {\bibfnamefont {R.~H.}\ \bibnamefont {Wechsler}}, \bibinfo {author} {\bibfnamefont {A.}~\bibnamefont {Muzzin}}, \bibinfo {author} {\bibfnamefont {C.}~\bibnamefont {Papovich}},\ and\ \bibinfo {author} {\bibfnamefont {M.}~\bibnamefont {Stefanon}},\ }\bibfield  {title} {\bibinfo {title} {Using {{Cumulative Number Densities To Compare Galaxies Across Cosmic Time}}},\ }\href {https://doi.org/10.1088/2041-8205/777/1/L10} {\bibfield  {journal} {\bibinfo  {journal} {Astrophys. J.}\ }\textbf {\bibinfo {volume} {777}},\ \bibinfo {pages} {L10} (\bibinfo {year} {2013})}\BibitemShut {NoStop}%
\bibitem [{\citenamefont {{Tacchella}}\ \emph {et~al.}(2018)\citenamefont {{Tacchella}}, \citenamefont {{Bose}}, \citenamefont {{Conroy}}, \citenamefont {{Eisenstein}},\ and\ \citenamefont {{Johnson}}}]{Tacchella2018a}%
  \BibitemOpen
  \bibfield  {author} {\bibinfo {author} {\bibfnamefont {S.}~\bibnamefont {{Tacchella}}}, \bibinfo {author} {\bibfnamefont {S.}~\bibnamefont {{Bose}}}, \bibinfo {author} {\bibfnamefont {C.}~\bibnamefont {{Conroy}}}, \bibinfo {author} {\bibfnamefont {D.~J.}\ \bibnamefont {{Eisenstein}}},\ and\ \bibinfo {author} {\bibfnamefont {B.~D.}\ \bibnamefont {{Johnson}}},\ }\bibfield  {title} {\bibinfo {title} {{A Redshift-independent Efficiency Model: Star Formation and Stellar Masses in Dark Matter Halos at z {\ensuremath{\gtrsim}} 4}},\ }\href {https://doi.org/10.3847/1538-4357/aae8e0} {\bibfield  {journal} {\bibinfo  {journal} {\apj}\ }\textbf {\bibinfo {volume} {868}},\ \bibinfo {eid} {92} (\bibinfo {year} {2018})},\ \Eprint {https://arxiv.org/abs/1806.03299} {arXiv:1806.03299 [astro-ph.GA]} \BibitemShut {NoStop}%
\bibitem [{\citenamefont {{Sabti}}\ \emph {et~al.}(2021)\citenamefont {{Sabti}}, \citenamefont {{Mu{\~n}oz}},\ and\ \citenamefont {{Blas}}}]{Sabti2021}%
  \BibitemOpen
  \bibfield  {author} {\bibinfo {author} {\bibfnamefont {N.}~\bibnamefont {{Sabti}}}, \bibinfo {author} {\bibfnamefont {J.~B.}\ \bibnamefont {{Mu{\~n}oz}}},\ and\ \bibinfo {author} {\bibfnamefont {D.}~\bibnamefont {{Blas}}},\ }\bibfield  {title} {\bibinfo {title} {{First constraints on small-scale non-Gaussianity from UV galaxy luminosity functions}},\ }\href {https://doi.org/10.1088/1475-7516/2021/01/010} {\bibfield  {journal} {\bibinfo  {journal} {\jcap}\ }\textbf {\bibinfo {volume} {2021}},\ \bibinfo {eid} {010} (\bibinfo {year} {2021})},\ \Eprint {https://arxiv.org/abs/2009.01245} {arXiv:2009.01245 [astro-ph.CO]} \BibitemShut {NoStop}%
\bibitem [{\citenamefont {{Barkana}}\ and\ \citenamefont {{Loeb}}(2005)}]{Barkana2005}%
  \BibitemOpen
  \bibfield  {author} {\bibinfo {author} {\bibfnamefont {R.}~\bibnamefont {{Barkana}}}\ and\ \bibinfo {author} {\bibfnamefont {A.}~\bibnamefont {{Loeb}}},\ }\bibfield  {title} {\bibinfo {title} {{Detecting the Earliest Galaxies through Two New Sources of 21 Centimeter Fluctuations}},\ }\href {https://doi.org/10.1086/429954} {\bibfield  {journal} {\bibinfo  {journal} {\apj}\ }\textbf {\bibinfo {volume} {626}},\ \bibinfo {pages} {1} (\bibinfo {year} {2005})},\ \Eprint {https://arxiv.org/abs/astro-ph/0410129} {arXiv:astro-ph/0410129 [astro-ph]} \BibitemShut {NoStop}%
\bibitem [{\citenamefont {{Mason}}\ \emph {et~al.}(2022)\citenamefont {{Mason}}, \citenamefont {{Mu{\~n}oz}}, \citenamefont {{Greig}}, \citenamefont {{Mesinger}},\ and\ \citenamefont {{Park}}}]{Mason2022}%
  \BibitemOpen
  \bibfield  {author} {\bibinfo {author} {\bibfnamefont {C.~A.}\ \bibnamefont {{Mason}}}, \bibinfo {author} {\bibfnamefont {J.~B.}\ \bibnamefont {{Mu{\~n}oz}}}, \bibinfo {author} {\bibfnamefont {B.}~\bibnamefont {{Greig}}}, \bibinfo {author} {\bibfnamefont {A.}~\bibnamefont {{Mesinger}}},\ and\ \bibinfo {author} {\bibfnamefont {J.}~\bibnamefont {{Park}}},\ }\bibfield  {title} {\bibinfo {title} {{21cmfish: Fisher-matrix framework for fast parameter forecasts from the cosmic 21-cm signal}},\ }\href@noop {} {\bibfield  {journal} {\bibinfo  {journal} {arXiv e-prints}\ ,\ \bibinfo {eid} {arXiv:2212.09797}} (\bibinfo {year} {2022})},\ \Eprint {https://arxiv.org/abs/2212.09797} {arXiv:2212.09797 [astro-ph.CO]} \BibitemShut {NoStop}%
\bibitem [{\citenamefont {{Pober}}\ \emph {et~al.}(2014)\citenamefont {{Pober}}, \citenamefont {{Liu}}, \citenamefont {{Dillon}}, \citenamefont {{Aguirre}}, \citenamefont {{Bowman}}, \citenamefont {{Bradley}}, \citenamefont {{Carilli}}, \citenamefont {{DeBoer}}, \citenamefont {{Hewitt}}, \citenamefont {{Jacobs}}, \citenamefont {{McQuinn}}, \citenamefont {{Morales}}, \citenamefont {{Parsons}}, \citenamefont {{Tegmark}},\ and\ \citenamefont {{Werthimer}}}]{Pober2014}%
  \BibitemOpen
  \bibfield  {author} {\bibinfo {author} {\bibfnamefont {J.~C.}\ \bibnamefont {{Pober}}}, \bibinfo {author} {\bibfnamefont {A.}~\bibnamefont {{Liu}}}, \bibinfo {author} {\bibfnamefont {J.~S.}\ \bibnamefont {{Dillon}}}, \bibinfo {author} {\bibfnamefont {J.~E.}\ \bibnamefont {{Aguirre}}}, \bibinfo {author} {\bibfnamefont {J.~D.}\ \bibnamefont {{Bowman}}}, \bibinfo {author} {\bibfnamefont {R.~F.}\ \bibnamefont {{Bradley}}}, \bibinfo {author} {\bibfnamefont {C.~L.}\ \bibnamefont {{Carilli}}}, \bibinfo {author} {\bibfnamefont {D.~R.}\ \bibnamefont {{DeBoer}}}, \bibinfo {author} {\bibfnamefont {J.~N.}\ \bibnamefont {{Hewitt}}}, \bibinfo {author} {\bibfnamefont {D.~C.}\ \bibnamefont {{Jacobs}}}, \bibinfo {author} {\bibfnamefont {M.}~\bibnamefont {{McQuinn}}}, \bibinfo {author} {\bibfnamefont {M.~F.}\ \bibnamefont {{Morales}}}, \bibinfo {author} {\bibfnamefont {A.~R.}\ \bibnamefont {{Parsons}}}, \bibinfo {author} {\bibfnamefont {M.}~\bibnamefont {{Tegmark}}},\ and\ \bibinfo {author} {\bibfnamefont {D.~J.}\
  \bibnamefont {{Werthimer}}},\ }\bibfield  {title} {\bibinfo {title} {{What Next-generation 21 cm Power Spectrum Measurements can Teach us About the Epoch of Reionization}},\ }\href {https://doi.org/10.1088/0004-637X/782/2/66} {\bibfield  {journal} {\bibinfo  {journal} {\apj}\ }\textbf {\bibinfo {volume} {782}},\ \bibinfo {eid} {66} (\bibinfo {year} {2014})},\ \Eprint {https://arxiv.org/abs/1310.7031} {arXiv:1310.7031 [astro-ph.CO]} \BibitemShut {NoStop}%
\bibitem [{\citenamefont {{Foreman-Mackey}}\ \emph {et~al.}(2013)\citenamefont {{Foreman-Mackey}}, \citenamefont {Hogg}, \citenamefont {Lang},\ and\ \citenamefont {Goodman}}]{Foreman-Mackey2013}%
  \BibitemOpen
  \bibfield  {author} {\bibinfo {author} {\bibfnamefont {D.}~\bibnamefont {{Foreman-Mackey}}}, \bibinfo {author} {\bibfnamefont {D.~W.}\ \bibnamefont {Hogg}}, \bibinfo {author} {\bibfnamefont {D.}~\bibnamefont {Lang}},\ and\ \bibinfo {author} {\bibfnamefont {J.}~\bibnamefont {Goodman}},\ }\bibfield  {title} {\bibinfo {title} {Emcee : {{The MCMC Hammer}}},\ }\href {https://doi.org/10.1086/670067} {\bibfield  {journal} {\bibinfo  {journal} {Publ. Astron. Soc. Pacific}\ }\textbf {\bibinfo {volume} {125}},\ \bibinfo {pages} {306} (\bibinfo {year} {2013})}\BibitemShut {NoStop}%
\bibitem [{\citenamefont {{DeBoer}}\ \emph {et~al.}(2017)\citenamefont {{DeBoer}}, \citenamefont {{Parsons}}, \citenamefont {{Aguirre}}, \citenamefont {{Alexander}}, \citenamefont {{Ali}}, \citenamefont {{Beardsley}}, \citenamefont {{Bernardi}}, \citenamefont {{Bowman}}, \citenamefont {{Bradley}}, \citenamefont {{Carilli}}, \citenamefont {{Cheng}}, \citenamefont {{de Lera Acedo}}, \citenamefont {{Dillon}}, \citenamefont {{Ewall-Wice}}, \citenamefont {{Fadana}}, \citenamefont {{Fagnoni}}, \citenamefont {{Fritz}}, \citenamefont {{Furlanetto}}, \citenamefont {{Glendenning}}, \citenamefont {{Greig}}, \citenamefont {{Grobbelaar}}, \citenamefont {{Hazelton}}, \citenamefont {{Hewitt}}, \citenamefont {{Hickish}}, \citenamefont {{Jacobs}}, \citenamefont {{Julius}}, \citenamefont {{Kariseb}}, \citenamefont {{Kohn}}, \citenamefont {{Lekalake}}, \citenamefont {{Liu}}, \citenamefont {{Loots}}, \citenamefont {{MacMahon}}, \citenamefont {{Malan}}, \citenamefont {{Malgas}}, \citenamefont {{Maree}}, \citenamefont
  {{Martinot}}, \citenamefont {{Mathison}}, \citenamefont {{Matsetela}}, \citenamefont {{Mesinger}}, \citenamefont {{Morales}}, \citenamefont {{Neben}}, \citenamefont {{Patra}}, \citenamefont {{Pieterse}}, \citenamefont {{Pober}}, \citenamefont {{Razavi-Ghods}}, \citenamefont {{Ringuette}}, \citenamefont {{Robnett}}, \citenamefont {{Rosie}}, \citenamefont {{Sell}}, \citenamefont {{Smith}}, \citenamefont {{Syce}}, \citenamefont {{Tegmark}}, \citenamefont {{Thyagarajan}}, \citenamefont {{Williams}},\ and\ \citenamefont {{Zheng}}}]{DeBoer2017}%
  \BibitemOpen
  \bibfield  {author} {\bibinfo {author} {\bibfnamefont {D.~R.}\ \bibnamefont {{DeBoer}}}, \bibinfo {author} {\bibfnamefont {A.~R.}\ \bibnamefont {{Parsons}}}, \bibinfo {author} {\bibfnamefont {J.~E.}\ \bibnamefont {{Aguirre}}}, \bibinfo {author} {\bibfnamefont {P.}~\bibnamefont {{Alexander}}}, \bibinfo {author} {\bibfnamefont {Z.~S.}\ \bibnamefont {{Ali}}}, \bibinfo {author} {\bibfnamefont {A.~P.}\ \bibnamefont {{Beardsley}}}, \bibinfo {author} {\bibfnamefont {G.}~\bibnamefont {{Bernardi}}}, \bibinfo {author} {\bibfnamefont {J.~D.}\ \bibnamefont {{Bowman}}}, \bibinfo {author} {\bibfnamefont {R.~F.}\ \bibnamefont {{Bradley}}}, \bibinfo {author} {\bibfnamefont {C.~L.}\ \bibnamefont {{Carilli}}}, \bibinfo {author} {\bibfnamefont {C.}~\bibnamefont {{Cheng}}}, \bibinfo {author} {\bibfnamefont {E.}~\bibnamefont {{de Lera Acedo}}}, \bibinfo {author} {\bibfnamefont {J.~S.}\ \bibnamefont {{Dillon}}}, \bibinfo {author} {\bibfnamefont {A.}~\bibnamefont {{Ewall-Wice}}}, \bibinfo {author} {\bibfnamefont {G.}~\bibnamefont
  {{Fadana}}}, \bibinfo {author} {\bibfnamefont {N.}~\bibnamefont {{Fagnoni}}}, \bibinfo {author} {\bibfnamefont {R.}~\bibnamefont {{Fritz}}}, \bibinfo {author} {\bibfnamefont {S.~R.}\ \bibnamefont {{Furlanetto}}}, \bibinfo {author} {\bibfnamefont {B.}~\bibnamefont {{Glendenning}}}, \bibinfo {author} {\bibfnamefont {B.}~\bibnamefont {{Greig}}}, \bibinfo {author} {\bibfnamefont {J.}~\bibnamefont {{Grobbelaar}}}, \bibinfo {author} {\bibfnamefont {B.~J.}\ \bibnamefont {{Hazelton}}}, \bibinfo {author} {\bibfnamefont {J.~N.}\ \bibnamefont {{Hewitt}}}, \bibinfo {author} {\bibfnamefont {J.}~\bibnamefont {{Hickish}}}, \bibinfo {author} {\bibfnamefont {D.~C.}\ \bibnamefont {{Jacobs}}}, \bibinfo {author} {\bibfnamefont {A.}~\bibnamefont {{Julius}}}, \bibinfo {author} {\bibfnamefont {M.}~\bibnamefont {{Kariseb}}}, \bibinfo {author} {\bibfnamefont {S.~A.}\ \bibnamefont {{Kohn}}}, \bibinfo {author} {\bibfnamefont {T.}~\bibnamefont {{Lekalake}}}, \bibinfo {author} {\bibfnamefont {A.}~\bibnamefont {{Liu}}}, \bibinfo
  {author} {\bibfnamefont {A.}~\bibnamefont {{Loots}}}, \bibinfo {author} {\bibfnamefont {D.}~\bibnamefont {{MacMahon}}}, \bibinfo {author} {\bibfnamefont {L.}~\bibnamefont {{Malan}}}, \bibinfo {author} {\bibfnamefont {C.}~\bibnamefont {{Malgas}}}, \bibinfo {author} {\bibfnamefont {M.}~\bibnamefont {{Maree}}}, \bibinfo {author} {\bibfnamefont {Z.}~\bibnamefont {{Martinot}}}, \bibinfo {author} {\bibfnamefont {N.}~\bibnamefont {{Mathison}}}, \bibinfo {author} {\bibfnamefont {E.}~\bibnamefont {{Matsetela}}}, \bibinfo {author} {\bibfnamefont {A.}~\bibnamefont {{Mesinger}}}, \bibinfo {author} {\bibfnamefont {M.~F.}\ \bibnamefont {{Morales}}}, \bibinfo {author} {\bibfnamefont {A.~R.}\ \bibnamefont {{Neben}}}, \bibinfo {author} {\bibfnamefont {N.}~\bibnamefont {{Patra}}}, \bibinfo {author} {\bibfnamefont {S.}~\bibnamefont {{Pieterse}}}, \bibinfo {author} {\bibfnamefont {J.~C.}\ \bibnamefont {{Pober}}}, \bibinfo {author} {\bibfnamefont {N.}~\bibnamefont {{Razavi-Ghods}}}, \bibinfo {author} {\bibfnamefont
  {J.}~\bibnamefont {{Ringuette}}}, \bibinfo {author} {\bibfnamefont {J.}~\bibnamefont {{Robnett}}}, \bibinfo {author} {\bibfnamefont {K.}~\bibnamefont {{Rosie}}}, \bibinfo {author} {\bibfnamefont {R.}~\bibnamefont {{Sell}}}, \bibinfo {author} {\bibfnamefont {C.}~\bibnamefont {{Smith}}}, \bibinfo {author} {\bibfnamefont {A.}~\bibnamefont {{Syce}}}, \bibinfo {author} {\bibfnamefont {M.}~\bibnamefont {{Tegmark}}}, \bibinfo {author} {\bibfnamefont {N.}~\bibnamefont {{Thyagarajan}}}, \bibinfo {author} {\bibfnamefont {P.~K.~G.}\ \bibnamefont {{Williams}}},\ and\ \bibinfo {author} {\bibfnamefont {H.}~\bibnamefont {{Zheng}}},\ }\bibfield  {title} {\bibinfo {title} {{Hydrogen Epoch of Reionization Array (HERA)}},\ }\href {https://doi.org/10.1088/1538-3873/129/974/045001} {\bibfield  {journal} {\bibinfo  {journal} {\pasp}\ }\textbf {\bibinfo {volume} {129}},\ \bibinfo {pages} {045001} (\bibinfo {year} {2017})},\ \Eprint {https://arxiv.org/abs/1606.07473} {arXiv:1606.07473 [astro-ph.IM]} \BibitemShut {NoStop}%
\bibitem [{\citenamefont {{Abdurashidova}}\ \emph {et~al.}(2022)\citenamefont {{Abdurashidova}}, \citenamefont {{Aguirre}}, \citenamefont {{Alexander}}, \citenamefont {{Ali}}, \citenamefont {{Balfour}}, \citenamefont {{Beardsley}}, \citenamefont {{Bernardi}}, \citenamefont {{Billings}}, \citenamefont {{Bowman}}, \citenamefont {{Bradley}}, \citenamefont {{Bull}}, \citenamefont {{Burba}}, \citenamefont {{Carey}}, \citenamefont {{Carilli}}, \citenamefont {{Cheng}}, \citenamefont {{DeBoer}}, \citenamefont {{Dexter}}, \citenamefont {{de Lera Acedo}}, \citenamefont {{Dibblee-Barkman}}, \citenamefont {{Dillon}}, \citenamefont {{Ely}}, \citenamefont {{Ewall-Wice}}, \citenamefont {{Fagnoni}}, \citenamefont {{Fritz}}, \citenamefont {{Furlanetto}}, \citenamefont {{Gale-Sides}}, \citenamefont {{Glendenning}}, \citenamefont {{Gorthi}}, \citenamefont {{Greig}}, \citenamefont {{Grobbelaar}}, \citenamefont {{Halday}}, \citenamefont {{Hazelton}}, \citenamefont {{Hewitt}}, \citenamefont {{Hickish}}, \citenamefont {{Jacobs}},
  \citenamefont {{Julius}}, \citenamefont {{Kern}}, \citenamefont {{Kerrigan}}, \citenamefont {{Kittiwisit}}, \citenamefont {{Kohn}}, \citenamefont {{Kolopanis}}, \citenamefont {{Lanman}}, \citenamefont {{La Plante}}, \citenamefont {{Lekalake}}, \citenamefont {{Lewis}}, \citenamefont {{Liu}}, \citenamefont {{MacMahon}}, \citenamefont {{Malan}}, \citenamefont {{Malgas}}, \citenamefont {{Maree}}, \citenamefont {{Martinot}}, \citenamefont {{Matsetela}}, \citenamefont {{Mesinger}}, \citenamefont {{Molewa}}, \citenamefont {{Morales}}, \citenamefont {{Mosiane}}, \citenamefont {{Murray}}, \citenamefont {{Neben}}, \citenamefont {{Nikolic}}, \citenamefont {{Nunhokee}}, \citenamefont {{Parsons}}, \citenamefont {{Patra}}, \citenamefont {{Pascua}}, \citenamefont {{Pieterse}}, \citenamefont {{Pober}}, \citenamefont {{Razavi-Ghods}}, \citenamefont {{Ringuette}}, \citenamefont {{Robnett}}, \citenamefont {{Rosie}}, \citenamefont {{Sims}}, \citenamefont {{Singh}}, \citenamefont {{Smith}}, \citenamefont {{Syce}}, \citenamefont
  {{Thyagarajan}}, \citenamefont {{Williams}}, \citenamefont {{Zheng}},\ and\ \citenamefont {{HERA Collaboration}}}]{2022ApJ...925..221A}%
  \BibitemOpen
  \bibfield  {author} {\bibinfo {author} {\bibfnamefont {Z.}~\bibnamefont {{Abdurashidova}}}, \bibinfo {author} {\bibfnamefont {J.~E.}\ \bibnamefont {{Aguirre}}}, \bibinfo {author} {\bibfnamefont {P.}~\bibnamefont {{Alexander}}}, \bibinfo {author} {\bibfnamefont {Z.~S.}\ \bibnamefont {{Ali}}}, \bibinfo {author} {\bibfnamefont {Y.}~\bibnamefont {{Balfour}}}, \bibinfo {author} {\bibfnamefont {A.~P.}\ \bibnamefont {{Beardsley}}}, \bibinfo {author} {\bibfnamefont {G.}~\bibnamefont {{Bernardi}}}, \bibinfo {author} {\bibfnamefont {T.~S.}\ \bibnamefont {{Billings}}}, \bibinfo {author} {\bibfnamefont {J.~D.}\ \bibnamefont {{Bowman}}}, \bibinfo {author} {\bibfnamefont {R.~F.}\ \bibnamefont {{Bradley}}}, \bibinfo {author} {\bibfnamefont {P.}~\bibnamefont {{Bull}}}, \bibinfo {author} {\bibfnamefont {J.}~\bibnamefont {{Burba}}}, \bibinfo {author} {\bibfnamefont {S.}~\bibnamefont {{Carey}}}, \bibinfo {author} {\bibfnamefont {C.~L.}\ \bibnamefont {{Carilli}}}, \bibinfo {author} {\bibfnamefont {C.}~\bibnamefont {{Cheng}}},
  \bibinfo {author} {\bibfnamefont {D.~R.}\ \bibnamefont {{DeBoer}}}, \bibinfo {author} {\bibfnamefont {M.}~\bibnamefont {{Dexter}}}, \bibinfo {author} {\bibfnamefont {E.}~\bibnamefont {{de Lera Acedo}}}, \bibinfo {author} {\bibfnamefont {T.}~\bibnamefont {{Dibblee-Barkman}}}, \bibinfo {author} {\bibfnamefont {J.~S.}\ \bibnamefont {{Dillon}}}, \bibinfo {author} {\bibfnamefont {J.}~\bibnamefont {{Ely}}}, \bibinfo {author} {\bibfnamefont {A.}~\bibnamefont {{Ewall-Wice}}}, \bibinfo {author} {\bibfnamefont {N.}~\bibnamefont {{Fagnoni}}}, \bibinfo {author} {\bibfnamefont {R.}~\bibnamefont {{Fritz}}}, \bibinfo {author} {\bibfnamefont {S.~R.}\ \bibnamefont {{Furlanetto}}}, \bibinfo {author} {\bibfnamefont {K.}~\bibnamefont {{Gale-Sides}}}, \bibinfo {author} {\bibfnamefont {B.}~\bibnamefont {{Glendenning}}}, \bibinfo {author} {\bibfnamefont {D.}~\bibnamefont {{Gorthi}}}, \bibinfo {author} {\bibfnamefont {B.}~\bibnamefont {{Greig}}}, \bibinfo {author} {\bibfnamefont {J.}~\bibnamefont {{Grobbelaar}}}, \bibinfo {author}
  {\bibfnamefont {Z.}~\bibnamefont {{Halday}}}, \bibinfo {author} {\bibfnamefont {B.~J.}\ \bibnamefont {{Hazelton}}}, \bibinfo {author} {\bibfnamefont {J.~N.}\ \bibnamefont {{Hewitt}}}, \bibinfo {author} {\bibfnamefont {J.}~\bibnamefont {{Hickish}}}, \bibinfo {author} {\bibfnamefont {D.~C.}\ \bibnamefont {{Jacobs}}}, \bibinfo {author} {\bibfnamefont {A.}~\bibnamefont {{Julius}}}, \bibinfo {author} {\bibfnamefont {N.~S.}\ \bibnamefont {{Kern}}}, \bibinfo {author} {\bibfnamefont {J.}~\bibnamefont {{Kerrigan}}}, \bibinfo {author} {\bibfnamefont {P.}~\bibnamefont {{Kittiwisit}}}, \bibinfo {author} {\bibfnamefont {S.~A.}\ \bibnamefont {{Kohn}}}, \bibinfo {author} {\bibfnamefont {M.}~\bibnamefont {{Kolopanis}}}, \bibinfo {author} {\bibfnamefont {A.}~\bibnamefont {{Lanman}}}, \bibinfo {author} {\bibfnamefont {P.}~\bibnamefont {{La Plante}}}, \bibinfo {author} {\bibfnamefont {T.}~\bibnamefont {{Lekalake}}}, \bibinfo {author} {\bibfnamefont {D.}~\bibnamefont {{Lewis}}}, \bibinfo {author} {\bibfnamefont
  {A.}~\bibnamefont {{Liu}}}, \bibinfo {author} {\bibfnamefont {D.}~\bibnamefont {{MacMahon}}}, \bibinfo {author} {\bibfnamefont {L.}~\bibnamefont {{Malan}}}, \bibinfo {author} {\bibfnamefont {C.}~\bibnamefont {{Malgas}}}, \bibinfo {author} {\bibfnamefont {M.}~\bibnamefont {{Maree}}}, \bibinfo {author} {\bibfnamefont {Z.~E.}\ \bibnamefont {{Martinot}}}, \bibinfo {author} {\bibfnamefont {E.}~\bibnamefont {{Matsetela}}}, \bibinfo {author} {\bibfnamefont {A.}~\bibnamefont {{Mesinger}}}, \bibinfo {author} {\bibfnamefont {M.}~\bibnamefont {{Molewa}}}, \bibinfo {author} {\bibfnamefont {M.~F.}\ \bibnamefont {{Morales}}}, \bibinfo {author} {\bibfnamefont {T.}~\bibnamefont {{Mosiane}}}, \bibinfo {author} {\bibfnamefont {S.~G.}\ \bibnamefont {{Murray}}}, \bibinfo {author} {\bibfnamefont {A.~R.}\ \bibnamefont {{Neben}}}, \bibinfo {author} {\bibfnamefont {B.}~\bibnamefont {{Nikolic}}}, \bibinfo {author} {\bibfnamefont {C.~D.}\ \bibnamefont {{Nunhokee}}}, \bibinfo {author} {\bibfnamefont {A.~R.}\ \bibnamefont
  {{Parsons}}}, \bibinfo {author} {\bibfnamefont {N.}~\bibnamefont {{Patra}}}, \bibinfo {author} {\bibfnamefont {R.}~\bibnamefont {{Pascua}}}, \bibinfo {author} {\bibfnamefont {S.}~\bibnamefont {{Pieterse}}}, \bibinfo {author} {\bibfnamefont {J.~C.}\ \bibnamefont {{Pober}}}, \bibinfo {author} {\bibfnamefont {N.}~\bibnamefont {{Razavi-Ghods}}}, \bibinfo {author} {\bibfnamefont {J.}~\bibnamefont {{Ringuette}}}, \bibinfo {author} {\bibfnamefont {J.}~\bibnamefont {{Robnett}}}, \bibinfo {author} {\bibfnamefont {K.}~\bibnamefont {{Rosie}}}, \bibinfo {author} {\bibfnamefont {P.}~\bibnamefont {{Sims}}}, \bibinfo {author} {\bibfnamefont {S.}~\bibnamefont {{Singh}}}, \bibinfo {author} {\bibfnamefont {C.}~\bibnamefont {{Smith}}}, \bibinfo {author} {\bibfnamefont {A.}~\bibnamefont {{Syce}}}, \bibinfo {author} {\bibfnamefont {N.}~\bibnamefont {{Thyagarajan}}}, \bibinfo {author} {\bibfnamefont {P.~K.~G.}\ \bibnamefont {{Williams}}}, \bibinfo {author} {\bibfnamefont {H.}~\bibnamefont {{Zheng}}},\ and\ \bibinfo {author}
  {\bibnamefont {{HERA Collaboration}}},\ }\bibfield  {title} {\bibinfo {title} {{First Results from HERA Phase I: Upper Limits on the Epoch of Reionization 21 cm Power Spectrum}},\ }\href {https://doi.org/10.3847/1538-4357/ac1c78} {\bibfield  {journal} {\bibinfo  {journal} {\apj}\ }\textbf {\bibinfo {volume} {925}},\ \bibinfo {eid} {221} (\bibinfo {year} {2022})},\ \Eprint {https://arxiv.org/abs/2108.02263} {arXiv:2108.02263 [astro-ph.CO]} \BibitemShut {NoStop}%
\bibitem [{\citenamefont {{Berkhout}}\ \emph {et~al.}(2024)\citenamefont {{Berkhout}}, \citenamefont {{Jacobs}}, \citenamefont {{Abdurashidova}}, \citenamefont {{Adams}}, \citenamefont {{Aguirre}}, \citenamefont {{Alexander}}, \citenamefont {{Ali}}, \citenamefont {{Baartman}}, \citenamefont {{Balfour}}, \citenamefont {{Beardsley}}, \citenamefont {{Bernardi}}, \citenamefont {{Billings}}, \citenamefont {{Bowman}}, \citenamefont {{Bradley}}, \citenamefont {{Bull}}, \citenamefont {{Burba}}, \citenamefont {{Carey}}, \citenamefont {{Carilli}}, \citenamefont {{Chen}}, \citenamefont {{Cheng}}, \citenamefont {{Choudhuri}}, \citenamefont {{DeBoer}}, \citenamefont {{de Lera Acedo}}, \citenamefont {{Dexter}}, \citenamefont {{Dillon}}, \citenamefont {{Dynes}}, \citenamefont {{Eksteen}}, \citenamefont {{Ely}}, \citenamefont {{Ewall-Wice}}, \citenamefont {{Fagnoni}}, \citenamefont {{Fritz}}, \citenamefont {{Furlanetto}}, \citenamefont {{Gale-Sides}}, \citenamefont {{Garsden}}, \citenamefont {{Gehlot}}, \citenamefont
  {{Ghosh}}, \citenamefont {{Glendenning}}, \citenamefont {{Gorce}}, \citenamefont {{Gorthi}}, \citenamefont {{Greig}}, \citenamefont {{Grobbelaar}}, \citenamefont {{Halday}}, \citenamefont {{Hazelton}}, \citenamefont {{Hewitt}}, \citenamefont {{Hickish}}, \citenamefont {{Huang}}, \citenamefont {{Josaitis}}, \citenamefont {{Julius}}, \citenamefont {{Kariseb}}, \citenamefont {{Kern}}, \citenamefont {{Kerrigan}}, \citenamefont {{Kim}}, \citenamefont {{Kittiwisit}}, \citenamefont {{Kohn}}, \citenamefont {{Kolopanis}}, \citenamefont {{Lanman}}, \citenamefont {{La Plante}}, \citenamefont {{Liu}}, \citenamefont {{Loots}}, \citenamefont {{Ma}}, \citenamefont {{MacMahon}}, \citenamefont {{Malan}}, \citenamefont {{Malgas}}, \citenamefont {{Malgas}}, \citenamefont {{Marero}}, \citenamefont {{Martinot}}, \citenamefont {{Mesinger}}, \citenamefont {{Molewa}}, \citenamefont {{Morales}}, \citenamefont {{Mosiane}}, \citenamefont {{Murray}}, \citenamefont {{Neben}}, \citenamefont {{Nikolic}}, \citenamefont {{Devi Nunhokee}},
  \citenamefont {{Nuwegeld}}, \citenamefont {{Parsons}}, \citenamefont {{Pascua}}, \citenamefont {{Patra}}, \citenamefont {{Pieterse}}, \citenamefont {{Qin}}, \citenamefont {{Rath}}, \citenamefont {{Razavi-Ghods}}, \citenamefont {{Riley}}, \citenamefont {{Robnett}}, \citenamefont {{Rosie}}, \citenamefont {{Santos}}, \citenamefont {{Sims}}, \citenamefont {{Singh}}, \citenamefont {{Storer}}, \citenamefont {{Swarts}}, \citenamefont {{Tan}}, \citenamefont {{Thyagarajan}}, \citenamefont {{van Wyngaarden}}, \citenamefont {{Williams}}, \citenamefont {{Zheng}},\ and\ \citenamefont {{Xu}}}]{2024arXiv240104304B}%
  \BibitemOpen
  \bibfield  {author} {\bibinfo {author} {\bibfnamefont {L.~M.}\ \bibnamefont {{Berkhout}}}, \bibinfo {author} {\bibfnamefont {D.~C.}\ \bibnamefont {{Jacobs}}}, \bibinfo {author} {\bibfnamefont {Z.}~\bibnamefont {{Abdurashidova}}}, \bibinfo {author} {\bibfnamefont {T.}~\bibnamefont {{Adams}}}, \bibinfo {author} {\bibfnamefont {J.~E.}\ \bibnamefont {{Aguirre}}}, \bibinfo {author} {\bibfnamefont {P.}~\bibnamefont {{Alexander}}}, \bibinfo {author} {\bibfnamefont {Z.~S.}\ \bibnamefont {{Ali}}}, \bibinfo {author} {\bibfnamefont {R.}~\bibnamefont {{Baartman}}}, \bibinfo {author} {\bibfnamefont {Y.}~\bibnamefont {{Balfour}}}, \bibinfo {author} {\bibfnamefont {A.~P.}\ \bibnamefont {{Beardsley}}}, \bibinfo {author} {\bibfnamefont {G.}~\bibnamefont {{Bernardi}}}, \bibinfo {author} {\bibfnamefont {T.~S.}\ \bibnamefont {{Billings}}}, \bibinfo {author} {\bibfnamefont {J.~D.}\ \bibnamefont {{Bowman}}}, \bibinfo {author} {\bibfnamefont {R.~F.}\ \bibnamefont {{Bradley}}}, \bibinfo {author} {\bibfnamefont {P.}~\bibnamefont
  {{Bull}}}, \bibinfo {author} {\bibfnamefont {J.}~\bibnamefont {{Burba}}}, \bibinfo {author} {\bibfnamefont {S.}~\bibnamefont {{Carey}}}, \bibinfo {author} {\bibfnamefont {C.~L.}\ \bibnamefont {{Carilli}}}, \bibinfo {author} {\bibfnamefont {K.-F.}\ \bibnamefont {{Chen}}}, \bibinfo {author} {\bibfnamefont {C.}~\bibnamefont {{Cheng}}}, \bibinfo {author} {\bibfnamefont {S.}~\bibnamefont {{Choudhuri}}}, \bibinfo {author} {\bibfnamefont {D.~R.}\ \bibnamefont {{DeBoer}}}, \bibinfo {author} {\bibfnamefont {E.}~\bibnamefont {{de Lera Acedo}}}, \bibinfo {author} {\bibfnamefont {M.}~\bibnamefont {{Dexter}}}, \bibinfo {author} {\bibfnamefont {J.~S.}\ \bibnamefont {{Dillon}}}, \bibinfo {author} {\bibfnamefont {S.}~\bibnamefont {{Dynes}}}, \bibinfo {author} {\bibfnamefont {N.}~\bibnamefont {{Eksteen}}}, \bibinfo {author} {\bibfnamefont {J.}~\bibnamefont {{Ely}}}, \bibinfo {author} {\bibfnamefont {A.}~\bibnamefont {{Ewall-Wice}}}, \bibinfo {author} {\bibfnamefont {N.}~\bibnamefont {{Fagnoni}}}, \bibinfo {author}
  {\bibfnamefont {R.}~\bibnamefont {{Fritz}}}, \bibinfo {author} {\bibfnamefont {S.~R.}\ \bibnamefont {{Furlanetto}}}, \bibinfo {author} {\bibfnamefont {K.}~\bibnamefont {{Gale-Sides}}}, \bibinfo {author} {\bibfnamefont {H.}~\bibnamefont {{Garsden}}}, \bibinfo {author} {\bibfnamefont {B.~K.}\ \bibnamefont {{Gehlot}}}, \bibinfo {author} {\bibfnamefont {A.}~\bibnamefont {{Ghosh}}}, \bibinfo {author} {\bibfnamefont {B.}~\bibnamefont {{Glendenning}}}, \bibinfo {author} {\bibfnamefont {A.}~\bibnamefont {{Gorce}}}, \bibinfo {author} {\bibfnamefont {D.}~\bibnamefont {{Gorthi}}}, \bibinfo {author} {\bibfnamefont {B.}~\bibnamefont {{Greig}}}, \bibinfo {author} {\bibfnamefont {J.}~\bibnamefont {{Grobbelaar}}}, \bibinfo {author} {\bibfnamefont {Z.}~\bibnamefont {{Halday}}}, \bibinfo {author} {\bibfnamefont {B.~J.}\ \bibnamefont {{Hazelton}}}, \bibinfo {author} {\bibfnamefont {J.~N.}\ \bibnamefont {{Hewitt}}}, \bibinfo {author} {\bibfnamefont {J.}~\bibnamefont {{Hickish}}}, \bibinfo {author} {\bibfnamefont
  {T.}~\bibnamefont {{Huang}}}, \bibinfo {author} {\bibfnamefont {A.}~\bibnamefont {{Josaitis}}}, \bibinfo {author} {\bibfnamefont {A.}~\bibnamefont {{Julius}}}, \bibinfo {author} {\bibfnamefont {M.}~\bibnamefont {{Kariseb}}}, \bibinfo {author} {\bibfnamefont {N.~S.}\ \bibnamefont {{Kern}}}, \bibinfo {author} {\bibfnamefont {J.}~\bibnamefont {{Kerrigan}}}, \bibinfo {author} {\bibfnamefont {H.}~\bibnamefont {{Kim}}}, \bibinfo {author} {\bibfnamefont {P.}~\bibnamefont {{Kittiwisit}}}, \bibinfo {author} {\bibfnamefont {S.~A.}\ \bibnamefont {{Kohn}}}, \bibinfo {author} {\bibfnamefont {M.}~\bibnamefont {{Kolopanis}}}, \bibinfo {author} {\bibfnamefont {A.}~\bibnamefont {{Lanman}}}, \bibinfo {author} {\bibfnamefont {P.}~\bibnamefont {{La Plante}}}, \bibinfo {author} {\bibfnamefont {A.}~\bibnamefont {{Liu}}}, \bibinfo {author} {\bibfnamefont {A.}~\bibnamefont {{Loots}}}, \bibinfo {author} {\bibfnamefont {Y.-Z.}\ \bibnamefont {{Ma}}}, \bibinfo {author} {\bibfnamefont {D.~H.~E.}\ \bibnamefont {{MacMahon}}}, \bibinfo
  {author} {\bibfnamefont {L.}~\bibnamefont {{Malan}}}, \bibinfo {author} {\bibfnamefont {C.}~\bibnamefont {{Malgas}}}, \bibinfo {author} {\bibfnamefont {K.}~\bibnamefont {{Malgas}}}, \bibinfo {author} {\bibfnamefont {B.}~\bibnamefont {{Marero}}}, \bibinfo {author} {\bibfnamefont {Z.~E.}\ \bibnamefont {{Martinot}}}, \bibinfo {author} {\bibfnamefont {A.}~\bibnamefont {{Mesinger}}}, \bibinfo {author} {\bibfnamefont {M.}~\bibnamefont {{Molewa}}}, \bibinfo {author} {\bibfnamefont {M.~F.}\ \bibnamefont {{Morales}}}, \bibinfo {author} {\bibfnamefont {T.}~\bibnamefont {{Mosiane}}}, \bibinfo {author} {\bibfnamefont {S.~G.}\ \bibnamefont {{Murray}}}, \bibinfo {author} {\bibfnamefont {A.~R.}\ \bibnamefont {{Neben}}}, \bibinfo {author} {\bibfnamefont {B.}~\bibnamefont {{Nikolic}}}, \bibinfo {author} {\bibfnamefont {C.}~\bibnamefont {{Devi Nunhokee}}}, \bibinfo {author} {\bibfnamefont {H.}~\bibnamefont {{Nuwegeld}}}, \bibinfo {author} {\bibfnamefont {A.~R.}\ \bibnamefont {{Parsons}}}, \bibinfo {author} {\bibfnamefont
  {R.}~\bibnamefont {{Pascua}}}, \bibinfo {author} {\bibfnamefont {N.}~\bibnamefont {{Patra}}}, \bibinfo {author} {\bibfnamefont {S.}~\bibnamefont {{Pieterse}}}, \bibinfo {author} {\bibfnamefont {Y.}~\bibnamefont {{Qin}}}, \bibinfo {author} {\bibfnamefont {E.}~\bibnamefont {{Rath}}}, \bibinfo {author} {\bibfnamefont {N.}~\bibnamefont {{Razavi-Ghods}}}, \bibinfo {author} {\bibfnamefont {D.}~\bibnamefont {{Riley}}}, \bibinfo {author} {\bibfnamefont {J.}~\bibnamefont {{Robnett}}}, \bibinfo {author} {\bibfnamefont {K.}~\bibnamefont {{Rosie}}}, \bibinfo {author} {\bibfnamefont {M.~G.}\ \bibnamefont {{Santos}}}, \bibinfo {author} {\bibfnamefont {P.}~\bibnamefont {{Sims}}}, \bibinfo {author} {\bibfnamefont {S.}~\bibnamefont {{Singh}}}, \bibinfo {author} {\bibfnamefont {D.}~\bibnamefont {{Storer}}}, \bibinfo {author} {\bibfnamefont {H.}~\bibnamefont {{Swarts}}}, \bibinfo {author} {\bibfnamefont {J.}~\bibnamefont {{Tan}}}, \bibinfo {author} {\bibfnamefont {N.}~\bibnamefont {{Thyagarajan}}}, \bibinfo {author}
  {\bibfnamefont {P.}~\bibnamefont {{van Wyngaarden}}}, \bibinfo {author} {\bibfnamefont {P.~K.~G.}\ \bibnamefont {{Williams}}}, \bibinfo {author} {\bibfnamefont {H.}~\bibnamefont {{Zheng}}},\ and\ \bibinfo {author} {\bibfnamefont {Z.}~\bibnamefont {{Xu}}},\ }\bibfield  {title} {\bibinfo {title} {{Hydrogen Epoch of Reionization Array (HERA) Phase II Deployment and Commissioning}},\ }\href {https://doi.org/10.48550/arXiv.2401.04304} {\bibfield  {journal} {\bibinfo  {journal} {arXiv e-prints}\ ,\ \bibinfo {eid} {arXiv:2401.04304}} (\bibinfo {year} {2024})},\ \Eprint {https://arxiv.org/abs/2401.04304} {arXiv:2401.04304 [astro-ph.IM]} \BibitemShut {NoStop}%
\bibitem [{\citenamefont {{Liu}}\ \emph {et~al.}(2009)\citenamefont {{Liu}}, \citenamefont {{Tegmark}}, \citenamefont {{Bowman}}, \citenamefont {{Hewitt}},\ and\ \citenamefont {{Zaldarriaga}}}]{2009MNRAS.398..401L}%
  \BibitemOpen
  \bibfield  {author} {\bibinfo {author} {\bibfnamefont {A.}~\bibnamefont {{Liu}}}, \bibinfo {author} {\bibfnamefont {M.}~\bibnamefont {{Tegmark}}}, \bibinfo {author} {\bibfnamefont {J.}~\bibnamefont {{Bowman}}}, \bibinfo {author} {\bibfnamefont {J.}~\bibnamefont {{Hewitt}}},\ and\ \bibinfo {author} {\bibfnamefont {M.}~\bibnamefont {{Zaldarriaga}}},\ }\bibfield  {title} {\bibinfo {title} {{An improved method for 21-cm foreground removal}},\ }\href {https://doi.org/10.1111/j.1365-2966.2009.15156.x} {\bibfield  {journal} {\bibinfo  {journal} {\mnras}\ }\textbf {\bibinfo {volume} {398}},\ \bibinfo {pages} {401} (\bibinfo {year} {2009})},\ \Eprint {https://arxiv.org/abs/0903.4890} {arXiv:0903.4890 [astro-ph.CO]} \BibitemShut {NoStop}%
\bibitem [{\citenamefont {{Datta}}\ \emph {et~al.}(2010)\citenamefont {{Datta}}, \citenamefont {{Bowman}},\ and\ \citenamefont {{Carilli}}}]{2010ApJ...724..526D}%
  \BibitemOpen
  \bibfield  {author} {\bibinfo {author} {\bibfnamefont {A.}~\bibnamefont {{Datta}}}, \bibinfo {author} {\bibfnamefont {J.~D.}\ \bibnamefont {{Bowman}}},\ and\ \bibinfo {author} {\bibfnamefont {C.~L.}\ \bibnamefont {{Carilli}}},\ }\bibfield  {title} {\bibinfo {title} {{Bright Source Subtraction Requirements for Redshifted 21 cm Measurements}},\ }\href {https://doi.org/10.1088/0004-637X/724/1/526} {\bibfield  {journal} {\bibinfo  {journal} {\apj}\ }\textbf {\bibinfo {volume} {724}},\ \bibinfo {pages} {526} (\bibinfo {year} {2010})},\ \Eprint {https://arxiv.org/abs/1005.4071} {arXiv:1005.4071 [astro-ph.CO]} \BibitemShut {NoStop}%
\bibitem [{\citenamefont {Sarkar}\ \emph {et~al.}(2022)\citenamefont {Sarkar}, \citenamefont {Flitter},\ and\ \citenamefont {Kovetz}}]{Sarkar:2022dvl}%
  \BibitemOpen
  \bibfield  {author} {\bibinfo {author} {\bibfnamefont {D.}~\bibnamefont {Sarkar}}, \bibinfo {author} {\bibfnamefont {J.}~\bibnamefont {Flitter}},\ and\ \bibinfo {author} {\bibfnamefont {E.~D.}\ \bibnamefont {Kovetz}},\ }\bibfield  {title} {\bibinfo {title} {{Exploring delaying and heating effects on the 21-cm signature of fuzzy dark matter}},\ }\href {https://doi.org/10.1103/PhysRevD.105.103529} {\bibfield  {journal} {\bibinfo  {journal} {Phys. Rev. D}\ }\textbf {\bibinfo {volume} {105}},\ \bibinfo {pages} {103529} (\bibinfo {year} {2022})},\ \Eprint {https://arxiv.org/abs/2201.03355} {arXiv:2201.03355 [astro-ph.CO]} \BibitemShut {NoStop}%
\bibitem [{\citenamefont {{Mason}}\ \emph {et~al.}(2023)\citenamefont {{Mason}}, \citenamefont {{Mu{\~n}oz}}, \citenamefont {{Greig}}, \citenamefont {{Mesinger}},\ and\ \citenamefont {{Park}}}]{Mason23}%
  \BibitemOpen
  \bibfield  {author} {\bibinfo {author} {\bibfnamefont {C.~A.}\ \bibnamefont {{Mason}}}, \bibinfo {author} {\bibfnamefont {J.~B.}\ \bibnamefont {{Mu{\~n}oz}}}, \bibinfo {author} {\bibfnamefont {B.}~\bibnamefont {{Greig}}}, \bibinfo {author} {\bibfnamefont {A.}~\bibnamefont {{Mesinger}}},\ and\ \bibinfo {author} {\bibfnamefont {J.}~\bibnamefont {{Park}}},\ }\bibfield  {title} {\bibinfo {title} {{21CMFISH: Fisher-matrix framework for fast parameter forecasts from the cosmic 21-cm signal}},\ }\href {https://doi.org/10.1093/mnras/stad2145} {\bibfield  {journal} {\bibinfo  {journal} {\mnras}\ }\textbf {\bibinfo {volume} {524}},\ \bibinfo {pages} {4711} (\bibinfo {year} {2023})},\ \Eprint {https://arxiv.org/abs/2212.09797} {arXiv:2212.09797 [astro-ph.CO]} \BibitemShut {NoStop}%
\bibitem [{\citenamefont {{Mesinger}}\ and\ \citenamefont {{Furlanetto}}(2007)}]{Mesinger2007}%
  \BibitemOpen
  \bibfield  {author} {\bibinfo {author} {\bibfnamefont {A.}~\bibnamefont {{Mesinger}}}\ and\ \bibinfo {author} {\bibfnamefont {S.}~\bibnamefont {{Furlanetto}}},\ }\bibfield  {title} {\bibinfo {title} {{Efficient Simulations of Early Structure Formation and Reionization}},\ }\href {https://doi.org/10.1086/521806} {\bibfield  {journal} {\bibinfo  {journal} {\apj}\ }\textbf {\bibinfo {volume} {669}},\ \bibinfo {pages} {663} (\bibinfo {year} {2007})},\ \Eprint {https://arxiv.org/abs/0704.0946} {arXiv:0704.0946 [astro-ph]} \BibitemShut {NoStop}%
\bibitem [{\citenamefont {{Mesinger}}\ \emph {et~al.}(2011)\citenamefont {{Mesinger}}, \citenamefont {{Furlanetto}},\ and\ \citenamefont {{Cen}}}]{2011MNRAS.411..955M}%
  \BibitemOpen
  \bibfield  {author} {\bibinfo {author} {\bibfnamefont {A.}~\bibnamefont {{Mesinger}}}, \bibinfo {author} {\bibfnamefont {S.}~\bibnamefont {{Furlanetto}}},\ and\ \bibinfo {author} {\bibfnamefont {R.}~\bibnamefont {{Cen}}},\ }\bibfield  {title} {\bibinfo {title} {{21CMFAST: a fast, seminumerical simulation of the high-redshift 21-cm signal}},\ }\href {https://doi.org/10.1111/j.1365-2966.2010.17731.x} {\bibfield  {journal} {\bibinfo  {journal} {\mnras}\ }\textbf {\bibinfo {volume} {411}},\ \bibinfo {pages} {955} (\bibinfo {year} {2011})},\ \Eprint {https://arxiv.org/abs/1003.3878} {arXiv:1003.3878 [astro-ph.CO]} \BibitemShut {NoStop}%
\bibitem [{\citenamefont {{Murray}}\ \emph {et~al.}(2020)\citenamefont {{Murray}}, \citenamefont {{Greig}}, \citenamefont {{Mesinger}}, \citenamefont {{Mu{\~n}oz}}, \citenamefont {{Qin}}, \citenamefont {{Park}},\ and\ \citenamefont {{Watkinson}}}]{Murray2021}%
  \BibitemOpen
  \bibfield  {author} {\bibinfo {author} {\bibfnamefont {S.}~\bibnamefont {{Murray}}}, \bibinfo {author} {\bibfnamefont {B.}~\bibnamefont {{Greig}}}, \bibinfo {author} {\bibfnamefont {A.}~\bibnamefont {{Mesinger}}}, \bibinfo {author} {\bibfnamefont {J.}~\bibnamefont {{Mu{\~n}oz}}}, \bibinfo {author} {\bibfnamefont {Y.}~\bibnamefont {{Qin}}}, \bibinfo {author} {\bibfnamefont {J.}~\bibnamefont {{Park}}},\ and\ \bibinfo {author} {\bibfnamefont {C.}~\bibnamefont {{Watkinson}}},\ }\bibfield  {title} {\bibinfo {title} {{21cmFAST v3: A Python-integrated C code for generating 3D realizations of the cosmic 21cm signal.}},\ }\href {https://doi.org/10.21105/joss.02582} {\bibfield  {journal} {\bibinfo  {journal} {The Journal of Open Source Software}\ }\textbf {\bibinfo {volume} {5}},\ \bibinfo {eid} {2582} (\bibinfo {year} {2020})},\ \Eprint {https://arxiv.org/abs/2010.15121} {arXiv:2010.15121 [astro-ph.IM]} \BibitemShut {NoStop}%
\bibitem [{\citenamefont {{Mu{\~n}oz}}\ \emph {et~al.}(2022)\citenamefont {{Mu{\~n}oz}}, \citenamefont {{Qin}}, \citenamefont {{Mesinger}}, \citenamefont {{Murray}}, \citenamefont {{Greig}},\ and\ \citenamefont {{Mason}}}]{Munoz2022}%
  \BibitemOpen
  \bibfield  {author} {\bibinfo {author} {\bibfnamefont {J.~B.}\ \bibnamefont {{Mu{\~n}oz}}}, \bibinfo {author} {\bibfnamefont {Y.}~\bibnamefont {{Qin}}}, \bibinfo {author} {\bibfnamefont {A.}~\bibnamefont {{Mesinger}}}, \bibinfo {author} {\bibfnamefont {S.~G.}\ \bibnamefont {{Murray}}}, \bibinfo {author} {\bibfnamefont {B.}~\bibnamefont {{Greig}}},\ and\ \bibinfo {author} {\bibfnamefont {C.}~\bibnamefont {{Mason}}},\ }\bibfield  {title} {\bibinfo {title} {{The impact of the first galaxies on cosmic dawn and reionization}},\ }\href {https://doi.org/10.1093/mnras/stac185} {\bibfield  {journal} {\bibinfo  {journal} {\mnras}\ }\textbf {\bibinfo {volume} {511}},\ \bibinfo {pages} {3657} (\bibinfo {year} {2022})},\ \Eprint {https://arxiv.org/abs/2110.13919} {arXiv:2110.13919 [astro-ph.CO]} \BibitemShut {NoStop}%
\bibitem [{\citenamefont {{Endsley}}\ \emph {et~al.}(2023)\citenamefont {{Endsley}}, \citenamefont {{Stark}}, \citenamefont {{Whitler}}, \citenamefont {{Topping}}, \citenamefont {{Johnson}}, \citenamefont {{Robertson}}, \citenamefont {{Tacchella}}, \citenamefont {{Alberts}}, \citenamefont {{Baker}}, \citenamefont {{Bhatawdekar}}, \citenamefont {{Boyett}}, \citenamefont {{Bunker}}, \citenamefont {{Cameron}}, \citenamefont {{Carniani}}, \citenamefont {{Charlot}}, \citenamefont {{Chen}}, \citenamefont {{Chevallard}}, \citenamefont {{Curtis-Lake}}, \citenamefont {{Danhaive}}, \citenamefont {{Egami}}, \citenamefont {{Eisenstein}}, \citenamefont {{Hainline}}, \citenamefont {{Helton}}, \citenamefont {{Ji}}, \citenamefont {{Looser}}, \citenamefont {{Maiolino}}, \citenamefont {{Nelson}}, \citenamefont {{Pusk{\'a}s}}, \citenamefont {{Rieke}}, \citenamefont {{Rieke}}, \citenamefont {{Rix}}, \citenamefont {{Sandles}}, \citenamefont {{Saxena}}, \citenamefont {{Simmonds}}, \citenamefont {{Smit}}, \citenamefont {{Sun}},
  \citenamefont {{Williams}}, \citenamefont {{Willmer}}, \citenamefont {{Willott}},\ and\ \citenamefont {{Witstok}}}]{Endsley2023}%
  \BibitemOpen
  \bibfield  {author} {\bibinfo {author} {\bibfnamefont {R.}~\bibnamefont {{Endsley}}}, \bibinfo {author} {\bibfnamefont {D.~P.}\ \bibnamefont {{Stark}}}, \bibinfo {author} {\bibfnamefont {L.}~\bibnamefont {{Whitler}}}, \bibinfo {author} {\bibfnamefont {M.~W.}\ \bibnamefont {{Topping}}}, \bibinfo {author} {\bibfnamefont {B.~D.}\ \bibnamefont {{Johnson}}}, \bibinfo {author} {\bibfnamefont {B.}~\bibnamefont {{Robertson}}}, \bibinfo {author} {\bibfnamefont {S.}~\bibnamefont {{Tacchella}}}, \bibinfo {author} {\bibfnamefont {S.}~\bibnamefont {{Alberts}}}, \bibinfo {author} {\bibfnamefont {W.~M.}\ \bibnamefont {{Baker}}}, \bibinfo {author} {\bibfnamefont {R.}~\bibnamefont {{Bhatawdekar}}}, \bibinfo {author} {\bibfnamefont {K.}~\bibnamefont {{Boyett}}}, \bibinfo {author} {\bibfnamefont {A.~J.}\ \bibnamefont {{Bunker}}}, \bibinfo {author} {\bibfnamefont {A.~J.}\ \bibnamefont {{Cameron}}}, \bibinfo {author} {\bibfnamefont {S.}~\bibnamefont {{Carniani}}}, \bibinfo {author} {\bibfnamefont {S.}~\bibnamefont {{Charlot}}},
  \bibinfo {author} {\bibfnamefont {Z.}~\bibnamefont {{Chen}}}, \bibinfo {author} {\bibfnamefont {J.}~\bibnamefont {{Chevallard}}}, \bibinfo {author} {\bibfnamefont {E.}~\bibnamefont {{Curtis-Lake}}}, \bibinfo {author} {\bibfnamefont {A.~L.}\ \bibnamefont {{Danhaive}}}, \bibinfo {author} {\bibfnamefont {E.}~\bibnamefont {{Egami}}}, \bibinfo {author} {\bibfnamefont {D.~J.}\ \bibnamefont {{Eisenstein}}}, \bibinfo {author} {\bibfnamefont {K.}~\bibnamefont {{Hainline}}}, \bibinfo {author} {\bibfnamefont {J.~M.}\ \bibnamefont {{Helton}}}, \bibinfo {author} {\bibfnamefont {Z.}~\bibnamefont {{Ji}}}, \bibinfo {author} {\bibfnamefont {T.~J.}\ \bibnamefont {{Looser}}}, \bibinfo {author} {\bibfnamefont {R.}~\bibnamefont {{Maiolino}}}, \bibinfo {author} {\bibfnamefont {E.}~\bibnamefont {{Nelson}}}, \bibinfo {author} {\bibfnamefont {D.}~\bibnamefont {{Pusk{\'a}s}}}, \bibinfo {author} {\bibfnamefont {G.}~\bibnamefont {{Rieke}}}, \bibinfo {author} {\bibfnamefont {M.}~\bibnamefont {{Rieke}}}, \bibinfo {author} {\bibfnamefont
  {H.-W.}\ \bibnamefont {{Rix}}}, \bibinfo {author} {\bibfnamefont {L.}~\bibnamefont {{Sandles}}}, \bibinfo {author} {\bibfnamefont {A.}~\bibnamefont {{Saxena}}}, \bibinfo {author} {\bibfnamefont {C.}~\bibnamefont {{Simmonds}}}, \bibinfo {author} {\bibfnamefont {R.}~\bibnamefont {{Smit}}}, \bibinfo {author} {\bibfnamefont {F.}~\bibnamefont {{Sun}}}, \bibinfo {author} {\bibfnamefont {C.~C.}\ \bibnamefont {{Williams}}}, \bibinfo {author} {\bibfnamefont {C.~N.~A.}\ \bibnamefont {{Willmer}}}, \bibinfo {author} {\bibfnamefont {C.}~\bibnamefont {{Willott}}},\ and\ \bibinfo {author} {\bibfnamefont {J.}~\bibnamefont {{Witstok}}},\ }\bibfield  {title} {\bibinfo {title} {{The Star-forming and Ionizing Properties of Dwarf z\raisebox{-0.5ex}\textasciitilde6-9 Galaxies in JADES: Insights on Bursty Star Formation and Ionized Bubble Growth}},\ }\href {https://doi.org/10.48550/arXiv.2306.05295} {\bibfield  {journal} {\bibinfo  {journal} {arXiv e-prints}\ ,\ \bibinfo {eid} {arXiv:2306.05295}} (\bibinfo {year} {2023})},\
  \Eprint {https://arxiv.org/abs/2306.05295} {arXiv:2306.05295 [astro-ph.GA]} \BibitemShut {NoStop}%
\bibitem [{\citenamefont {Foreman-Mackey}(2016)}]{corner}%
  \BibitemOpen
  \bibfield  {author} {\bibinfo {author} {\bibfnamefont {D.}~\bibnamefont {Foreman-Mackey}},\ }\bibfield  {title} {\bibinfo {title} {corner.py: Scatterplot matrices in python},\ }\href {https://doi.org/10.21105/joss.00024} {\bibfield  {journal} {\bibinfo  {journal} {The Journal of Open Source Software}\ }\textbf {\bibinfo {volume} {1}},\ \bibinfo {pages} {24} (\bibinfo {year} {2016})}\BibitemShut {NoStop}%
\bibitem [{\citenamefont {P{\'e}rez}\ and\ \citenamefont {Granger}(2007)}]{Perez2007a}%
  \BibitemOpen
  \bibfield  {author} {\bibinfo {author} {\bibfnamefont {F.}~\bibnamefont {P{\'e}rez}}\ and\ \bibinfo {author} {\bibfnamefont {B.~E.}\ \bibnamefont {Granger}},\ }\bibfield  {title} {\bibinfo {title} {{{IPython}}: {{A}} system for interactive scientific computing},\ }\href {https://doi.org/10.1109/MCSE.2007.53} {\bibfield  {journal} {\bibinfo  {journal} {Comput. Sci. Eng.}\ }\textbf {\bibinfo {volume} {9}},\ \bibinfo {pages} {21} (\bibinfo {year} {2007})}\BibitemShut {NoStop}%
\bibitem [{\citenamefont {Hunter}(2007)}]{Hunter2007a}%
  \BibitemOpen
  \bibfield  {author} {\bibinfo {author} {\bibfnamefont {J.~D.}\ \bibnamefont {Hunter}},\ }\bibfield  {title} {\bibinfo {title} {Matplotlib: {{A 2D}} graphics environment},\ }\href {https://doi.org/10.1109/MCSE.2007.55} {\bibfield  {journal} {\bibinfo  {journal} {Comput. Sci. Eng.}\ }\textbf {\bibinfo {volume} {9}},\ \bibinfo {pages} {99} (\bibinfo {year} {2007})}\BibitemShut {NoStop}%
\bibitem [{\citenamefont {Van Der~Walt}\ \emph {et~al.}(2011)\citenamefont {Van Der~Walt}, \citenamefont {Colbert},\ and\ \citenamefont {Varoquaux}}]{VanderWalt2011a}%
  \BibitemOpen
  \bibfield  {author} {\bibinfo {author} {\bibfnamefont {S.}~\bibnamefont {Van Der~Walt}}, \bibinfo {author} {\bibfnamefont {S.~C.}\ \bibnamefont {Colbert}},\ and\ \bibinfo {author} {\bibfnamefont {G.}~\bibnamefont {Varoquaux}},\ }\bibfield  {title} {\bibinfo {title} {The {{NumPy}} array: {{A}} structure for efficient numerical computation},\ }\href {https://doi.org/10.1109/MCSE.2011.37} {\bibfield  {journal} {\bibinfo  {journal} {Comput. Sci. Eng.}\ }\textbf {\bibinfo {volume} {13}},\ \bibinfo {pages} {22} (\bibinfo {year} {2011})}\BibitemShut {NoStop}%
\bibitem [{\citenamefont {Oliphant}(2007)}]{Oliphant2007a}%
  \BibitemOpen
  \bibfield  {author} {\bibinfo {author} {\bibfnamefont {T.~E.}\ \bibnamefont {Oliphant}},\ }\bibfield  {title} {\bibinfo {title} {{{SciPy}}: {{Open}} source scientific tools for {{Python}}},\ }\href {https://doi.org/10.1109/MCSE.2007.58} {\bibfield  {journal} {\bibinfo  {journal} {Comput. Sci. Eng.}\ }\textbf {\bibinfo {volume} {9}},\ \bibinfo {pages} {10} (\bibinfo {year} {2007})}\BibitemShut {NoStop}%
\bibitem [{\citenamefont {Robitaille}\ \emph {et~al.}(2013)\citenamefont {Robitaille}, \citenamefont {Tollerud}, \citenamefont {Greenfield}, \citenamefont {Droettboom}, \citenamefont {Bray}, \citenamefont {Aldcroft}, \citenamefont {Davis}, \citenamefont {Ginsburg}, \citenamefont {{Price-Whelan}}, \citenamefont {Kerzendorf}, \citenamefont {Conley}, \citenamefont {Crighton}, \citenamefont {Barbary}, \citenamefont {Muna}, \citenamefont {Ferguson}, \citenamefont {Grollier}, \citenamefont {Parikh}, \citenamefont {Nair}, \citenamefont {G{\"u}nther}, \citenamefont {Deil}, \citenamefont {Woillez}, \citenamefont {Conseil}, \citenamefont {Kramer}, \citenamefont {Turner}, \citenamefont {Singer}, \citenamefont {Fox}, \citenamefont {a.~Weaver}, \citenamefont {Zabalza}, \citenamefont {Edwards}, \citenamefont {Azalee~Bostroem}, \citenamefont {Burke}, \citenamefont {Casey}, \citenamefont {Crawford}, \citenamefont {Dencheva}, \citenamefont {Ely}, \citenamefont {Jenness}, \citenamefont {Labrie}, \citenamefont {Lim},
  \citenamefont {Pierfederici}, \citenamefont {Pontzen}, \citenamefont {Ptak}, \citenamefont {Refsdal}, \citenamefont {Servillat},\ and\ \citenamefont {Streicher}}]{Robitaille2013}%
  \BibitemOpen
  \bibfield  {author} {\bibinfo {author} {\bibfnamefont {T.~P.}\ \bibnamefont {Robitaille}}, \bibinfo {author} {\bibfnamefont {E.~J.}\ \bibnamefont {Tollerud}}, \bibinfo {author} {\bibfnamefont {P.}~\bibnamefont {Greenfield}}, \bibinfo {author} {\bibfnamefont {M.}~\bibnamefont {Droettboom}}, \bibinfo {author} {\bibfnamefont {E.}~\bibnamefont {Bray}}, \bibinfo {author} {\bibfnamefont {T.}~\bibnamefont {Aldcroft}}, \bibinfo {author} {\bibfnamefont {M.}~\bibnamefont {Davis}}, \bibinfo {author} {\bibfnamefont {A.}~\bibnamefont {Ginsburg}}, \bibinfo {author} {\bibfnamefont {A.~M.}\ \bibnamefont {{Price-Whelan}}}, \bibinfo {author} {\bibfnamefont {W.~E.}\ \bibnamefont {Kerzendorf}}, \bibinfo {author} {\bibfnamefont {A.}~\bibnamefont {Conley}}, \bibinfo {author} {\bibfnamefont {N.}~\bibnamefont {Crighton}}, \bibinfo {author} {\bibfnamefont {K.}~\bibnamefont {Barbary}}, \bibinfo {author} {\bibfnamefont {D.}~\bibnamefont {Muna}}, \bibinfo {author} {\bibfnamefont {H.}~\bibnamefont {Ferguson}}, \bibinfo {author}
  {\bibfnamefont {F.}~\bibnamefont {Grollier}}, \bibinfo {author} {\bibfnamefont {M.~M.}\ \bibnamefont {Parikh}}, \bibinfo {author} {\bibfnamefont {P.~H.}\ \bibnamefont {Nair}}, \bibinfo {author} {\bibfnamefont {H.~M.}\ \bibnamefont {G{\"u}nther}}, \bibinfo {author} {\bibfnamefont {C.}~\bibnamefont {Deil}}, \bibinfo {author} {\bibfnamefont {J.}~\bibnamefont {Woillez}}, \bibinfo {author} {\bibfnamefont {S.}~\bibnamefont {Conseil}}, \bibinfo {author} {\bibfnamefont {R.}~\bibnamefont {Kramer}}, \bibinfo {author} {\bibfnamefont {J.~E.~H.}\ \bibnamefont {Turner}}, \bibinfo {author} {\bibfnamefont {L.}~\bibnamefont {Singer}}, \bibinfo {author} {\bibfnamefont {R.}~\bibnamefont {Fox}}, \bibinfo {author} {\bibfnamefont {B.}~\bibnamefont {a.~Weaver}}, \bibinfo {author} {\bibfnamefont {V.}~\bibnamefont {Zabalza}}, \bibinfo {author} {\bibfnamefont {Z.~I.}\ \bibnamefont {Edwards}}, \bibinfo {author} {\bibfnamefont {K.}~\bibnamefont {Azalee~Bostroem}}, \bibinfo {author} {\bibfnamefont {D.~J.}\ \bibnamefont {Burke}},
  \bibinfo {author} {\bibfnamefont {A.~R.}\ \bibnamefont {Casey}}, \bibinfo {author} {\bibfnamefont {S.~M.}\ \bibnamefont {Crawford}}, \bibinfo {author} {\bibfnamefont {N.}~\bibnamefont {Dencheva}}, \bibinfo {author} {\bibfnamefont {J.}~\bibnamefont {Ely}}, \bibinfo {author} {\bibfnamefont {T.}~\bibnamefont {Jenness}}, \bibinfo {author} {\bibfnamefont {K.}~\bibnamefont {Labrie}}, \bibinfo {author} {\bibfnamefont {P.~L.}\ \bibnamefont {Lim}}, \bibinfo {author} {\bibfnamefont {F.}~\bibnamefont {Pierfederici}}, \bibinfo {author} {\bibfnamefont {A.}~\bibnamefont {Pontzen}}, \bibinfo {author} {\bibfnamefont {A.}~\bibnamefont {Ptak}}, \bibinfo {author} {\bibfnamefont {B.}~\bibnamefont {Refsdal}}, \bibinfo {author} {\bibfnamefont {M.}~\bibnamefont {Servillat}},\ and\ \bibinfo {author} {\bibfnamefont {O.}~\bibnamefont {Streicher}},\ }\bibfield  {title} {\bibinfo {title} {Astropy: {{A}} community {{Python}} package for astronomy},\ }\href {https://doi.org/10.1051/0004-6361/201322068} {\bibfield  {journal} {\bibinfo
  {journal} {Astron. Astrophys.}\ }\textbf {\bibinfo {volume} {558}},\ \bibinfo {pages} {A33} (\bibinfo {year} {2013})}\BibitemShut {NoStop}%
\end{thebibliography}%

\end{document}